**Distinguishing the aftershock sequences of the DPRK5 and DPRK6 underground nuclear tests**

Kitov, I.O.

**Abstract**

The rate of aftershocks in the sequence initiated by the DPRK underground tests has been increasing since January 2021. In total, 22 reliable aftershocks were detected between January 13 and October 1, 2021. Their characteristics are similar to the aftershocks in one of two clusters: 1) the fifth DPRK (DPRK5) test ($m_b$(IDC)=5.09) conducted on September 9, 2016, which induced the first DPRK aftershock in the sequence detected at 1:50:48 UTC on September 11, 2016; 2) the sixth DPRK (DPRK6) explosion ($m_b$(IDC)=6.07), which generates its aftershock sequence with characteristics significantly different from the aftershocks in the DPRK5 sequence. The length, intensity, and alternating character of these sequences suggest specific mechanisms of energy release likely associated with the interaction of the damaged zones of the DPRK5 and DPRK6 and the collapse of their cavities with progressive propagation of the collapsing chimneys to the free surface. According to the depth estimates based on the moment tensor modelling, the DPRK5 and DPRK6 were conducted at practically the same depths. The difference in magnitudes suggests that their damaged zones differ by a factor of 2 or more. The first aftershock of the DPRK6 ($m_b$(IDC=4.12) 8.5 minutes after the test is evidence of the cavity collapse and creation of a chimney, which did not reach the surface. The activity in 2021 indicates that the chimney collapse is not finished yet. One can expect more aftershocks in the near future, likely ended with the chimney reaching the free surface.

## Introduction

Between January and October 2021, more than 20 low magnitude seismic events were found by the International Data Centre (IDC) of the CTBTO which had locations close to the known aftershocks of the underground nuclear tests conducted by the DPRK. Table 1 lists dates and IDC locations for all six announced tests in the DPRK. The biggest measured aftershock in the whole sequence, likely manifesting the cavity collapse, occurred 8.5 min after the sixth DPRK explosion and had a body wave magnitude of 4.11. It was detected by many IMS stations at distances from 4 to 60 degrees, including several 3-C stations. However, almost all induced seismic events generated in the aftershock zone of the DPRK test site are so weak (magnitude below 3) that only two seismic stations of the International Monitoring System (IMS) USRK and KSRS can detect the $L_g$-waves and low-amplitude $P_n$-waves, some of them at the level of ambient seismic noise. Both stations are seismic arrays and have higher sensitivity and azimuth-slowness resolution than standard 3-C stations. For consistency, all seismic parameters in this study are estimated from data at these two stations, even if other stations may contribute to some studied events.

Table 1. Parameters of six DPRK test estimated by the IDC.

| # | Date | Time | Lat, deg | Lon, deg | Ndef | $m_b$ | Ms | ML |
|---|---|---|---|---|---|---|---|---|
| 1 | Oct 9, 2006 | 01:35:27.575 | 41.312 | 129.02 | 22 | 4.08 | - | 3.89 |
| 2 | May 25, 2009 | 00:54:42.802 | 41.311 | 129.05 | 72 | 4.51 | 3.56 | 4.27 |
| 3 | Feb 12, 2013 | 02:57:50.799 | 41.301 | 129.07 | 110 | 4.92 | 3.95 | 4.52 |
| 4 | Jan 6, 2016 | 01:30:00.494 | 41.304 | 129.05 | 102 | 4.82 | 3.92 | 4.61 |
| 5 | Sep 9, 2016 | 00:30:00.874 | 41.299 | 129.05 | 120 | 5.09 | 4.17 | 4.29 |
| 6 | Sep 3, 2017 | 03:30:01.080 | 41.320 | 129.03 | 189 | 6.07 | 4.91 | 5.17 |

Ndef - number of defining phases in the IDC solution

The induced DPRK events detected in 2021 represent an unusual observation for the aftershock sequences of the historical underground nuclear explosions with body wave magnitudes of 5 to 6 [Kitov and Kuznetsov, 1990; Adushkin and Spivak, 1993]. The length (5 years after the fifths DPRK test conducted on September 9, 2016, and 4 years after the DPRK6 conducted on September 3, 2017), and activity (~100 reliable event hypotheses detected at regional distances) of the DPRK aftershock sequence is a challenge for physical interpretation. This study is devoted to a consistent recovery of the sequence observed after September 9, 2016, including the smallest possible events, to the estimation of its principal characteristics, and to discrimination of the aftershocks associated with the hypocentre zones of the DPRK5 or DPRK6.

The possibility to divide the overall sequence into two clusters was first reported in [Kitov and Rozhkov, 2017; Kitov *et al*., 2018]. A tentative physical interpretation of the observed split into two sequences is based on the release of gravitational energy driven by the interaction of the damage zones of the DPRK5 and DPRK6, which have different linear sizes and the depths of burial near 1 km [Kitov *et al*., 2021]. This interaction makes the processes of energy release, as expressed by seismic events, continue for many years. The total energy released in the sequence so far may indicate that the process is not finished. Ultimately, the collapsing chimney of the DPRK5 and DPRK6 may reach the surface and create a crater in the years to come. Potential enhancement to the monitoring of the underground nuclear tests is the main goal of this study for the CTBTO. For the broader monitoring community, the study of the whole DPRK aftershock sequence, including the aftershocks of the DPRK3 and DPRK4 [Kitov *et al.,* 2021], is not only of theoretical interest to the unusual physical phenomenon but is also important for various empirical studies like atmospheric transport modelling, ATM, analysis of radionuclide detections, and satellite imaging of the Punggye-ri Nuclear Test Site.

## DPRK underground nuclear tests

All five underground nuclear tests conducted by the DPRK since February 12, 2009, were located within a circle of ~2.0 km in diameter. This extraordinary accuracy of the relative location was achieved by the waveform cross-correlation method [*e.g*., Bobrov *et al*., 2017b]. The DPRK1 was conducted in a different mountain approximately 2.5 km far from the DPRK2. When interpreting the events within the DPRK test site, we also use our estimates of the depth of burial [Rozhkov *et al*., 2018] and relative magnitude [Bobrov *et al*., 2017a]. For the DPRK5 and DPRK6, the depth of burial is estimated between 1 km and 1.5 km. The relative magnitude estimates for these two explosions are close to the IDC estimates in Table 1.

The estimates of ML and Ms in Table 1 reveal an important observation that the $m_b$/ML ratio is lower for the DPRK6 (1.17) than for the DPRK5 (1.19), and the $m_b$/Ms ratio is larger: 1.24 and 1.22, respectively. This difference indicates that the DPRK5 generates relatively larger in amplitude regional share waves at higher frequencies, and the LR-waves of relatively lower amplitude at the period of ~20 s. The more effective emission of the regional share waves by the DPRK5 may indicate that the damage process was more intensive. The DPRK6 was conducted very close to the DPRK5 and the volume damaged by the DPRK5 did not participate in the generation of high-frequency regional waves. The degree of damages is also important for the cavity collapse and the chimney progress to the surface. The DPRK6 cavity collapse was practically immediate, and the DPRK3 through DPRK5 cavity collapses were not observed within several months from the mainshock [Kitov *et al*., 2021].

By origin, the DPRK aftershocks also have to be very close in space. Some of them were detected by many stations of the International Monitoring System (IMS) and corresponding event hypotheses were built by the International Data Centre (IDC). The two closest regional arrays of the primary IMS seismic network USRK and KSRS always report detections of the signals generated by the DPRK aftershocks. After the DPRK4, we have been using a detector based on waveform cross-correlation, WCC, as an optimal method for repeated events' detection [Adushkin *et al*., 2015; Adushkin *et al*., 2017].

The WCC-based detection procedure is using the time series of cross-correlation coefficient, *CC*, at individual sensors of array stations instead of the original waveforms used in standard IDC processing [Coyne *et al*., 2012]. This detection method is so sensitive that we were able to find the first and weak (ML~2.1) aftershock of the DPRK5 on September 11, 2016 [Adushkin *et al*., 2017]. With the increasing number of aftershocks found after the DPRK6 test, we obtained an opportunity to develop and test a multi-master phase association method, which significantly improves the detection sensitivity and reliability of event hypotheses [Kitov and Rozhkov, 2017; Kitov *et al*., 2018]. A prototype multi-master method was tested in a preliminary version of routine processing in 2018 [Kitov et al., 2018]. An enhanced version of the multi-master method [Adushkin *et al*., 2021] is used in this study.

## Detection and Local Association

To recover the DPRK aftershock sequence and to obtain a consistent set of principal parameters describing all found seismic events, we have reprocessed waveforms at IMS seismic stations USRK and KSRS for the period between 09.09.2016 (DPRK5), and October 1, 2021. The period between 01.01.2009 and 09.09.2016 is studied in a companion paper using the same approach, and several aftershocks were found after the DPRK3 and DPRK4 [Kitov et al., 2021]. To calculate CC-traces, we use 57 multi-channel waveform templates as obtained from 29 master-events: 6 DPRK underground nuclear tests and 23 aftershocks found in the previous study between September 2016 and April 2018 [Kitov et al., 2018; Kitov and Rozhkov, 2019; Kitov et al., 2021]. The DPRK1 was conducted on October 9, 2006, when station USRK was not operational. The list of master events is given in Table 2. The overall quality of a template can be illustrated by the number of other templates from the master event set correlating with it above some threshold, as column "Total" in Table 2 demonstrates. The templates at stations KSRS and USRK may have different qualities, as corresponding columns show. The highest quality is 54 correlating templates from 57. This is an indication that some waveform templates are likely poor. We intentionally included them in the master event list for the assessment of the relative importance of similarity in template shapes and signal quality. The signals with large SNR are likely better correlating, but this correlation degrades with the distance between the master and sought events. Poor signals from practically collocated events may correlate better than the good signals from remote events. A high correlation of the poor template with a sought signal would indicate the repeating character of their source.

Before the start of full reprocessing, we conducted an extensive tuning procedure using the aftershock and detection lists from the previous study [Kitov *et al*., 2019; Kitov *et al*., 2021] and confirmed that the updated set of detections provides higher detection reliability with a slightly lower sensitivity which is compensated by the second high-resolution processing step aimed at short time intervals around the event hypotheses found in the first stage [Kitov *et al*., 2021]. The detection reliability is measured as a fall in the false detection rate and accompanied fall in the rate of false event creation for a given set of event definition criteria. In the updated set of parameters, a different short-term average STA=0.8 s is used instead of the previous value of 0.3 s [Kitov *et al.*, 2018]. The original long-term average LTA=60 s is replaced by 120 s. The STA window leads the LTA window by half of its length as adopted by the IDC [Coyne *et al.*, 2012]. The change in STA results in a dramatic change in the detection threshold for routine processing.

Table 2. Master events. The DPRK1 was not processed and has no estimates obtained by cross correlation in columns. Explosions have extension "0". Extensions "1", "2", ... are the index number for a given day. Origin times (also epoch) are obtained from the cross correlation solution for a given master.

| | id | Origin time | Origin epoch time | ***RM*** | KSRS | USRK | Total |
|---|---|---|---|---|---|---|---|
| 1 | 2006282_0 | 01:35:28 | - | - | - | - | - |
| 2 | 2009145_0 | 00:54:43 | 1243212880.87 | 5.00 | 17 | 25 | 42 |
| 3 | 2013043_0 | 02:57:51 | 1360637869.09 | 5.31 | 17 | 23 | 40 |
| 4 | 2016006_0 | 01:30:00 | 1452043798.30 | 5.23 | 17 | 19 | 36 |
| 5 | 2016253_0 | 00:29:58 | 1473380998.67 | 5.35 | 25 | 27 | 52 |
| 6 | 2016255_1 | 01:50:48 | 1473558648.01 | 2.86 | 22 | 23 | 45 |
| 7 | 2017246_0 | 03:30:00 | 1504409398.92 | 5.94 | 25 | 22 | 47 |
| 8 | 2017246_1 | 03:38:31 | 1504409910.45 | 4.11 | 23 | 14 | 37 |
| 9 | 2017246_2 | 09:31:28 | 1504431087.60 | 2.66 | 22 | 22 | 44 |
| 10 | 2017266_1 | 04:42:58 | 1506141777.71 | 3.06 | 22 | 19 | 41 |
| 11 | 2017266_2 | 08:29:14 | 1506155354.55 | 3.60 | 25 | 26 | 51 |
| 12 | 2017285_1 | 16:41:07 | 1507826465.96 | 3.27 | 27 | 27 | 54 |
| 13 | 2017304_1 | 10:20:13 | 1509445213.19 | 2.52 | 12 | 23 | 35 |
| 14 | 2017335_1 | 22:45:54 | 1512168354.43 | 2.91 | 24 | 26 | 50 |
| 15 | 2017339_1 | 14:40:52 | 1512484851.85 | 3.14 | 23 | 24 | 47 |
| 16 | 2017340_1 | 16:20:05 | 1512577204.78 | 2.6 | 11 | 18 | 29 |
| 17 | 2017343_1 | 06:08:40 | 1512799720.05 | 2.79 | 19 | 23 | 42 |
| 18 | 2017343_2 | 06:13:32 | 1512800011.25 | 3.41 | 16 | 24 | 40 |
| 19 | 2017343_3 | 06:40:00 | 1512801600.08 | 3.11 | 24 | 27 | 51 |
| 20 | 2018036_1 | 10:32:30 | 1517826749.75 | 2.67 | 17 | 27 | 44 |
| 21 | 2018036_2 | 20:07:30 | 1517861249.34 | 2.79 | 17 | 24 | 41 |
| 22 | 2018036_3 | 21:57:35 | 1517867855.28 | 2.96 | 12 | 25 | 37 |
| 23 | 2018037_1 | 04:49:36 | 1517892575.86 | 2.65 | 8 | 23 | 31 |
| 24 | 2018037_2 | 10:12:30 | 1517911950.12 | 2.58 | 8 | 23 | 31 |
| 25 | 2018037_3 | 10:53:52 | 1517914432.06 | 3.02 | 26 | 27 | 53 |
| 26 | 2018038_1 | 21:46:23 | 1518039982.91 | 3.34 | 27 | 27 | 54 |
| 27 | 2018039_1 | 17:39:17 | 1518111556.77 | 2.74 | 23 | 26 | 49 |
| 28 | 2018112_1 | 19:25:08 | 1524425108.31 | 2.70 | 23 | 27 | 50 |
| 29 | 2018112_2 | 19:31:18 | 1524425477.81 | 2.98 | 27 | 27 | 54 |

We use a multi-length approach for cross-correlation windows, from 20 s to 120 s with a 20 s increment to address the possibility of different seismic phases to define the level of similarity: the DPRK nuclear tests have relatively short but high-amplitude $P_n$-phase and $P_g$-phase but a relatively low-amplitude and lengthy $L_g$-wave. The DPRK aftershocks usually have only the $L_g$-phase visible, and the $P_n$-phase is very close in amplitude to the ambient seismic noise. The changing cross-correlation window length allows exercising the similarity of various parts of the signals from the DPRK explosions and aftershocks and to partially avoid noise influence.

Waveform filtering is an additional noise suppression technique aimed at increasing CC-detection sensitivity. Five 4-order Butterworth band-pass filters are introduced: 1.0 Hz to 2.0 Hz, 1.5 Hz to 3.0 Hz, 2.0 Hz to 4.0 Hz, 3.0 Hz to 6.0 Hz, 4.0 Hz to 8.0 Hz. Using the best signals from the DPRK explosions and aftershocks in Table 2, sets of permanent multi-channel waveform templates is prepared for each of five filters. The total length of all templates is 205 s and includes a 5 s lead before the $P_n$-wave arrival time at the central element of an array (KSRS or USRK) and 200 s of the signal. The lead allows accommodating possible arrival mistiming as well as slightly earlier arrivals at array elements, which are closer to the source than the central station. The template length of 200 s includes all regular regional seismic phases, from $P_n$ to $R_g$, at distances from the DPRK test site to USRK and KSRS: $3.61^o$ and $3.97^o$, respectively.

The first aftershock was found on 11.09.2016 by the waveform cross-correlation method based on templates from the DPRK4 test conducted on 06.01.2016 [Adushkin *et al.*, 2017]. This example shows that the high-quality signals from the DRPK explosions can be effective as templates for finding weak aftershocks. In Figure 1, several waveform templates from six DPRK explosions are presented. The first, low-pass, filter reveals the decrease in the $L_g$-wave amplitude relative to the $P_n$- and/or $P_g$-wave amplitude from the smallest to the biggest test at both stations. This effect could be caused by the decrease in the corner frequency of the explosion source function below 1 Hz for the DPRK6, by diminishing tectonic energy release in the mountain after several repeating tests with an increasing yield which is accompanied by the increasing isotropic component in the MT-solution [Tape and Tape, 2015; Alvizuri and Tape, 2018], by the change in the depth of burial affecting the generation of the $L_g$-wave efficiency, and other factors. The change in $P_n/L_g$ amplitude ratio is also frequency-dependent, as Figure 2 demonstrates where the same original waveforms were filtered between 3 Hz and 6 Hz. The $L_g$-wave has almost disappeared for all DPRK underground tests. At the same time, the relative amplitude of the $P_n$-wave has increased and the first 20 s of the signal contain a larger part of the seismic energy at these frequencies. The observed change in the relative amplitudes is the reason for the introduction of cross-correlation windows with different lengths. We do not know in advance which part of the 200 s long template will be the most efficient in finding signals from the DPRK aftershocks.

The full evolution of the explosion signal shape for 5 filters is shown in Figure 3, where the DPRK6 is used as an example. Stations KSRS and USRK are different, with the relative amplitudes of the $P_n$- and $P_g$-waves also depending on the frequency content. The coda wave has a total duration of more than 200 s if to compare its amplitude with that of the ambient noise as observed in the 5 s interval before the $P_n$-wave arrival. The cross-correlation window length, CWL, of 200 s could help obtain a better result for some signals. For weak aftershocks, however, the coda is likely deep in the ambient noise at all frequencies and too long CC-windows do not provide any additional resolution [Kitov *et al.*, 2021]. Figure 4 presents the case of one of the biggest aftershocks of the DPRK6 (see Table 2) that occurred on October 12, 2017, which generated clear $P_n$- and $L_g$-waves. The difference with the signals in Figure 3 is significant. The $L_g$-wave is the largest at all frequencies. The $P_n$-wave is better seen at higher frequencies at both stations. The length of signals above the ambient noise is approximately 100 s. It is difficult to foresee which CC-window length will be the most effective for the aftershocks signals finding by the DPRK6 templates - short high-frequency $P_n$-wave or longer and low-frequency window of ~100 s.

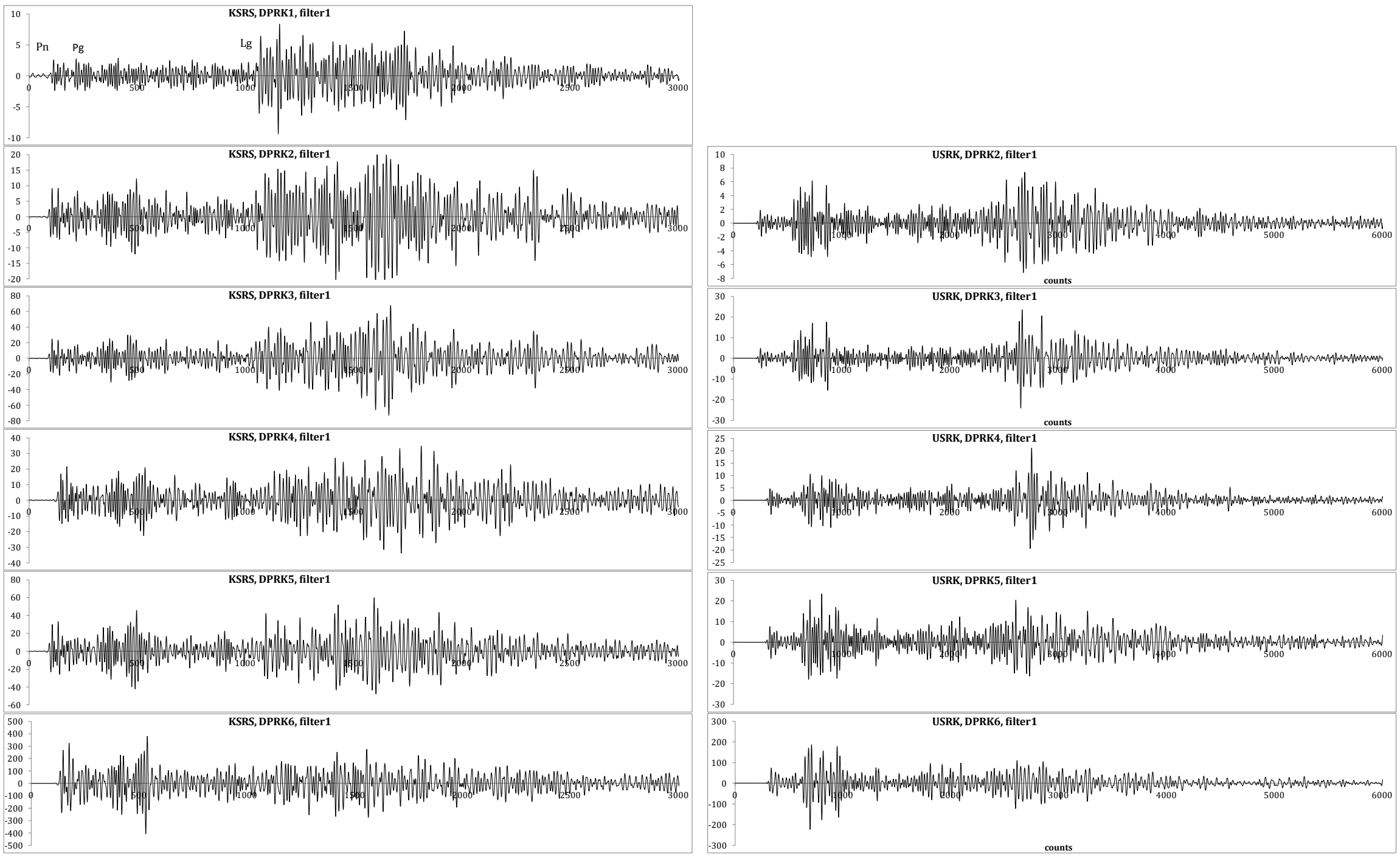

Figure 1. Waveform templates from the DPRK explosions recorded at central channels of IMS stations KSRS (left) and USRK (right). The low-pass filter (#1) results are shown to illustrate the difference and similarity of the seismic phases $P_n$, $P_g$, and $L_g$ depending on the source. Time counts are used instead of time to highlight the difference in the sampling rate: 20 Hz at KSRS and 40 Hz at USRK. Station USRK has no record from the DPRK1.

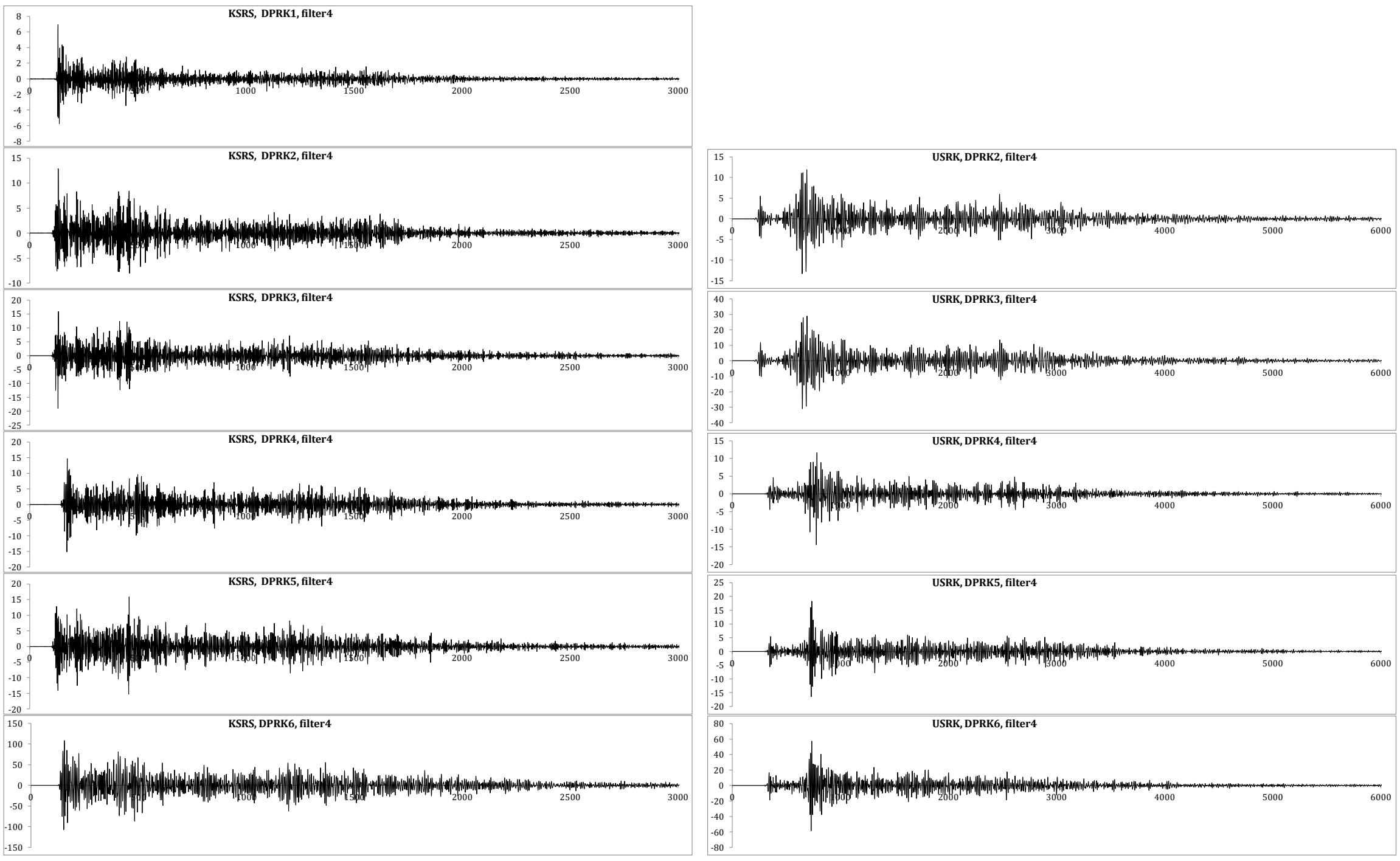

Figure 2. Waveform templates from the DPRK explosions recorded at central channels of IMS stations KSRS (left) and USRK (right). The high-pass filter (#4) results are shown to illustrate the difference and similarity of the seismic phases $P_n$, $P_g$, and $L_g$ depending on the source and station.

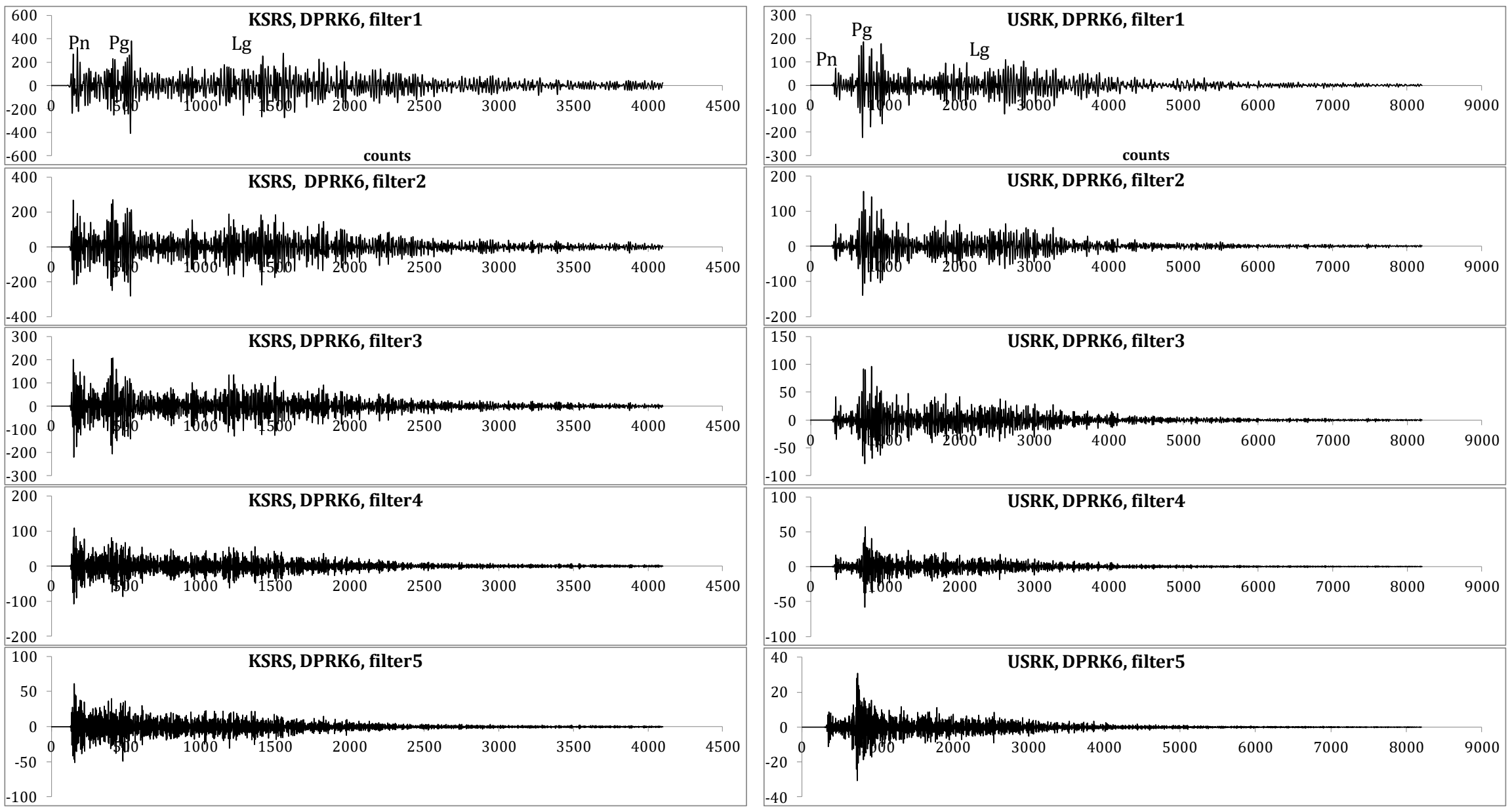

Figure 3. Waveform templates from the DPRK6 explosion, recorded at central channels of IMS stations KSRS (left) and USRK (right). All five filters are shown to illustrate the difference and similarity of the signals depending on the frequency content.

The DPRK aftershocks also have varying signal shapes and spectral content as illustrated by the comparison of Figure 4 and Figure 5 where the event detected on 13.06.2021 is presented. The relative amplitudes of the $P_n$- and $L_g$-waves are different for all filters and at both stations. This is an illustration of the fact that the sources of aftershocks may have different locations and source mechanisms. The complexity of the geological structure and the mechanical properties of the damaged rock near the source define the key part of the seismic transfer function between the elastic source and the seismic station. The other segments of the paths from the DPRK test site to corresponding stations are practically the same for all aftershocks induced by the underground tests. The DPRK3 and DPRK4 have also generated their aftershock sequences with signals well correlating with those generated by the DRPK5 and DRPK6 [Kitov *et al.*, 2021]. There were no aftershocks of the DPRK2 found by the multi-master method, which can find extremely weak aftershocks of the DRPK explosions. All these observations are evidences in favour of the absence of natural seismicity around the DPRK tests site and that all 100+ reliable event hypotheses belong to the population of the DPRK aftershocks. The qualitative visual difference between the signals from the DPRK aftershocks can be converted into quantitative parameters of similarity using the matched filter approach.

The aftershock found on June 13, 2021, and presented in Figure 5 occurred approximately four years after the DPRK6. The DPRK aftershock activity in 2021 was at a very high level, with events as large as those observed in a few weeks after the DPRK6 (Figure 4). This is an outstanding observation that needs a physical explanation. In Figures 6 and 7, two recent aftershocks that occurred on 24.06.2021 and 01.09.2021 are presented with less prominent signals in all frequency bands. The direct visual comparison is not possible and the only way to assess the signals' similarity is to calculate cross-correlation coefficients, *CC*, in all frequency bands and CWLs and compare the parameters best expressing the level of similarity. The absolute *CC* value depends on the frequency content and CWL, and thus, is not the optimal parameter to characterize the similarity of two signals, especially when they are hiding in the ambient seismic noise. The signal-to-noise ratio calculated for the aggregate CC-trace, SNRcc, plays the role of the similarity measure in this study. This parameter is also used for the detection of signals hidden in the ambient noise.

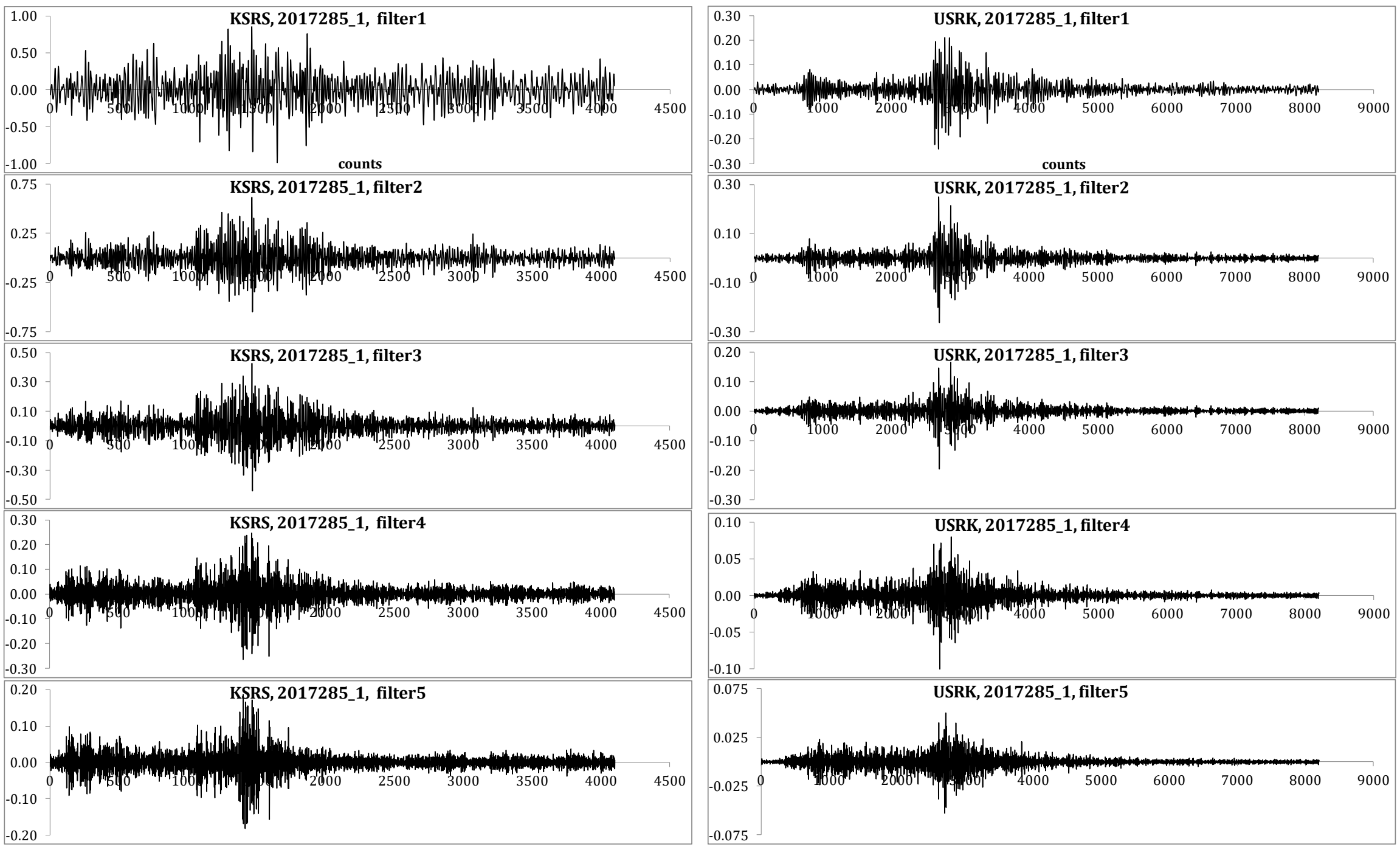

Figure 4. Waveform templates from the aftershock 2017285_1 (DPRK6 sequence) recorded at central channels of IMS stations KSRS (left) and USRK (right).

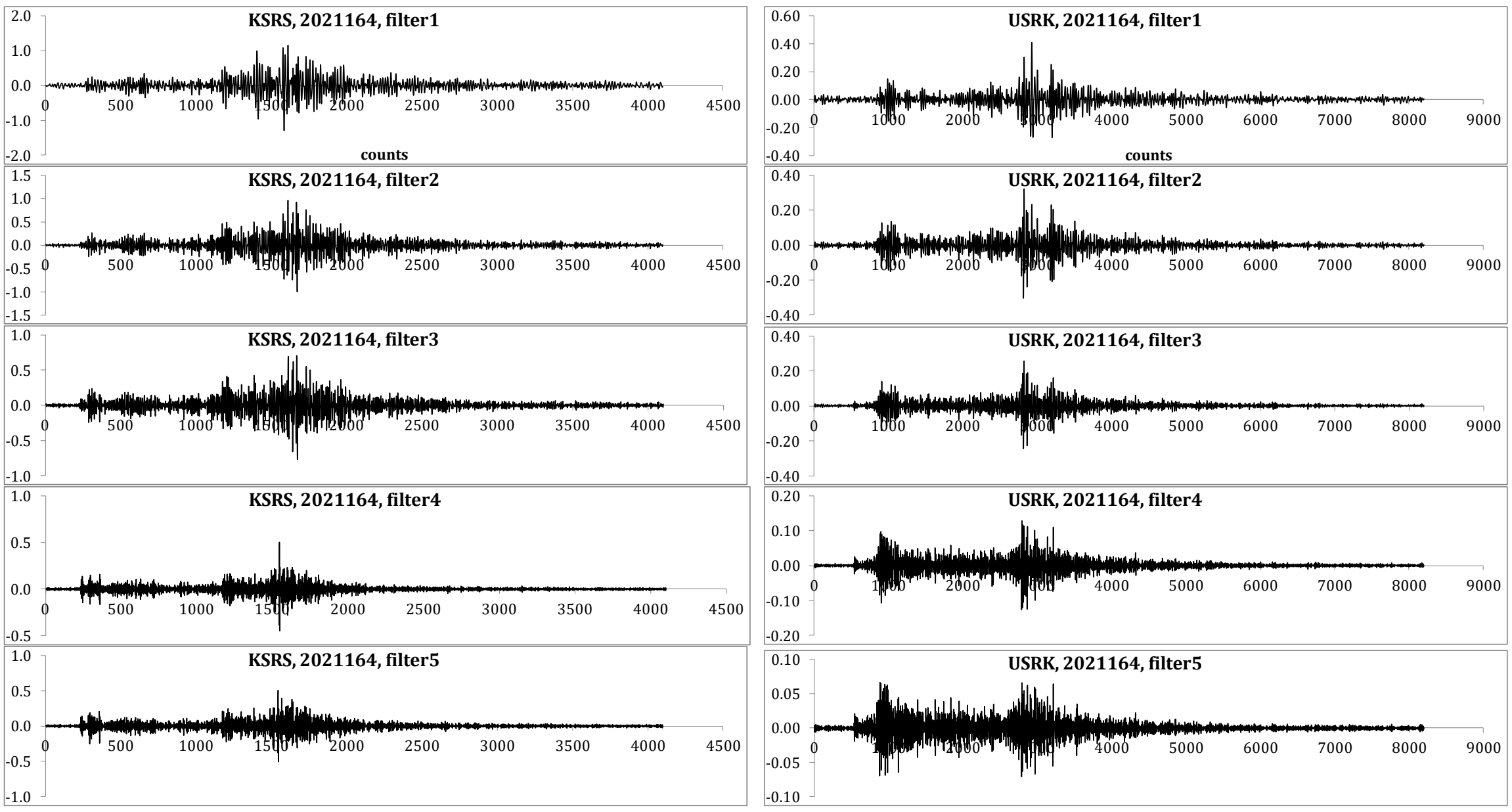

Figure 5. Signals from the aftershock 2021164_1 (DPRK6 sequence) recorded at central channels of IMS stations KSRS (left) and USRK (right).

We use the matched filter detection procedure, which is based on the conversion of the original multichannel waveforms into an aggregate trace of cross-correlation coefficients. The CC-traces at each individual channel of an array station or 3-C station, as well as the aggregate CC-trace are calculated according to the procedure described in [Bobrov *et al.*, 2014]. In some aspects, the CC-detection procedure is similar to the standard IDC detection procedure [Coyne *et al*., 2012]. For example, the first step is to filter the original multi-channel raw waveforms using one of the filters predefined in the template creation procedure. The next step is cross-correlation specific: for each individual channel of an array, we calculate a *CC* time series using the pre-calculated waveform templates at corresponding channels. The SNRcc is calculated at the aggregate CC-trace as averaged over all individual channels of the array or 3-C station. The SNRcc is equal to the ratio of the short-term average, STA, and the long-term average, LTA, with the lengths selected in an extended tuning procedure [Kitov *et al*., 2021]. To avoid the influence of the sought signal on the SNRcc estimate, the LTA value is frozen when detection is declared and this fixed value is used within the 2·CWL interval.

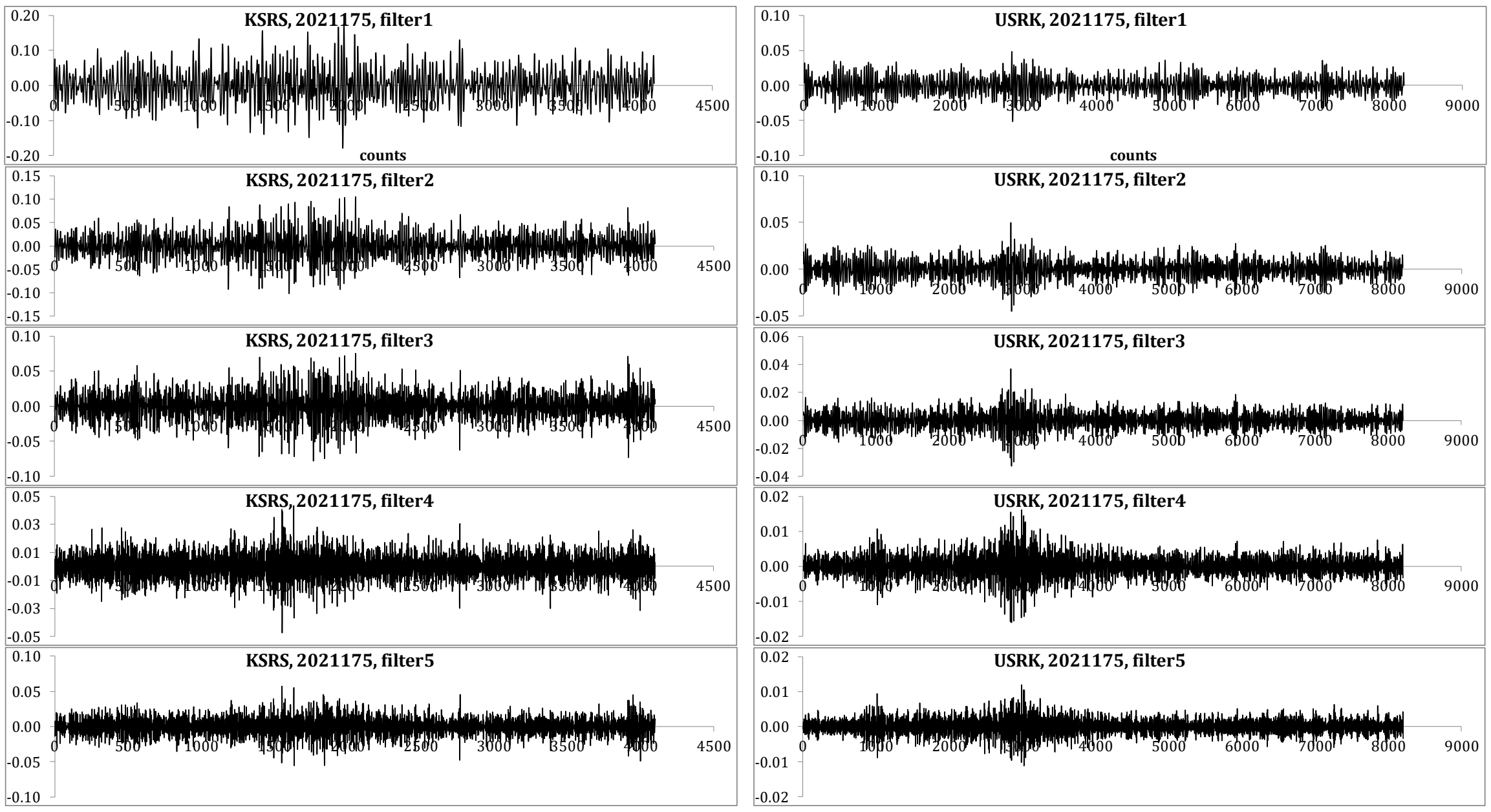

Figure 6. Signals from the aftershock 2021175_1 (DPRK5 sequence) recorded at central channels of IMS stations KSRS (left) and USRK (right).

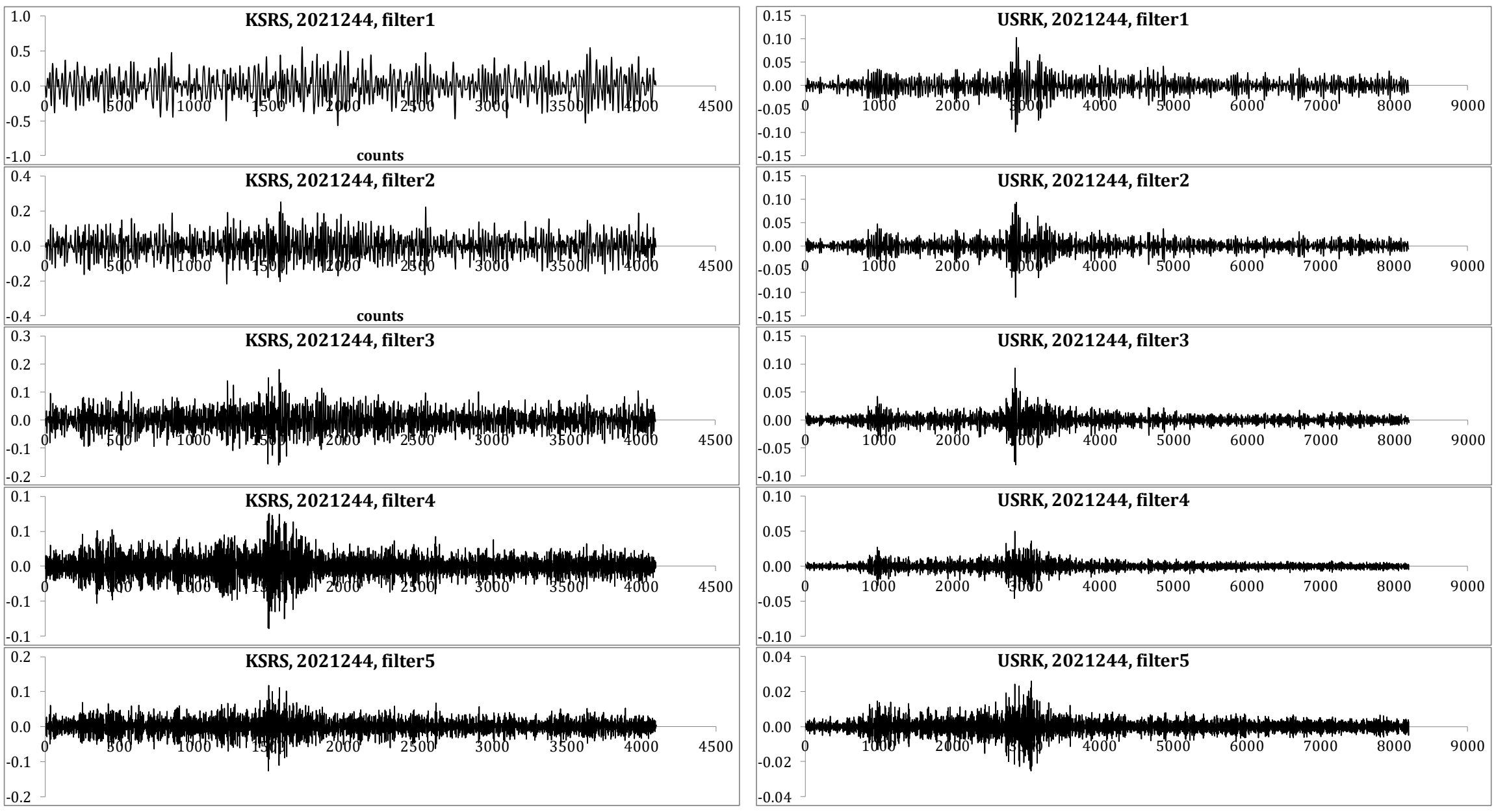

Figure 7. Signals from the aftershock 2021244_1 (DPRK5 sequence).

There are no time shifts between the channels in the averaging procedure since all the individual CC-traces have to be synchronized in time by the requirement of almost perfect collocation of the master and the sought events. This procedure is repeated for all filters and CWLs. For a given time count, the maximum SNRcc value among all filter/CWL combinations is saved as the final SNRcc. For a given template, the SNRcc time series consists of the maximum SNRcc values as the standard SNR time series consists of the maximum SNR values among all filters and beams. The same selection procedure is applied to the largest absolute *CC* value among all filter/CWL combinations. The individual CC-traces are used for the final arrival time estimates and can also be used in various applications, *e.g.* the FK-analysis. The individual SNRcc-traces are used in the CC-detection procedure.

In many aspects, the CC-detection procedure is different from physical signal detection. At the CC-traces, the sought signal can manifest itself at almost the same time as the tail of the waveform template touches the sought signal, *i.e.* at time $t_0$-$T$, where $t_0$ is the arrival time of the physical signal and $T$ is the length of the cross-correlation window. This effect is observed due to the signal/template coherence and the absence of coherence between the template and the ambient seismic noise. When a template meets the sought signal, the *CC*-trace starts to increase in amplitude above the CC-trace obtained for the ambient seismic noise. This effect makes the CC-detection based on the STA/LTA ratio a more complex procedure compared to standard IDC detection. One needs to select the correct parameter and procedure for the onset time estimate, as usually based on thorough statistics obtained in the routine seismological practice. The CC-detector has a very limited set of studies in recent years, and thus, is not based on a hundred-year-long history of measurements. Any new study may provide input to the CC-detection statistics and a general understanding of the accuracy and uncertainty associated with the CC-detection process and its results.

There are various cases associated with CC-detection. When the ambient noise level is low, the *CC* has to gradually grow before the sought signal and the template are synchronized with the *CC* reaching its peak value. Then the *CC* has to decrease before the template leaves the sought signal at $t_0$+$T$. For the autocorrelation case, the arrival time unambiguously corresponds to *CC*=1.0 and the arrival time estimation is an easy task. When the sought signal has low amplitude (in the worst case the signal is below the noise level) and it is significantly different from the template, many local *CC* maxima are observed, and it is difficult to select the proper one for the arrival time because of potential noise influence. To accommodate this possibility into the CC-detection procedure, we use SNRcc as an integral characteristic of the similarity between the signal and template. The SNRcc is supposed to have an absolute peak near the actual arrival time of the sought signal.

By definition, the matched filter detector is based on cross-correlation between two signals and represents the optimal linear filter for maximizing the signal-to-noise ratio in the presence of additive stochastic noise. In seismic applications, there are two deviations from the theoretical matched filter consideration: 1) the waveform template and the sought signal are always different even for highly repeatable events; 2) seismic noise might not be fully stochastic, *i.e.* the signals constituting the noise may have characteristics similar to both or one of the two signals. Figure 8 depicts the maximum SNRcc and *CC* traces for several aftershocks found in 2021. For station KSRS, the SNRcc peaks are well synchronized with the CC-peaks. KSRS is a large aperture array, effectively suppressing the uncorrelated noise in the original and CC-traces. The peak SNRcc value, selected among all 29 templates and all filter/CWL combinations, reaches ~30 for the event observed on 01.09.2021 with the corresponding $CC$≈0.4. Station USRK is a small aperture array that is not so effective in noise suppression. The maximum CC-trace is obtained at the low-frequency filter, as Figure 9 demonstrates. The peak $CC$ is a high-frequency burst on a low-frequency background. It is also important that the peak $CC$ and peak SNRcc may not coincide, and one has to decide which peak defines the onset time.

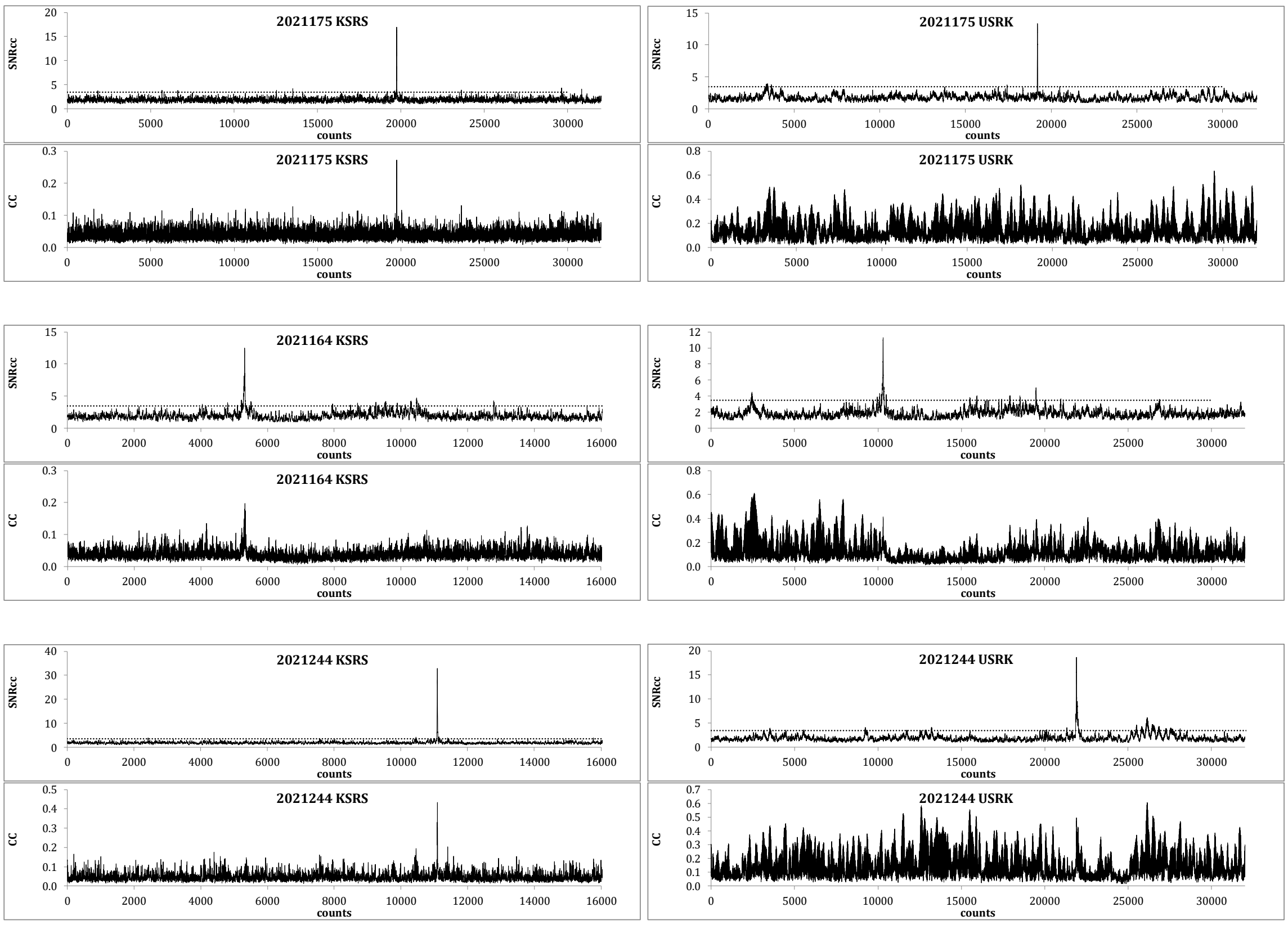

Figure 8. Maximum SNRcc and absolute CC traces for 3 aftershocks observed in 2021.

The difference between the template and the sought signal is unavoidable, and the level of similarity varies between the $CC$≈1.0 and $CC$ approaching zero. The latter case is of special importance for CC-detection of the ultra-weak signals in this study. It is slightly counterintuitive that reliable detection is possible for $CC\sim10^{-6}$ or even lower, even though this situation is commonplace in the Fourier transform used in the signal filtering or spectral estimates. For detection, the absolute $CC$ value of the sought signal does not matter - the difference between the signal and noise $CC$ is important. Therefore, the second deviation from the theoretical matched filter procedure is likely an influential feature for the detection of ultra-weak signals.

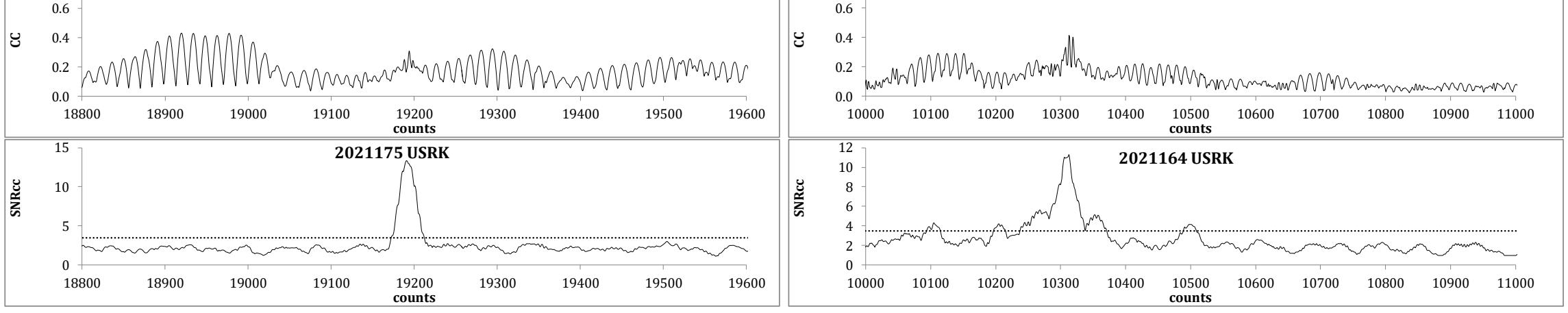

Figure 9. Maximum SNRcc and absolute CC traces at USRK, illustrating the CC traces corresponding to the peak SNRcc.

The signal/noise coherence makes the quantitative similarity measure as expressed by the cross-correlation coefficient to be less reliable. We use SNRcc to characterize the similarity between two signals: it considers both the similarity between signals and suppression of uncorrelated noise, as we suggest that the ambient noise has no significant component coherent with any of the found aftershocks. There are a few exceptions to the incoherent noise assumption, all related to higher aftershock activity near the DPRK explosions' epicentres. The immediate aftershock (8.5 min) of the DPRK6 is observed in the coda waves generated by the mainshock. For this case, the emission from the mainshock and corresponding seismic noise at USRK and KSRS is clearly coherent with the signal from the aftershock [Kitov *et al.*, 2021] This aftershock is by far the biggest among the DPRK aftershocks. There were also episodes, when several DPRK aftershocks were observed within a few minutes. The first events in these short series could generate seismic noise, reducing the CC-detector resolution.

A detection hypothesis is tested when the running SNRcc value in the maximum SNRcc time series for a given master exceeds the predefined threshold. Then the estimate of the arrival time and other defining parameters starts. One has to consider the structure of the maximum SNRcc trace, which may jump between all available detection channels defined by filters and CWLs. For a unique filter/CWL combination that triggered the detection hypothesis, the SNRcc has to gradually grow from the detection threshold to the peak value near the physical signal arrival time. Therefore, the peak SNRcc has to be within the corresponding CWL. There is another possibility - to use the maximum SNRcc trace over all filters and window lengths within some interval after the detection time, which may vary between the shortest and longest CWL. We use the first option, but also check the result of detection by the second option to gather more statistics on the CC-detector. In the standard detection procedure adopted in the IDC for array stations, the detection threshold depends on the detection beam parameters: filter, station-event azimuth, and slowness. The filter characteristics mainly define the detection threshold, as the microseismic noise level usually peaks near 1 Hz and then rapidly falls with frequency. For the WCC processing, the detection threshold is the same for all filters. The threshold fine-tuning needs extensive statistics.

The maximum SNRcc peak defines the centre of the interval, where we search for the maximum *CC* in the time series defined by the filter/CWL combination that triggered the detection process. The width of this interval may vary depending on the STA value and filter. We use the ±1 s window, which accommodates the difference in the SNRcc and *CC* peak times observed in our study and is well within the width of the phase association window ±3 s. This is an important correction of the arrival time because the basic assumption is that the synchronization of signal and template results in the highest *CC* value. The CC-detection process is characterized by the non-physical effect of the template length on the causality principle - the detection starts long before the physical signal arrives. Therefore, it has to be a still period after each detection of the length larger than the CWL for the next detection. This condition allows avoiding the influence of the previous detection on the CC-trace near the next detection. In addition, the LTA window length has to be considered to avoid the mutual influence of the subsequent signals on the SNRcc. The DPRK aftershocks are not so often and close to each other to be affected by the still period between the subsequent arrivals. For the aftershock sequences of catastrophic earthquakes, such an assumption is likely not appropriate.

The SNRcc is used in automatic detection as a parameter related to the probability of a signal to be associated with an event within the zone around the locations of the DPRK master events. The initial processing phase is aimed at finding all possible aftershocks during the period between September 9, 2016, and October 1, 2021. The routine WCC processing focused on the DPRK aftershocks will be stopped when the aftershock activity fades away after the final DPRK5 and DPRK6 chimney collapse, with or without the surface craters' creation. Considering the length of the studied interval and constraints of available computing resources, we used a reduced but efficient version of detection parameters: only one cross-correlation window length (120 s) and three (#2 to #4) filters. These are the parameters quasi-optimal for the detection of an aftershock signal by another aftershock template [Kitov *et al.*, 2021]. Figure 10 shows an example of the SNRcc frequency distributions as obtained by 28 templates at stations USRK (no template for 2006282_0) and 29 templates at stations KSRS on May 13, 2021. In these distributions, all SNRcc estimates are used, *i.e.* 3,456,000 readings (86400 s times 40 Hz) for USRK and 1,728,000 readings (86400 s times 20 Hz) for KSRS. For almost all templates, an exponential roll-off is observed with the increasing SNRcc. This is a sign of the stochastic behaviour of the cross-correlation coefficient, as one could expect when no signals from the monitored area are present in the data. In the detection procedure, we redefine all SNRcc values below 1.0 to 1.0. This is the reason why all distributions have peaks at SNRcc=1.

There were no DPRK aftershocks detected on May 13, 2021, by the cross-correlation technique, and all SNRcc values are most likely not related to seismic activity in the zone of DPRK master events. For a few templates, the SNRcc distributions have slightly thicker tails than those predicted by the exponential roll-off. This effect is likely related to side lobes of the CC-detector sensitive to strong signals not associated with the DPRK zone. Therefore, these are not valid detections for the DPRK-related events. Having the SNRcc frequency distributions in Figure 10, we have to define such a detection threshold which does not allow the false event rate above some reasonable value (say, once per month or once per year) with the event definition criteria (EDC) discussed in the companion paper [Kitov *et al.*, 2021], where the detection tuning procedure is also described in detail. With all these constraints, the detection threshold for the initial processing stage is 3.5 for all templates and the EDC is slightly looser than in the final processing. Figure 11 illustrates the day-to-day variability in the PDFs for all templates at a given station. The exponential roll-off between SNRcc 2.0 and 3.5 is observed on both days. Station KSRS has thicker tails, likely associated with a higher side lobes sensitivity of the *CC*-detector, larger aperture, and lower sampling rate.

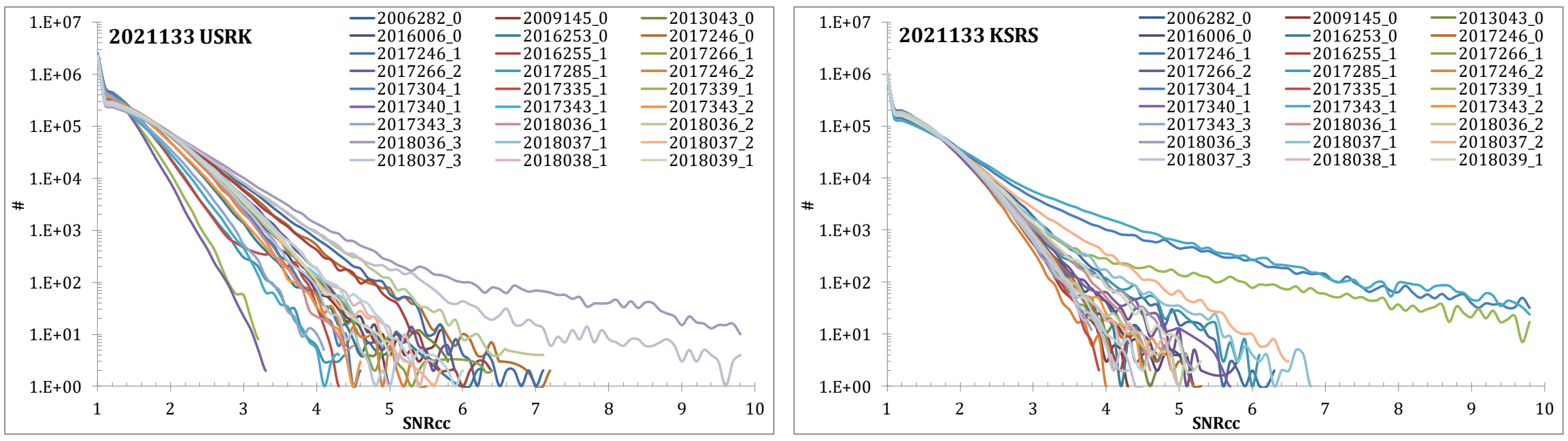

Figure 10. Frequency distribution of SNRcc values obtained for 29 master events for one day 2021133 (May 13, 2021) at stations USRK and KSRS.

Association of several seismic phases detected by the WCC method with a unique event is assuming that one arrival of a regular seismic phase at one station is counted once [Bobrov *et al.*, 2014; Bobrov *et al.*, 2016; Kitov *et al.*, 2019]. This is a physically and statistically reasonable assumption. The probability of an event hypothesis (the combination of hypocentre and origin time) depends on the probability of several phases to have close origin times, *i.e*. their arrival times less the travel times between the event hypothesis and station. Therefore, one phase can be counted only once. To ensure the predefined level of reliability of the events in the Reviewed Event Bulletin, REB, which is the final product of the IDC interactive analysis, the principal event definition criterion consists in the requirement of association of the first P-phases detected at 3 or more primary seismic stations. The quality of each associated phase, and thus, its input to the joint probability, can be estimated from such parameters as travel time residual, slowness and azimuth. For the WCC processing, an additional value for the hypothesis may introduce the estimates of regular SNR and the SNRcc. All event defining parameters involved in the event hypothesis estimation have to be within predefined tolerances as obtained from the extensive statistics associated with the broader seismological practice.

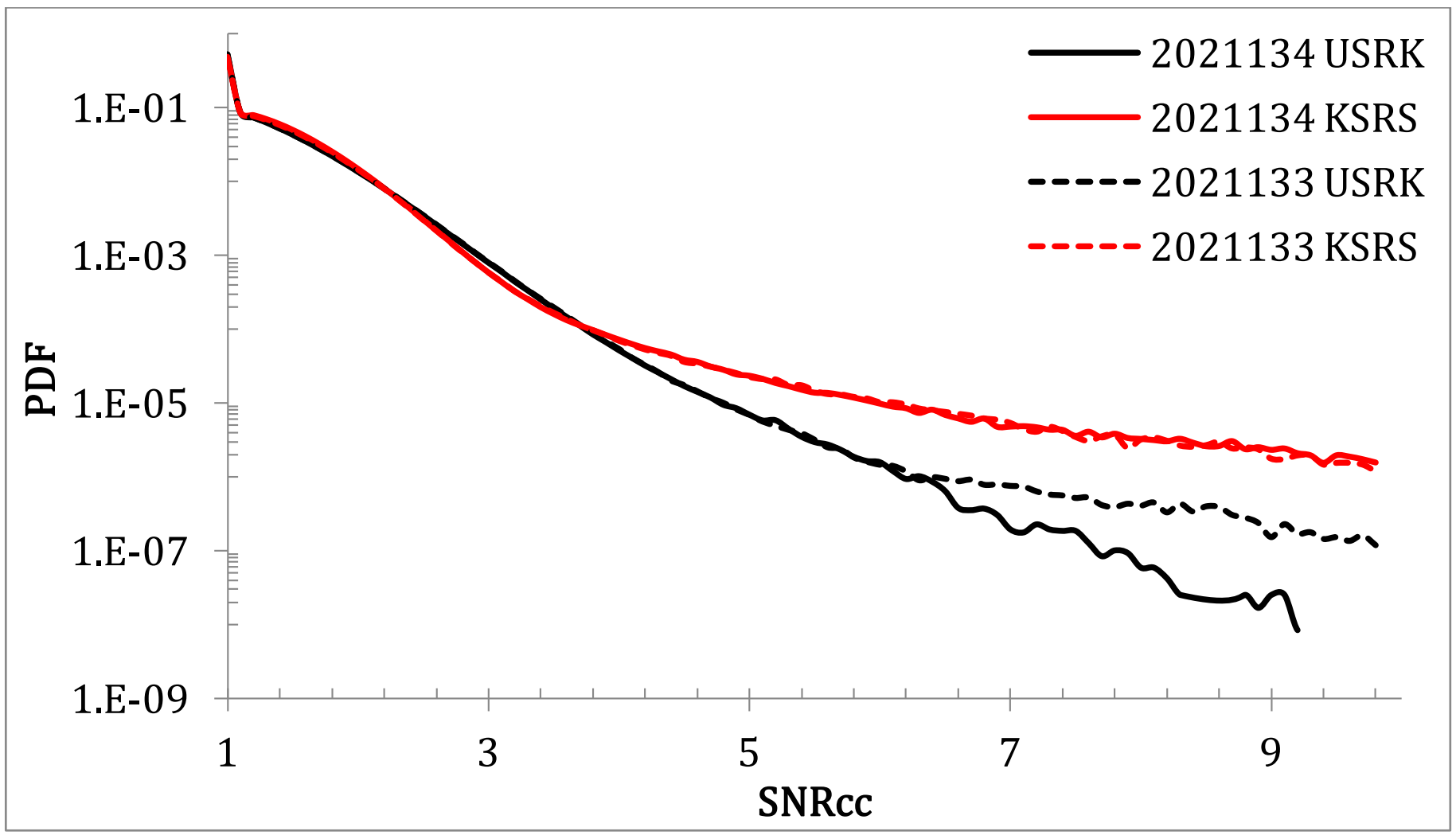

Figure 11. Probability distribution functions, PDF, obtained from the frequency distribution of the total number of SNRcc values for all masters on May 13 and May 14 (2021134) at stations USRK and KSRS.

The DPRK aftershock processing has some specific features in both detection and phase association. Only two stations are available for data processing, and the third station requirement would reduce the number of found aftershocks by almost two orders of magnitude. Two-station event hypotheses are not reliable. Many mining areas within a few hundred of kilometre from the DPRK test site may generate regional phases similar to those generated by the DPRK aftershocks. For low-amplitude signals, it is difficult to distinguish between the DPRK and mining events, even at such large-aperture seismic arrays as KSRS. There exists a strong criterion of the DPRK aftershocks, however. The arrival time difference at stations USRK and KSRS of 5.5 s estimated from the DPRK explosions gives a chance to distinguish between the DPRK aftershocks from the mining events or natural events out of the DPRK test site.

For repeated events, one can count the same physical signal several times if it is detected by templates from close but different events [Kitov *et al*., 2018; Adushkin *et al*., 2021]. The intuition behind this procedure is a higher similarity of signals from spatially close events than the similarity of remote events, especially at array stations or seismic antennas. Moreover, a slight variation in the template shapes between master events can be useful for discrimination if this variation has a physical origin. The number of detections obtained by a given set of templates at a given station is a good measure of the reliability of the corresponding physical signal, which is possibly hidden in the ambient seismic noise and not seen by analysts. The share of successful templates with detections within the predefined arrival time interval is somewhat similar to the rate of successful detections of reflected signals for a given radar pulse rate as the defining parameter of the radar performance. The difference in the templates' shape is somewhat similar to the change in the pulse frequency content or pulse width. Therefore, we consider the multi-master association, based on two stations and dozens of templates generated by different master events, as a version of radar, with the CC-detections representing the matched reflections.

For the set of master events and related templates listed in Table 2, one can find an optimal set of detection and association parameters that balances the rates of false/true detections and false/true event hypotheses. The known start of the DPRK aftershock sequence allows defining a quasi-optimal set of detection and association parameters using a predefined probability of a false event. The period before 11.09.2016 was studied in a companion paper and an extensive tuning procedure gave two sets of parameters: for routine processing and for the high-resolution processing applied to one-hour intervals, which included the event hypotheses found in the routine stage [Kitov *et al*., 2021]. When defining the false event rate, we allowed one false event to be created per month in the routine regime during the period without aftershock activity. There were no aftershocks found by the multi-master method between 01.01.2009 and 12.02.2013 (DPRK3) and this interval can serve as a reference for the assessment of the false event rate. One year after the DPRK2 was processed in the high-resolution regime. No aftershocks were found, and the absence of valid event hypotheses confirms the assumption that only false events can be created during this period.

There are two sorts of false events – the result of an occasional association of stochastically distributed false detections and those created by detection of non-relevant strong signals from larger events due to non-zero side sensitivity of the *CC*-detector. The former sort is usually characterized by low SNRcc values and low relative magnitudes, as introduced in [Bobrov *et al*., 2014, Bobrov *et al*., 2017a]. For random *CC*-detections obtained by 57 templates, the probability of several arrivals to have close origin times depends on the detection rate. The higher is the detection threshold, the lower is the false detection rate. The larger is the number of arrivals in a given origin time window needed to build an event hypothesis, the lower is the joint probability of such an outcome. The rate of this sort of false detections and false events we can control. The second sort of false detections is also characterized by low SNRcc values, but they have much larger relative magnitudes, which are proportional to the logarithm of the ratio of RMS amplitudes in the CC-window. Such detections are mainly driven by large-amplitude teleseismic P-waves, and thus, are predominantly related to the templates of the DPRK explosions with impulsive $P_n$-waves. We remove such false detections/events from consideration as irrelevant. They do not affect the false event rate calculations.

When the final list of detections is available for further analysis, the Local Association procedure is applied. All arrival times obtained by the WCC method are projected back to their origin times using the master/station travel times estimated with an accuracy of 0.001 s. For the DPRK explosions and the biggest aftershocks, the travel times are known from the REB. For the smaller aftershocks in Table 2, we correlate their templates with the waveforms of the DPRK5 and calculate the travel time corrections as the difference between the DPRK5 arrival time in the REB and the arrival time obtained by a given template. Effectively, these corrections reduce the travel times of the aftershocks to the DPRK5 travel times. This procedure also compensates the uncertainty in the arrival times estimated for the small aftershocks as related to the absence of clear $P_n$-wave signals, which are often hidden in the ambient noise, and to the variation in the $L_g$-wave shape (see Figures 1 through 7) making the cross-correlation estimates less accurate.

All origin times at both stations are ordered according to their absolute values and the number of detections in the next 8 s (the length of association window) after each arrival, moving from the first one to the last. In Figure 12, the number of detections, NDT, generated by the 57 templates in the 8 s intervals starting at subsequent origin times from 0 to 86392 is estimated for the two most recent days with found aftershocks: 2021253 and 2021254. An additional counting rule is applied: each station has an input of 3 or more detections, *i.e.* the minimum Nass=6. Two aftershocks were found on September 10, 2021, and one aftershock on September 11, 2021, as the 8 s segments with NDT>35 suggest. There are several intervals with 20>NDT≥11, which have to be tested in the high-resolution regime.

The association window has to accommodate the scattering of arrival times obtained by 57 templates, which is related to the uncertainty in the arrival times and to the change in travel times to USRK and KSRS for event hypotheses in the area of possible aftershocks. When the number of detections in an eight-second-wide window is above the predefined threshold, a relative location procedure is applied as a part of the Local Association [Bobrov *et al.*, 2017b; Kitov *et al.*, 2019]. A predefined set of virtual locations (say, a rectangular grid with a 100 m step within a circle of 2 km in radius) instead of the actual master location is used. This procedure is a standard grid search, minimizing the RMS origin time residual. For the virtual nodes, the travel times to USRK and KSRS are calculated using the empirical travel time for the real master/station pair corrected for the master-node distance and the theoretical slowness of the $P_n$-wave. When the event hypothesis is not collocated with the master event, one may reduce the scattering of the origin times in the association window. The node with the minimum RMS origin time residual pretends to be an event hypothesis location relative to the master event, in case it also provides the highest number of associated templates.

The origin time residual for individual detection has to be less than 3 s. In standard seismic location adopted by the IDC, the allowed travel time residual, which is another representation of the origin time residual, depends on seismic phase and varies from 2 s for the P-phase to 3.5 s for the secondary share wave phases like $S_n$ and $L_g$. In the DPRK aftershocks association, the maximum travel time residual is 3 c. Therefore, in the association process, two parameters are tuned when moving over the virtual grid – the number of detections within ±3 s from their average origin time and the RMS origin time residual in this group. The grid node with the largest number of associated detections and lowest origin time scattering is the preferred relative location of the event hypothesis. The event hypotheses with Nass≥11 found in the routine processing are then reprocessed in the high-resolution mode. Only the event hypotheses with Nass≥20 are promoted to the cross-correlation standard event list, XSEL, which is similar to other IDC products of automatic processing: SEL1, SEL2, and SEL3. The full set of event definition criteria for the DPRK aftershocks is presented in [Kitov *et al.*, 2021]. The main criteria are as follows: 1) a valid event hypothesis has at least 11 associated templates; 2) input of each station has to be more than 3 detections; 3) the origin time residual has to be within the uncertainty range of 3 s.

In the final, high-resolution, processing stage, the set of defining parameters for detection and association is used to estimate the parameters related to cross-correlation: arrival times,

SNRcc, relative magnitudes and relative location for the event hypotheses in the XSEL. All filters and CC-window lengths are used for detection. Some event hypotheses from the routine stage are false and thus rejected. The best hypotheses from the routine stage can be significantly improved as defined by the number of associated detections. The local association procedure was updated in accordance with the new distribution of detection parameters and the minimum Nass=20. This threshold removes all possible random false events, and the event hypotheses associated with the side lobes of the CC-detector are rejected by their relative magnitudes.

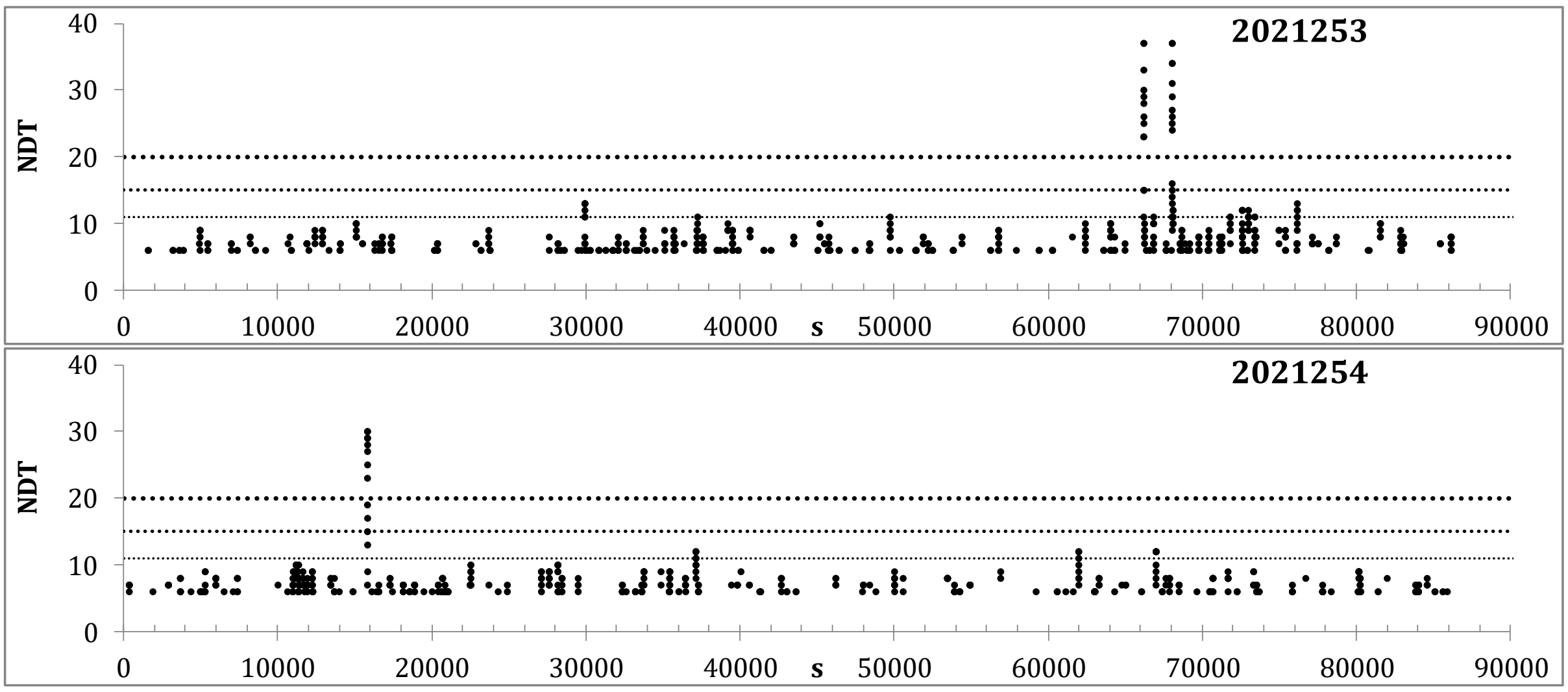

Figure 12. The number of detections, NDT, in the running eight-second-intervals with 1 s step for two days with the most recent aftershocks. There are two potential aftershocks on 10.09.2021 (2021253) and 11.09.2021

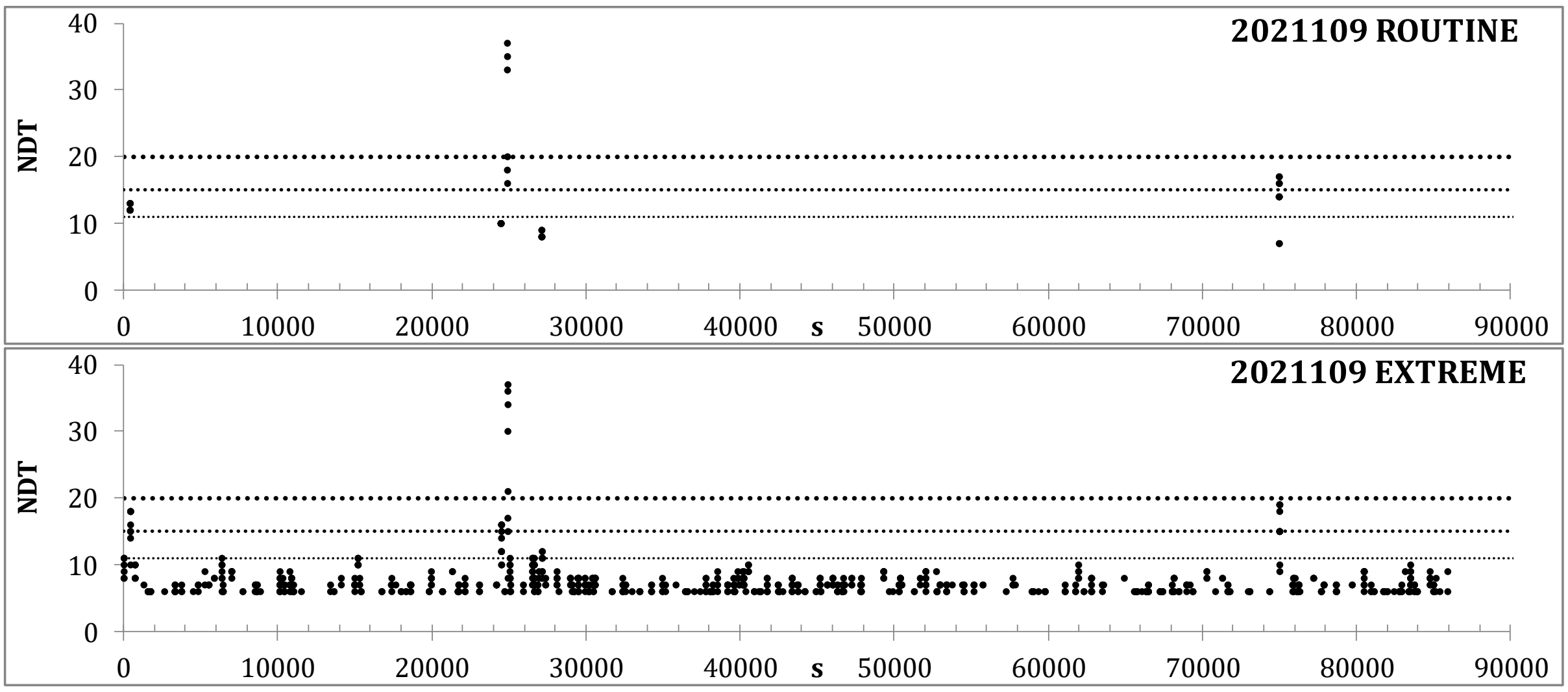

Figure 13. The number of detections, NDT, obtained by all 57 templates in the running eight-second-intervals with 1 s step as estimated for 2021109 in the routine and high-resolution (extreme) processing mode.

In the upper panel of Figure 13, the number of detections, NDT, generated by the 57 templates in the 8 s intervals is estimated at 19.04.2021. In the routine regime, one aftershock was discovered by the multi-master method on April 19, 2021. There are several intervals with NDT≥11, which may generate event hypotheses to be tested in the high-resolution regime. The result obtained using the high-resolution set of control parameters is shown in the lower panel. No additional XSEL events with Nass≥20 were created, but the number of 8 s segments with NDT≥11has increased.

There is a simple way to visualize the process of Local Association. Figure 14 depicts the SNRcc for all detections obtained on 12.10.2017 as a function of time. The origin times are obtained with the empirical travel times for the corresponding master events. We distinguish detections at KSRS and USRK to illustrate the rule that both stations have to provide significant input to a true event. There is one clear surge in the SNRcc synchronized at two stations. The largest SNRcc values at KSRS and USRK belong to autocorrelation, but the second-best detections also have SNRcc around 50. This is an extremely high-value considering the highest SNRcc between the DPRK explosions of 70 to 90 (excluding autocorrelation). The number of master events with the SNRcc>20 suggests their close spatial proximity and the similarity in the physical conditions of the $P_n$- and $L_g$-wave generation.

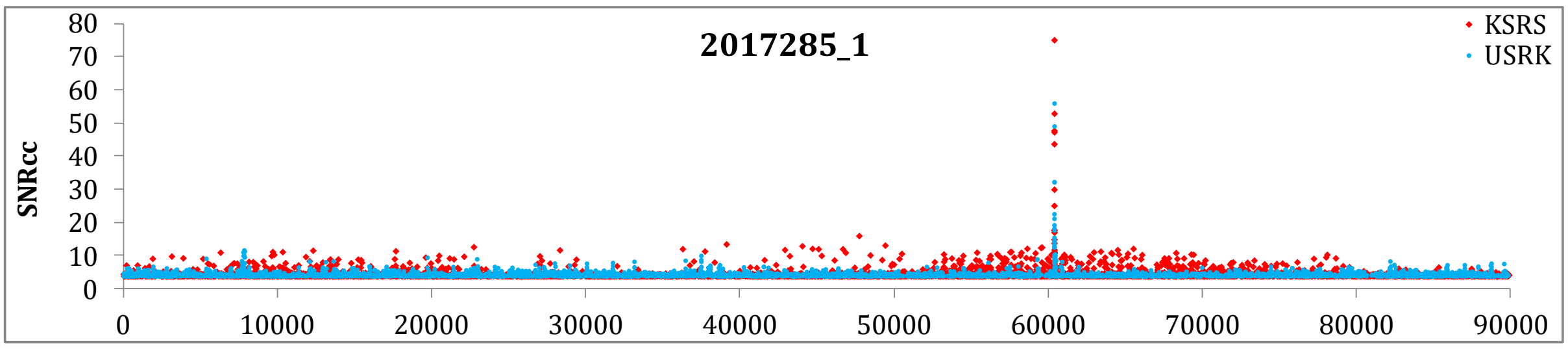

Figure 14. Distribution of SNRcc at stations KSRS and USRK as a function of the elapsed time for 2017285. The aftershock 2017285_1 is seen as synchronized increase in the SNRcc values

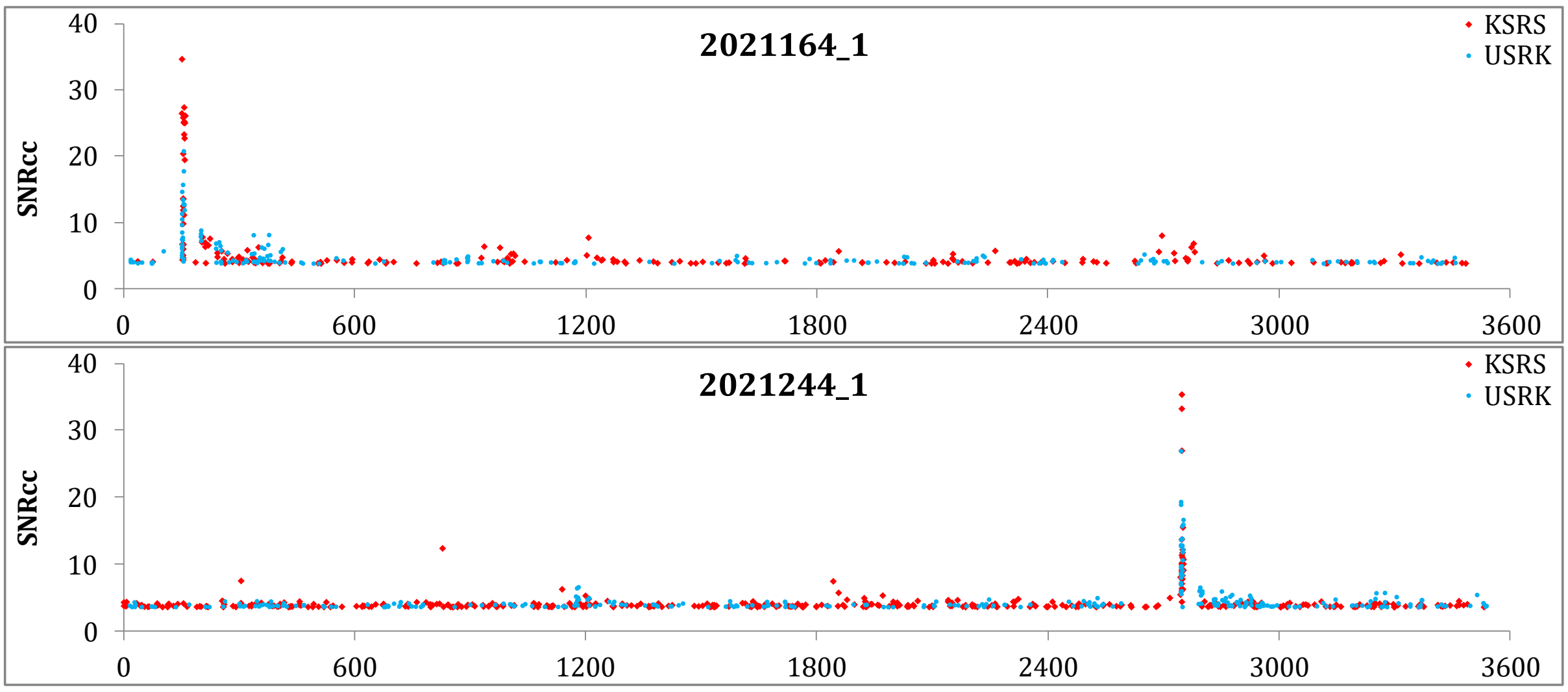

Figure 15. Distribution of SNRcc at stations KSRS and USRK as a function of the elapsed time for 2021164 and 2021244 in the one-hour intervals with the corresponding aftershocks.

A closer view of the SNRcc distribution as a function of time is presented in Figure 15 for two aftershocks found in 2021. These examples demonstrate the quality of event hypotheses and their uniqueness for the studied days. There are also some strings of higher SNRcc at one station, but there is no one synchronized surge in the number of detections in short time intervals at two stations. Both distributions also demonstrate gaps before and after the event hypotheses. This observation illustrates the condition of a still period after detection. For a random detection distribution, the 57 templates generate a more or less stationary flux of detection. The detections synchronized with a valid event have to kill other detections within 120 s to 200 s.

The WCC processing includes standard procedures: detection, Local Association, and Conflict Resolution. The latter has specific features for cross-correlation when many CC-detections belong to the same physical signal. In the Local Association, the virtual location grid presents a simple example of Conflict Resolution in action. Moving a virtual event location over the grid, one creates a set of solutions with changing Nass and RMS origin time residual as two defining parameters. If Nass has just one maximum value for a given grid, conflict resolution is not needed and the event hypothesis with the max Nass is saved for further analysis. When two or more nodes have the same maximum Nass value, the node with the lowermost RMS origin time residual is chosen as the final solution.

When two or more neighbouring association windows create several event hypotheses competing for the same detections (see Figures 12 and 13) one has to define the rule for automatic decision. Following the IDC practice, the event with the largest Nass (or with the lowermost RMS origin time residual when several hypotheses have the same max Nass) wins the conflict and takes all detections associated with it. All other pretending event hypotheses lose these detections and their solutions are recalculated accordingly. Some of them have to be rejected according to the EDC Nass≥11. Some event hypotheses losing one or more detections may have Nass≥11, and thus, are valid according to the adopted EDC. In Conflict Resolution, we start with the biggest event hypothesis and determine all detections associated with it. These detections are removed from the detection list, and a new configuration of event hypotheses is calculated. The biggest event in the new set of hypotheses wins the competition for the associated detections. This process is repeated before all conflicts between event hypotheses are resolved.

The WCC processing has specific detection and phase association features when applied to the DPRK aftershocks. These features are needed to find ultra-weak signals deep in the ambient noise and associate them with the event hypotheses located within a small footprint around the DPRK test site. The multi-master method is an important extension in the WCC processing, allowing higher resolution and increased reliability of the event hypotheses compared with the standard cross-correlation detection/association base on individual master events. The best masters in the master event set could find most of the DPRK aftershocks in the final XSEL. The reliability of the smallest found aftershocks would be lower, however, because of the increasing number of false events, which, in turn, should be suppressed by the increasing detection threshold resulting in a reduced resolution. The multi-master approach is similar to radar operation: each detection of a reflected signal has individual value as can be treated as a finding of the sought object, but the repeating detections increase the reliability of the reflecting object finding. We retain the number of master events at the same level since the start of DPRK processing and have no doubts that no aftershocks or natural events are missing, at least of the size of the smallest event in the XSEL.

Briefly, the setting for the WCC reprocessing of the period between September 9, 2016, and October 1, 2021, is as follows. At the routine stage, the STA=0.8s, LTA=120s, filters from #2 to #4 are used, CWL=120 s, and the SNRcc detection threshold is set to 3.5. The Nass threshold is set to 11 from 57 associated templates, with at least 4 detections from one station.

The event hypotheses matching the EDC are promoted to the high-resolution processing with STA=0.5 s, LTA=120 s, CWL from 20 s to 120 s with a 20 s step, all five filters used, and the SNRcc detection threshold is retained to 3.5. The EDC are the same as for the routine processing, but only the events with Nass≥20 are promoted to the XSEL. Some event hypotheses not formally matching the EDC, but considered as real, are saved in the pre-XSEL for further study. Since August 2021, the routine processing has been upgraded to a level similar to the high-resolution one due to the importance of accurate and expedite estimates. In the upgraded routine mode, the SNRcc threshold is 3.6. Other detection parameters are the same as in the high-resolution regime.

**Results**

The largest aftershocks originally found between September 11, 2016, and May 2018 and reported in [Kitov *et al.*, 2018] were confirmed in the two-stage data reprocessing. Their parameters were re-estimated in the high-resolution regime. Some of these aftershocks were also reported by other research groups [*e.g.*, Schaff *et al.*, 2018; Tian *et al.*, 2018] and found in the IDC routine processing [Kitov *et al.*, 2021]. In this study, many aftershocks were found between June 2018 and October 2021, and a few new aftershocks were found between the DPRK5 and May 2018 in addition to those already reported [Kitov *et al.*, 2018]. Eighty-nine reliable aftershocks of the DPRK5 and DPRK6 are presented in Table S1 (Supplement 1) together with their parameters obtained by the cross-correlation method. The number of associated detections (templates) at each station and the sum of these individual inputs is presented in the last three columns of Table S1 and illustrate the reliability of these aftershocks. The larger is the Nass value, the higher the reliability of the event hypothesis. The signals from both DPRK tests, which induced the aftershock sequence, do not demonstrate the best correlation with their aftershocks due to the difference in source functions for the P- and S-waves (see Figures 1 through 7). The relative magnitude (RM) is calculated using the logarithm of the ratio of the RMS amplitudes of master and slave events in the CC-window for the detection filter/CWL combination. The DPRK5 and DPRK6 relative magnitudes are also calculated by this procedure. They are different from the corresponding $m_b$ estimates. For an array, the RMS amplitude is calculated using all individual channels filtered by the detection filter. As an alternative, one could calculate the RMS amplitude using the beam steered to the master event, as was used in the P/S spectral amplitude estimates [Kitov and Rozhkov, 2017].

In the companion paper, nine aftershocks of the DPRK3 and DPRK4 were detected [Kitov *et al.*, 2021]. The biggest aftershocks of the DPRK3 and DPRK4 may manifest the final cavity or chimney collapse. For the DPRK3 it happened ~450 days after the event and the last DPRK4 aftershock occurred ~180 days after the explosion. The same collapse processes accompanied by aftershocks are observed after the DPRK5 and DPRK6, and one can expect the final cavity/chimney collapse to stop the aftershock activity. Four years after the DPRK6, seismic activity is still observed within the DPRK test site. Summing the aftershocks before and after the DPRK5, one obtains ~100 events found by the WCC method using the multi-master approach. The DRPK aftershock activity is likely not finished, and we expect new events in the near future. The recovery of the DPRK aftershocks and the estimation of their parameters may shed some light on the processes related to the underground nuclear tests' source functions, the embedding rock properties, and the inhomogeneous geological structure. The aftershocks are also used for remote monitoring of the nuclear test sites [Adushkin *et al.*, 2017] and for the purposes of the on-site inspection.

There are several aftershocks almost matching the EDC, which are not formally included in the XSEL. We retain them in an additional list similar to the Late Event Bulletin (LEB), where the IDC saves the event hypotheses not formally matching the EDC of the REB, but physically viable and potentially valid in the complete seismic network of the International Monitoring System. Future changes in the current EDC version adopted by the IDC are also possible. Hence, saving the aftershocks almost matching the EDC for the DPRK aftershocks may be helpful for further analysis with new data and by advanced methods.

The lower similarity between the signals generated by the DPRK underground explosion and their aftershocks is a well-known feature and can be accurately explained by the difference in source mechanisms. There is a striking difference between the signals generated by aftershocks, however, which needs a thorough quantitative study and physical explanation. The difference can be expressed using various parameters of similarity based on the SNRcc measured in the high-resolution regime. In Table S1, eighty-nine DPRK aftershocks are distributed over three groups: a) related to the DPRK5; b) related to the DPRK6; c) poor events which are difficult to interpret unambiguously. Column 3 lists the assignment to one of these three groups: 5 (DPRK5), 6 (DPRK6), and P (poor).

For reprocessing, we used 29 master events: 6 DPRK explosions and 23 aftershocks observed between September 2016 and April 2018 [Kitov *et al*., 2018; Kitov *et al*., 2021]. There are 29 templates at KSRS and 28 templates at USRK. These 57 templates were used in the multi-master method to find and characterize aftershocks. For the analysis of the source region for each aftershock, we used two subsets of master events as related to DPRK5 and DPRK6. Six explosions are excluded from this analysis as well as the immediate aftershock of the DPRK6 (id=2017246_1), which is likely related to the DPRK6 cavity collapse and has signals similar to those generated by the explosions. This aftershock is also merged with strong coherent noise from the DPRK6.

In Figures S1 and S2, the SNRcc values are listed as obtained by 22 templates at two stations, USRK and KSRS, respectively. These templates are split into two groups: 10 templates (from 20172661 to 20173433) are related to the cluster of the DPRK6 aftershocks and 12 templates (from 20162551 to 20181122) belong to the DPRK5 cluster. Two clusters were initially identified in an iterative process, and the final sets of templates for the DPRK5 and DPRK6 have been obtained in 2018 [Kitov *et al*., 2018]. It has not been changed ever since and used as a reference for the cluster identification for all new aftershocks. The aftershocks found by these templates are listed in row 1 and named according to the Julian day and the index number in that day: 2017246_2 is the second aftershock observed on September 3, 2017. The SNRcc represents the level of similarity of the associated signals. If a signal is not detected by a given template, then the maximum SNRcc value is determined within ± 3 s of the corresponding origin time. It has to be lower than the detection threshold, but in rare cases, the detection procedure can miss good detection within the association window when two or more detections with different filter/CWL combinations are competing, and the winning detection is out of the association window due to the microseismic noise influence [Kitov *et al*., 2021].

The colour formatting in Figures S1 and S2 is applied to each column separately. This formatting allows highlighting the aftershocks better correlating with a given cluster in relative terms. The green cells indicate higher relative similarity, and red cells show a lower similarity of a given aftershock with the corresponding master/template. For example, panel

a) in Figure S1 demonstrates that, at station USRK, the events in the bottom part better correlate (green cells) with the immediate aftershock of the DPRK5: 20162551, and other templates in the DPRK5 group. All 32 aftershocks in panel a) belong to the DPRK5 cluster, as measured at station USRK. Autocorrelation is excluded from the further analysis, and the corresponding cells are empty. The same panel in Figure S2 confirms the DRPK 5 identification by data at station KSRS. Panels b) in Figures S1 and S2 illustrate the higher correlation between 46 aftershocks related to the DPRK6 cluster with the DPRK6 templates at stations USRK and KSRS. Panel c) shows relatively reliable aftershocks without significant difference in correlation between events associated with the DPRK5 and DPRK6. These aftershocks are characterized by lower SNRcc values, which make the DPRK5/DPRK6 discrimination a difficult task.

The relative similarity between aftershocks associated with the DPRK5 and DPRK6 as well as the absolute level of SNRcc is practically identical at USRK and KSRS. Figure S3 presents a compilation of the corresponding panels in Figures S1 and S2 with the colour formatting of the columns representing both stations: USRK - upper part and KSRS – lower part. Panel a) in Figure S3 represents the DPRK5 related aftershocks, and panel b) – the DPRK6 aftershocks. For a given cluster, SNRcc values are very close at KSRS and USRK despite the principal difference in the aperture and number of sensors (3 km and 9 sensors for USRK and 15 km and 19 sensors for KSRS). The number of channels is likely compensated by the sampling rate (40 Hz for USRK and 20 Hz for KSRS) since the total template length is a product of the number of channels, length of cross-correlation window, and sampling rate (frequency band). The independence on aperture is likely related to the high coherency of the signals over the spatial distribution of individual sensors within 15 km. One can conclude that both stations indicate the same aftershock identification. This observation favours the aftershock difference in their respective source areas. This is not a station-specific observation, and thus, the clustering is not the propagation path effect. Figures S1 through S3 illustrate the clustering effect in a qualitative form and highlight the principal parameters for quantitative analysis.

The following parameters can be used to characterize the quantitative difference between the DPRK5 and DPRK6 sequences: maximum SNRcc among all templates associated with DPRK5 and DPRK6, the average SNRcc for a given group of templates, and the sum of these parameters at USRK and KSRS. The maximum SNRcc within a given group of templates may reveal the spatial closeness and shape similarity, both important for the SNRcc estimate. The mean value within the same group demonstrates the affinity of the group members. The sum of these parameters may increase the difference between two groups of templates and assist a better identification.

Figure 16 presents the maximum SNRcc values among all templates associated with two different clusters: DPRK5 and DPRK6 for the aftershocks associated with the DPRK5 (upper panels) and DPRK6 (lower panels), as listed in Figures S1 and S2, respectively. This parameter identifies the template with the highest similarity with a given aftershock in the sequence corresponding to the DPRK5 (DPRK6) (autocorrelation is excluded). Two IMS stations are presented in Figure 16: USRK (left) and b) KSRS (right). For a given aftershock, the maximum SNRcc values obtained by templates related to the DPRK5 at both stations are much larger than those for the templates related to the DPRK6, and vice versa.

Figure 17 is similar to Figure 16 and presents the mean SNRcc for the DPRK5 and DPRK6 clusters. In both cases, the difference between max/mean SNRcc values illustrates a clear separation into two clusters - there is always a master event which is likely closer in space and has signals closer in shape, as expresses by the SNRcc, in one of the two groups at both stations. This master event identifies the cluster for each new aftershock in the joint DPRK aftershock sequence. The group affinity has to support this choice. Only poor aftershocks with low SNRcc values for the associated templates may have swapping identity between stations and parameters. This is the reason they are excluded from the clusters. These poor events do belong to one of the two clusters, however. Using an extended set of templates as obtained from the largest aftershocks found since May 2018, one could identify the clusters for these poor events. This result will not change the main conclusion of this study - there are two spatially separated clusters of the aftershock activity, likely related to the collapse of the underground cavities created by the DPRK5 and DPRK6 explosions.

The sum of the maximum and mean SNRcc at two stations is depicted in Figure 18. The maximum and the mean SNRcc both confirm the consistency of aftershock association with one of two clusters as determined by two sets of templates. The separation between two aftershocks clusters is convincing for all measures presented in Figures 16 through 18. It does not mean, however, that the SNRcc between two events in different clusters is always lower than the SNRcc within one group. The quality of templates, *e.g.* the amplitude of the $L_g$-phase above the ambient noise, varies in a relatively wide range and the SNRcc definitely depends on the proportion of noise and signal. The range of quality variation is approximately the same in both template groups.

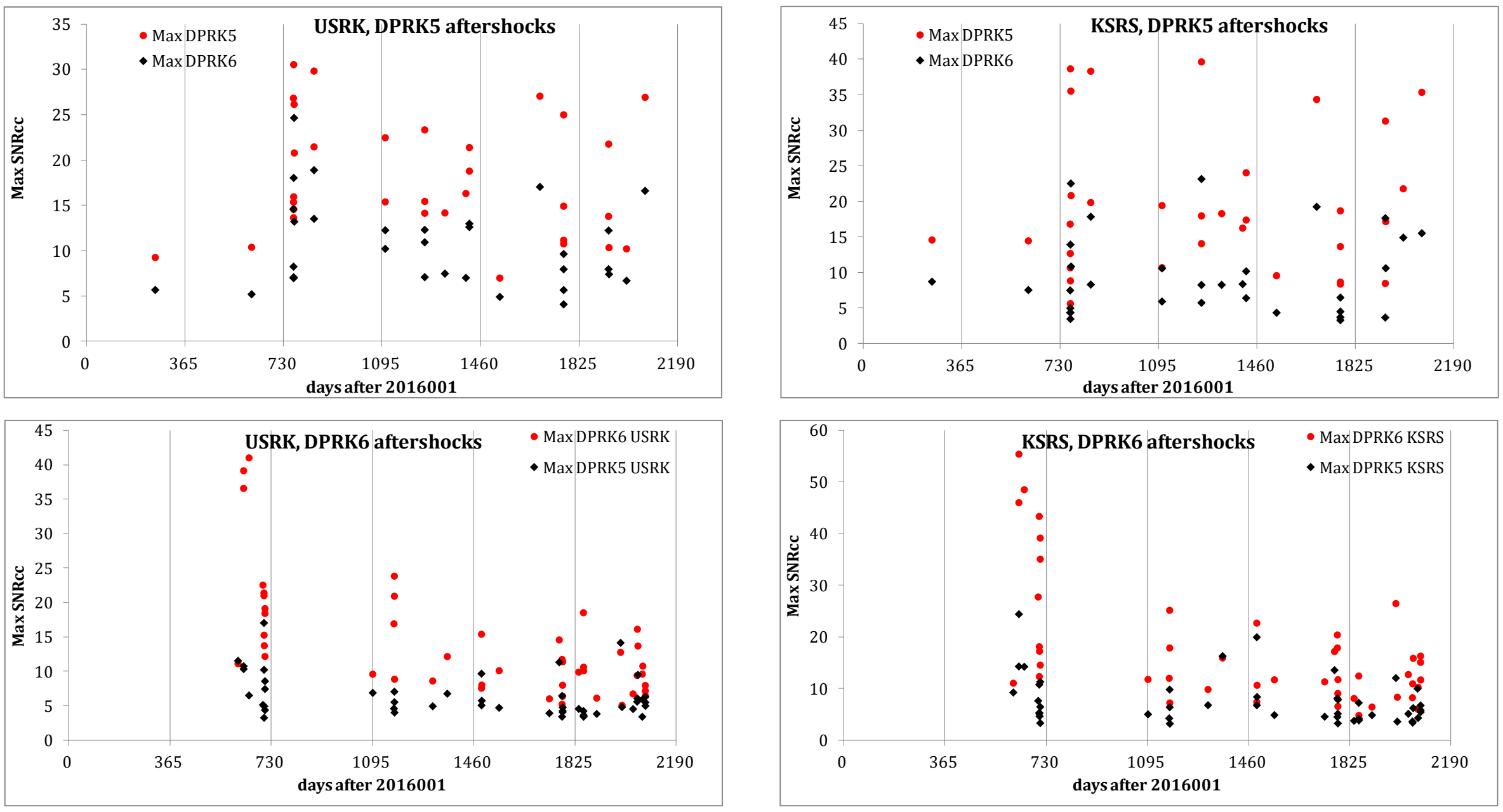

Figure 16. Maximum SNRcc values among all templates associated with two different clusters: DPRK5 and DPRK6 for the aftershocks associated with DPRK5 (upper panels) DPRK6 (lower panels). Two stations are presented, USRK (left) and KSRS (right). For the DPRK5 aftershocks, the maximum SNRcc obtained by templates related to the DPRK5 are much larger than those related to the DPRK6. For the DPRK6 aftershocks, the DPRK6 templates work better. The difference between max SNRcc values illustrates the separation into two clusters.

There is an important difference between the decay in max SNRcc for the DPRK5 and DPRK aftershocks. Figures 19 and 20 present the evolution of the max SNRcc at stations KSRS and USRK for the DPRK5 and DPRK6  aftershocks, as obtained separately with the DPRK6 and DPRK5 templates. The linear regression lines show that the DPRK6 aftershock sequence decays with time for both the DPRK6 and DPRK5 templates. The DPRK6 sequence is likely fading away, even though the number of the DPRK6 aftershocks in 2021 is larger than in 2018. The smaller aftershocks may manifest the final stage of the cavity and chimney collapse process or the propagation of the chimney to the surface, i.e. to the regions with lower tectonic stresses and the narrowing chimney. The DPRK5 sequence has fewer events but their strength, as expressed by the number of associated templates, Nass, and the SNRcc values, has not been changing from the very beginning. There were only 5 DPRK5 aftershocks in 2021, but 4 of them were larger than the majority of the DPRK6 aftershocks. Both stations reveal the same pattern, suggesting that these decay effects are related to the epicentral areas of the DPRK5 and DPRK6.

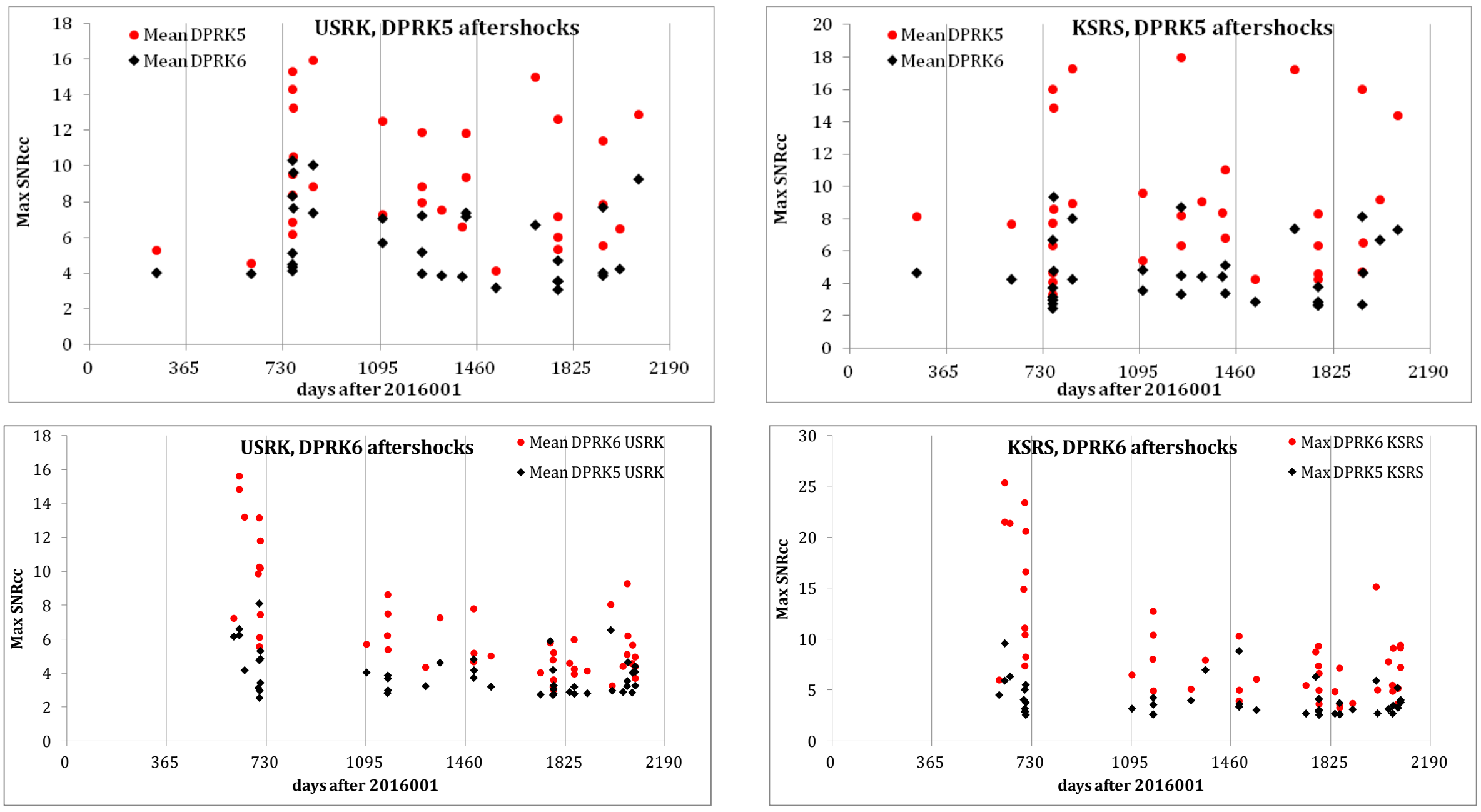

Figure 17. Mean SNRcc values as averaged over the templates associated with two different clusters: DPRK5 and DPRK6 for the aftershocks associated with DPRK5 (upper panels) DPRK6 (lower panels). Two stations are presented, USRK (left) and KSRS (right). For the DPRK5 aftershocks, the mean SNRcc values obtained for the templates related to the DPRK5 are much larger than those related to the DPRK6. For the DPRK6 aftershocks, the DPRK6 templates work better. The difference between the mean SNRcc values illustrates the separation into two clusters.

We defined several parameters to separate the DPRK5 and DPRK6 aftershocks. The total number of aftershocks in these two sequences is 78. The duration and characteristics of the DPRK aftershock activity make this case a very specific one. It is highly unlikely that such a behaviour may be repeated in the future in some other place. Therefore, a strict statistical assessment of the discrimination procedure would be excessive. A simple discrimination graph may be useful for a better understanding of the difference in the distribution of the DPRK5 and DPRK6 parameters. Figure 21 presents the difference between the sum of the max SNRcc at KSRS and USRK for the DPRK5 and DPRK6 templates. as well as the difference of the corresponding mean SNRcc values. The decision line is the x-axis. The events above the decision line belong to the DPRK5 cluster, and those below the line are the DPRK6 aftershocks. The poor events are also shown. They are closer to the decision line than the majority of the identified aftershock, but some DPRK 6 aftershocks are also close to the decision line. The poor events include 3 examples of the LEB-like hypotheses with Nass<20. The decision that the event is poor is taken when there are contradicting identification by various parameters or very low Nass around 20, as illustrated in Figure 22. For example, the identification changes with the parameter (max and mean SNRcc) at two stations for 2018161_1. The events 2019011_1, 2020095_1, and 2020105_1 are considered poor and are also not included in the XSEL according to low Nass<20. Despite the low Nass, these events have relatively high max and mean SNRcc and are likely real aftershocks.

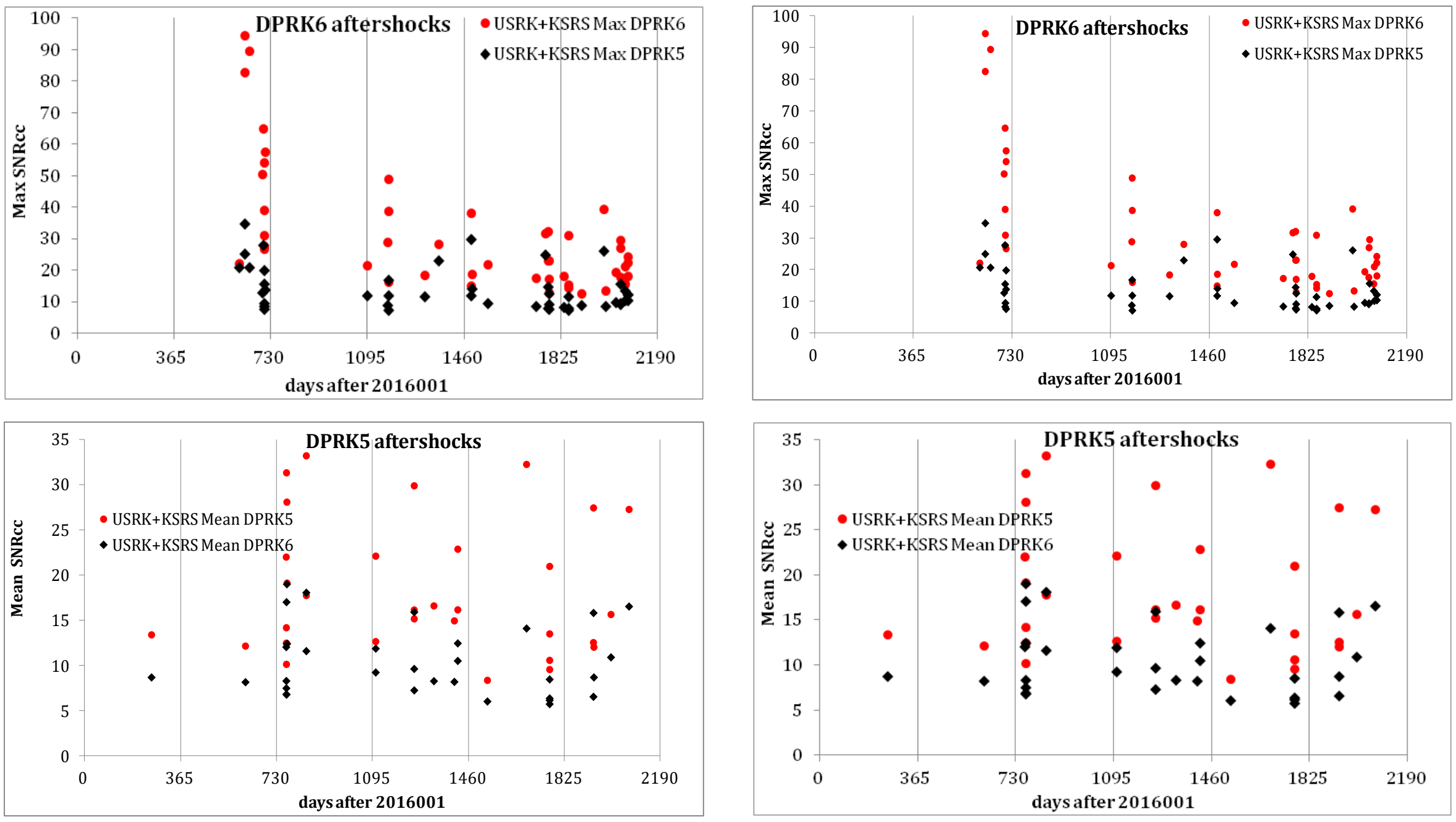

Figure 18. The sum of the maximum (upper panels) and mean (lower panels) SNRcc values at stations USRK and KSRS.

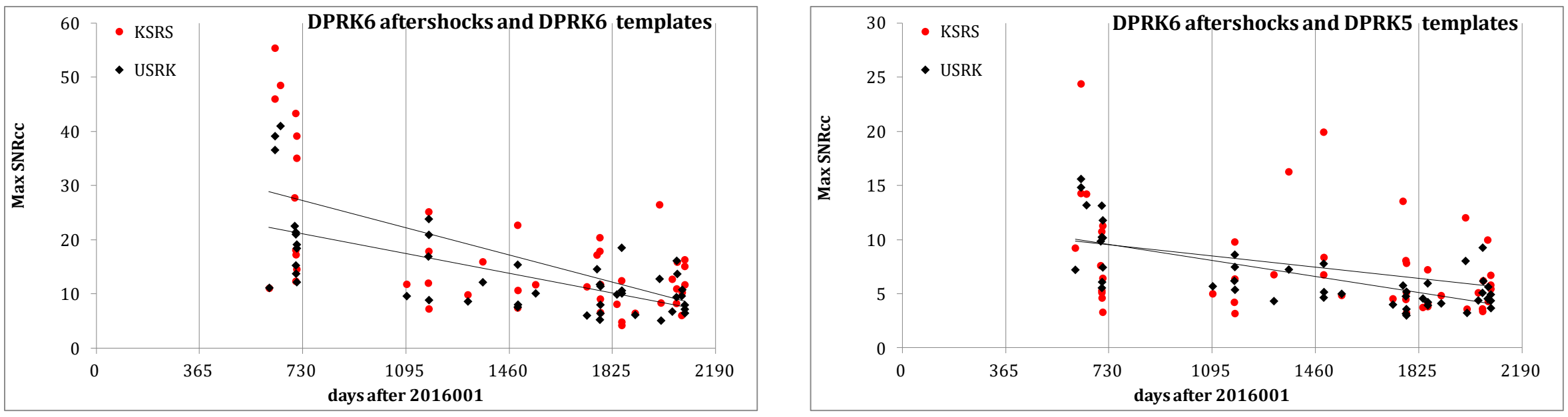

Figure 19. Evolution of max SNRcc for the DPRK6 aftershocks at stations KSRS and USRK as a function of time. Left panel: DPRK6 templates. Right panel: DPRK5 templates.

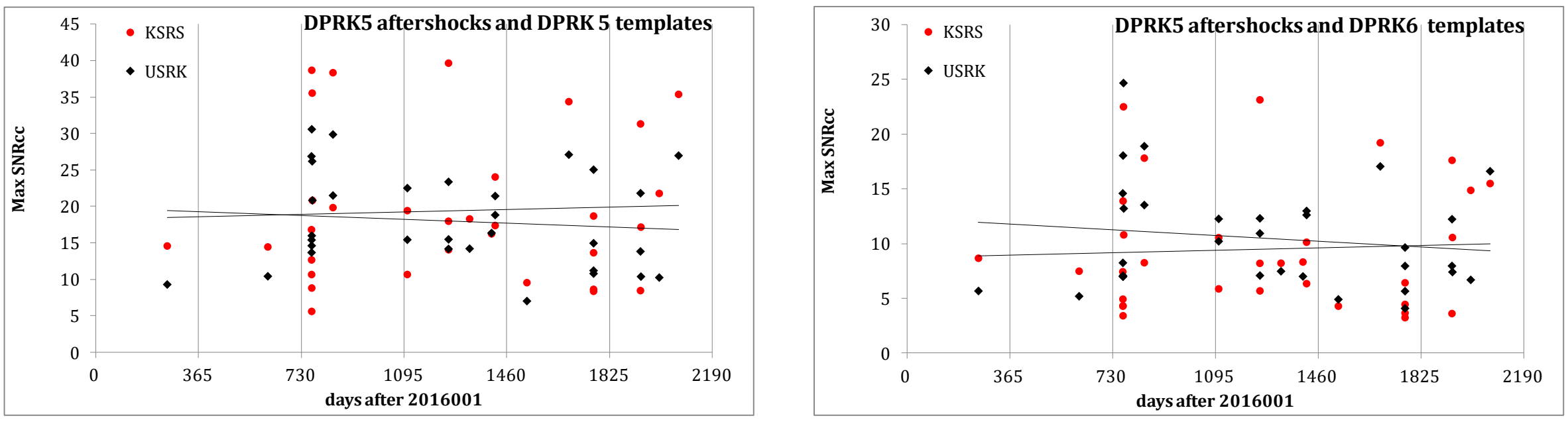

Figure 20. Evolution of max SNRcc for the DPRK5 aftershocks at stations KSRS and USRK as a function of time. Left panel: DPRK5 templates. Right panel: DPRK6 templates.

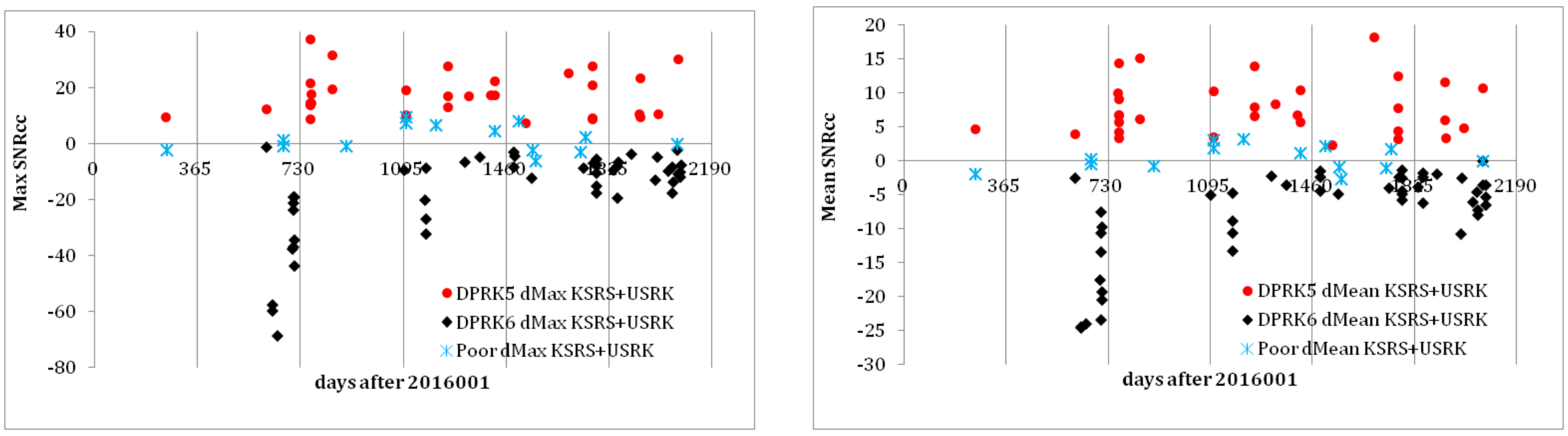

Figure 21. The decision line for the difference between the sum of the max (left) and the mean (right) SNRcc values.

| | 2016257_1 | 2017303_1 | 2017304_1 | 2018161_1 | 2019011_1 | 2019011_4 | 2019117_1 | 2019323_3 | 2020046_1 | 2020095_1 | 2020105_1 | 2020263_1 | 2020282_1 | 2021242_2 |
|---|---|---|---|---|---|---|---|---|---|---|---|---|---|---|
| Nass | 25 | 27 | 20 | 25 | 18 | 20 | 20 | 28 | 25 | 18 | 19 | 20 | 28 | 22 |
| USRK | 7.44 | 6.98 | 4.34 | 5.56 | 6.55 | 5.91 | 5.40 | 9.72 | 3.82 | 5.57 | 5.45 | 4.36 | 10.57 | 6.75 |
| KSRS | 4.67 | 4.36 | 3.79 | 4.65 | 3.27 | 3.70 | 2.69 | 3.33 | 6.13 | 6.55 | 8.99 | 6.02 | 3.09 | 4.42 |
| SUM MAX DPRK6 | 12.11 | 11.34 | 8.13 | 10.20 | 9.82 | 9.60 | 8.09 | 13.05 | 9.95 | 12.12 | 14.44 | 10.38 | 13.66 | 11.16 |
| USRK | 5.37 | 5.05 | 5.26 | 3.88 | 12.31 | 10.95 | 6.95 | 8.98 | 3.76 | 5.37 | 3.28 | 3.99 | 8.99 | 3.71 |
| KSRS | 4.37 | 5.49 | 4.04 | 5.32 | 7.05 | 5.73 | 7.73 | 8.58 | 14.07 | 4.55 | 4.73 | 3.46 | 6.96 | 7.22 |
| SUM MAX DPRK5 | 9.73 | 10.54 | 9.30 | 9.20 | 19.36 | 16.68 | 14.69 | 17.56 | 17.83 | 9.92 | 8.02 | 7.46 | 15.96 | 10.93 |
| USRK | 5.19 | 4.18 | 3.28 | 3.29 | 4.04 | 3.56 | 2.60 | 6.08 | 2.80 | 3.57 | 3.58 | 2.84 | 4.91 | 3.67 |
| KSRS | 3.71 | 2.76 | 2.51 | 3.01 | 2.59 | 2.74 | 2.39 | 2.91 | 3.31 | 3.48 | 4.66 | 3.72 | 2.53 | 3.24 |
| SUM AVE DPRK6 | 8.90 | 6.93 | 5.79 | 6.30 | 6.63 | 6.29 | 4.99 | 8.99 | 6.11 | 7.05 | 8.24 | 6.55 | 7.43 | 6.92 |
| USRK | 3.94 | 3.06 | 3.17 | 2.72 | 5.90 | 4.97 | 4.03 | 5.48 | 2.81 | 3.42 | 2.73 | 2.89 | 5.33 | 2.60 |
| KSRS | 2.94 | 3.45 | 2.86 | 2.82 | 3.76 | 3.15 | 4.11 | 4.60 | 5.44 | 2.76 | 2.80 | 2.56 | 3.87 | 4.21 |
| SUM AVE DPRK5 | 6.88 | 6.50 | 6.03 | 5.54 | 9.66 | 8.11 | 8.14 | 10.08 | 8.25 | 6.18 | 5.53 | 5.45 | 9.20 | 6.81 |

Figure 22. The estimates of the max and mean SNRcc for poor aftershocks. An event is considered as poor according to low Nass (*e.g.*, 2019011_1 and 2019011_4) or/and the association changing with parameter (*e.g.*, 2019323_3 and 2020282_1)

Figures 23 and 24 depict the evolution of two aftershock sequences related to the DPRK5 and DPRK6 in one graph. There are two measures presented – the relative magnitude, ***RM***, and the number of associated templates, Nass. The over aftershock sequence starts on 2016255 and extends into the second half of 2021. The vertical lines are shown with a quarterly and yearly spacing. The aftershocks of the same sequence usually group in time, as it happened in April 2021, and do not merge with the other cluster within one-week intervals except one or two cases, which have to be closer investigated. The interaction between clusters is not immediate, but rather related to the propagation of stress relaxation waves after aftershocks. As a consequence, we observe two alternating sequences with varying spacing between individual bursts. The poor events are not specifically related to the prominent outbursts. Some of them are rather isolated events. The external influence on both sequences is also possible – a larger part of aftershocks since 2019 is observed in winter and spring. The vertical lines represent the years after 01.01.2016. The first DPRK5 aftershock was detected on 11.09.2016. In wintertime, additional snow load may affect the stress distribution. In spring, the snow melts, and water may penetrate the cracks and reduce the critical stress threshold. In 2021, however, a larger part of the aftershocks were observed in summer and fall.

The usage of two processing regimes is justified by the change in Nass presented in the last column of Table S1. For several event hypotheses created in the routine regime, the increase in the number of associated templates reach 8. One event obtained 11 new detections. The effect of the high-resolution processing has been growing in time with the overall decay in the SNRcc values observed for the DPRK6 aftershocks (see Figure19). The event hypotheses with 20≥Nass≥14 can be promoted to the XSEL as reliable members. There are no hypotheses with Nass<14 which match the EDC after the high-resolution processing. There are some event hypotheses which lost a few associated templates. The conflict between many filter/CWL combinations may result in suppression of valid detections by false detections when the SNRcc threshold is reduced to the level corresponding to the high rate of false detection [Kitov *et al.*, 2021]. And this is another reason not to use the high-resolution mode in routine processing. The upgraded routine processing provides practically the same sensitivity and resolution as the high-resolution mode. As a result, the change in Nass is 0 for the most recent period, and only one event has lost one associated detection.

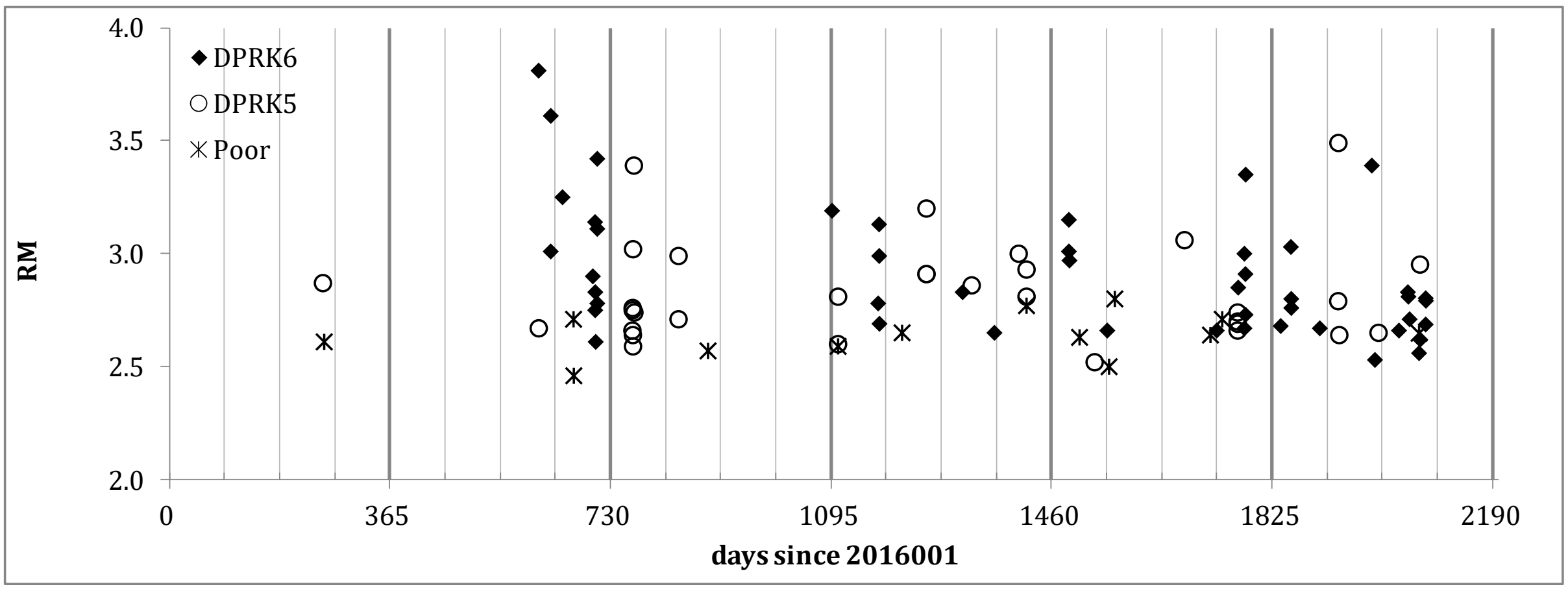

Figure 23. Relative magnitudes of the aftershocks.

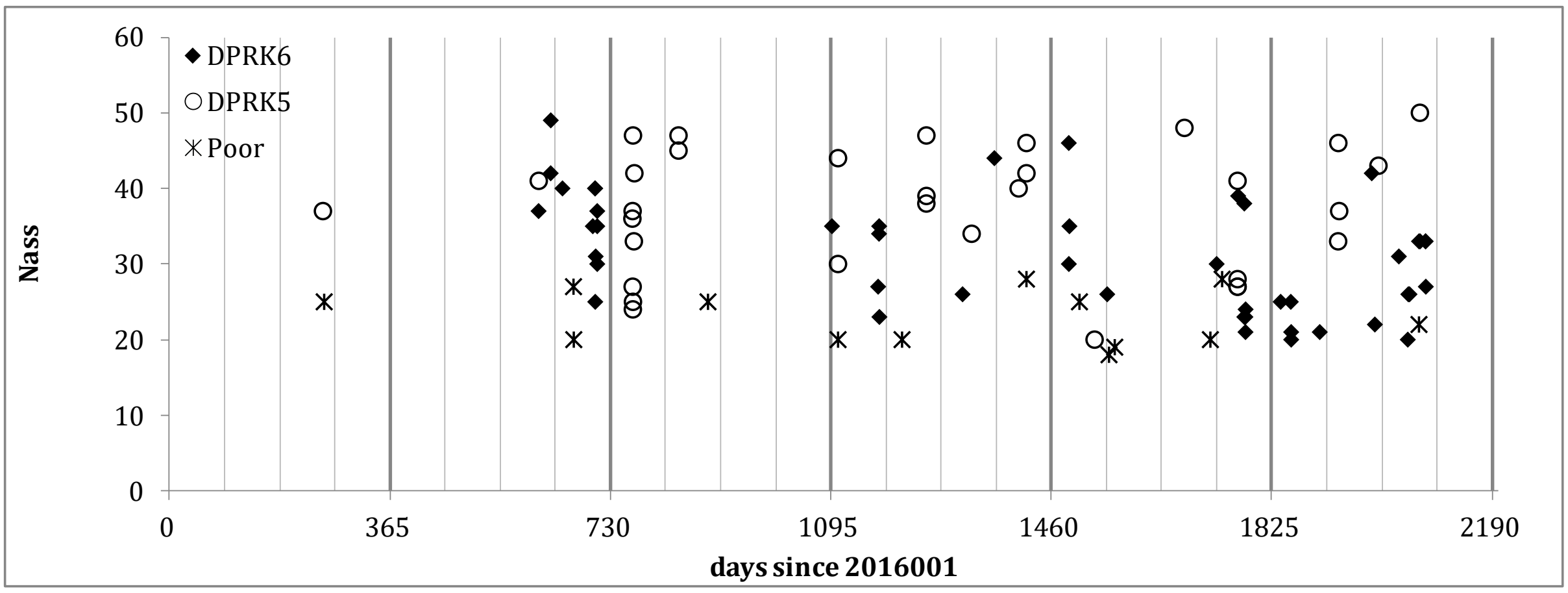

Figure 24. The number of associated templates at two stations.

The relative aftershock magnitudes do not demonstrate a significant decrease with time. This might be associated with the bias introduced by the ambient noise. Cross-correlation can find very low-amplitude signals, and then the noise defines the saturation of the magnitude estimates. Despite the decay in the SNRcc observed in detections associated with the DPRK6 sequence, the number of associated templates does not demonstrate any significant change with time and illustrates the power of the multi-master method. The number of associated templates is a robust parameter for the assessment of an event hypothesis probability. The radar-like approach has demonstrated its high sensitivity and resolution when applied to small aftershocks of the DPRK underground tests. This approach provides significant improvement in the method of remote monitoring of low-magnitude aftershocks we developed in [Adushkin *et al*., 2017].

## Discussion

There were no DPRK aftershocks detected by the multi-master method in the routine processing regime between January 1, 2009, and February 12, 2013, as shown in [Kitov *et al*., 2021]. The threshold of Nass=11 was obtained using the rate of the false event less than 1 per month during this period. The absence of reliable aftershock-like events before the DPRK3 is strong evidence that all 89 events in Table S1 are caused by the DPRK underground explosions, as well as the nine aftershocks of the DPRK3 and DPRK4. They are not natural tectonic events in that sense but are rather related to the sporadic relaxation of the gravitational and tectonic energy in seismic events, which is facilitated by the reduced critical stresses in the rocks damaged by the DPRK explosions. The source mechanisms and source functions of these aftershocks can be similar to those in shallow earthquakes.

In 2021, more than 20 new aftershocks of the DPRK explosions were found, and the elapsed time of 4 years from the DPRK6 represents a challenge for interpretation. The aftershock sequences of such a length are not reported for the events with body wave magnitude around 6 in hard rock. One of the possible mechanisms behind these observations is associated with the interaction of the zones of radial cracks generated by the DPRK5 and DPRK6. According to the scaling law, $R_e \sim 100 \cdot Y^{1/3}$, *i.e.,* the characteristic linear size, $R_e$, (elastic source radius or the damage zone radius) is proportional to the cube root of yield in *kt*. The depth difference between DPRK5 and DPRK6 is not larger than a few hundred of meters [Kitov *et al*., 2021]. Considering the body wave magnitudes of both explosions and the range of the corresponding yields, the distance between the DPRK5 and DPRK6 hypocentres should not be less than 800 m to 1.5 km. Figure 25 displays the relative locations of the DPRK underground explosions as estimated by pair-wise relative location based of waveform cross-correlation of the signals at IMS seismic stations [Bobrov *et al*., 2017b]. The DPRK5 is the reference event placed to (0,0) point. The circles with the elastic radii of the DPRK5 and DPRK6 are shown by dotted lines: $R_e$=300 m and $R_e$=700 m, respectively. The DPRK6 position is a guess from the arrival time difference at KSRS and USRK [Kitov *et al*., 2021]: 1000 m to the west and 0 m to the north. The accurate relative location based on cross-correlation was not possible in the DPRK6 hypocentre. The size of the DPRK6 elastic source is not negligible for the precision requirement of the travel time difference of ~0.005 s that corresponds to 40 m for the $P_n$-wave velocity of 8 km/s. With the uncertainty in the relative locations and elastic radii, one may suggest that the damaged zones and chimneys may influence each other by changing stresses and deformations at distances larger than their elastic radii.

This mechanism of interaction suggests that there are two individual clusters near the hypocentres of the DPRK5 and DPRK6, and likely some events in the zone between these two hypocentres. The possibility to identify the difference between the DPRK5 and DPRK6 aftershocks is based on the availability of a good aftershock from the DPRK5 detected by cross-correlation on September 11, 2016. This is a template, discriminating (see Figure 21) the events from DPRK5 (high similarity) and DPRK6 (lower similarity). The events that occurred in a few weeks after the DPRK6 can be used as its templates. The difference in waveform templates between two clusters with spacing from 1 km to 2 km is likely related to local conditions of signal generation. The most prominent phase from these aftershocks is $L_g$ and this phase mainly defines the cross-correlation coefficients observed in the study. The propagation paths to USRK and KSRS are identical for all aftershocks except the near-source region, which also suffered significant structural changes induced by high-amplitude shock waves. The difference between P-waves from the DPRK explosions used for relative location is lower because the outgoing waves are almost vertical. The $L_g$-waves from the DPRK explosions have slightly lowered similarity due to the same local effects.

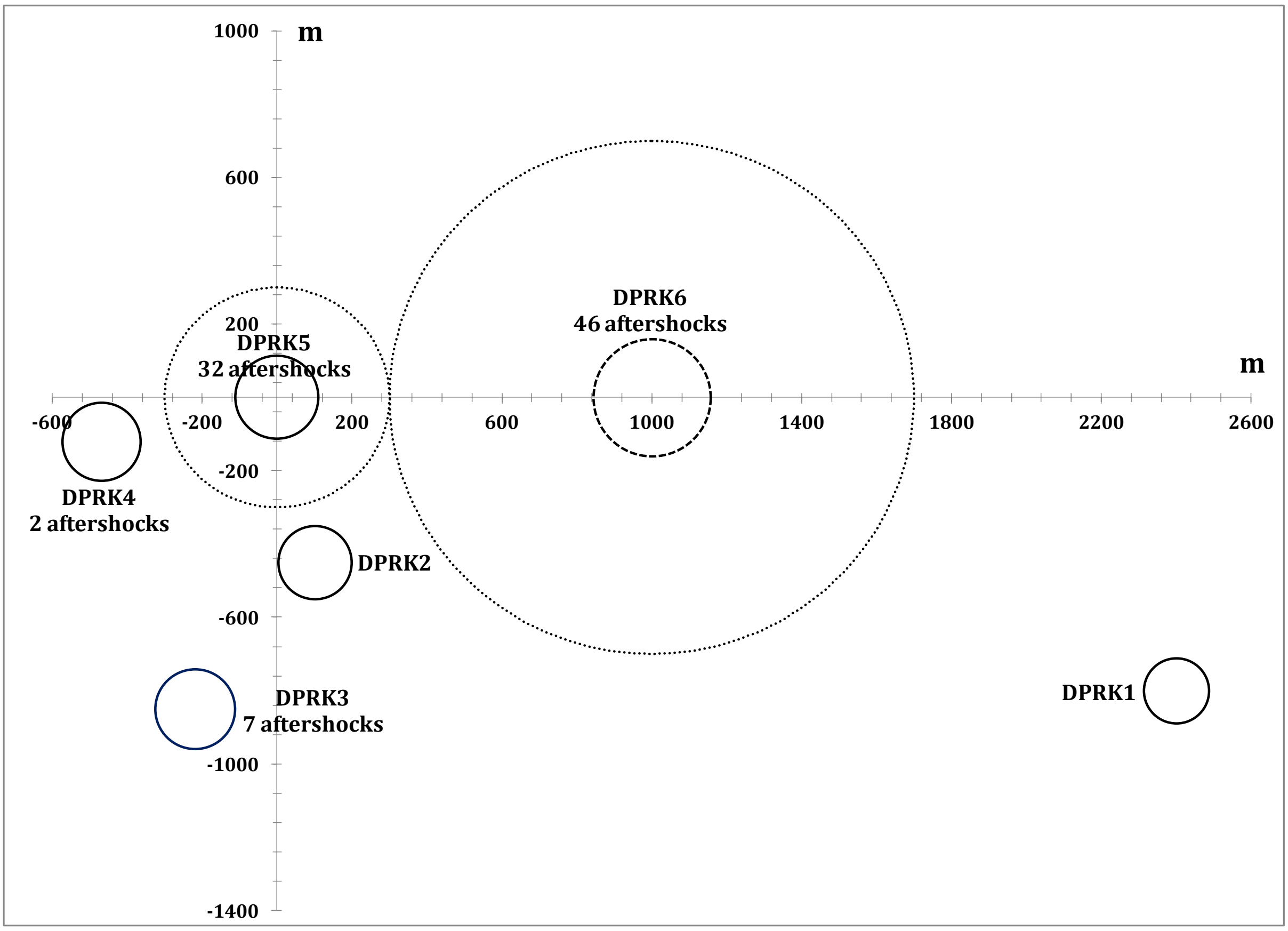

Figure 25. Relative location of the DPRK underground explosions, as estimated by pair-wise relative location based on waveform cross-correlation of the signals at IMS seismic stations. The circles with the elastic radii of the DPRK5 and DPRK6 are shown by dotted lines. Their radii are scaled according to the $r \sim Y^{1/3}$ law: 300 m and 700 m, respectively The DPRK6 position is a guess from the arrival time difference at KSRS and USRK. The relative location based on cross-correlation failed.

The source mechanism of the aftershocks is likely the same and should be similar to the usual double-couple mechanism of earthquakes. The methods of moment tensor estimates are not effective for two-station solutions. More stations in different azimuths are needed to reveal the source mechanism of the DPRK aftershocks. It would be helpful for understanding of the difference between the DPRK5 and DPRK6 sequences. On the other hand, the sources of the aftershocks with magnitudes between 2 and 3 have linear sizes of a few tens of meters and the co-seismic phase duration is likely less than 0.1 s. In that sense, all aftershocks are point sources with δ-time functions. Therefore, there is no big difference in relative P- and S-wave emission between aftershock, except the radiation pattern term. Stations KSRS and USRK are situated in the opposite directions, and the source directivity term is hard to estimate from their data.

The study of the difference between the DPRK5 and DPRK6 aftershocks is in the very beginning. There are also DPRK3 and DPRK4 aftershocks with signals well correlated with the last two explosions conducted at the DPRK test site. The relative location of the DPRK6 has to be accurately estimated for the analysis of mechanical interaction between the underground cavities and collapsing chimneys. The mechanics of the interaction process accompanied by the seismic energy emission is important for the theoretical consideration as well as for practical applications related to the safety and containment problems.

Further development of the multi-master method based on waveform cross-correlation is needed for the enhancement of the CTBT seismic monitoring and for the On-site Inspection planning. Weak aftershocks may not be found at regional distances from the underground explosion in the different geological environments or conducted with evasion measures like the underground cavity decoupling. It could be helpful to use synthetic seismograms calculated for a regional velocity structure or to modify the explosion waveforms at regional stations according to the difference in source mechanisms of a shallow earthquake and explosion.

## References

Adushkin V.V. and A.A. Spivak. 1993. Geomechanics of Large-Scale Explosions. Nedra, Moscow. [in Russian].

Adushkin V.V., I.O. Kitov, N. L. Konstantinovskaya, K.S. Nepeina, M.A. Nesterkina, and I.A. Sanina. 2015. Detection of Ultraweak Signals on the Mikhnevo Small-Aperture Seismic Array by Using Cross-Correlation of Waveforms. Doklady Earth Sci. 460 (2), 189–191.

Adushkin V.V., D.I. Bobrov, I.O. Kitov, M.V. Rozhkov, and I.A. Sanina. 2017. Remote detection of aftershock activity as a new method of seismic monitoring. Doklady Earth Sciences. 473 , Part 1, pp. 303–307.

Adushkin V.V., I.O. Kitov, and I.A. Sanina. 2021. Clustering of aftershock activity of underground explosions in the DPRK. Doklady Earth Sciences. [in press]

Alvizuri C., Tape C. 2018. Full Moment Tensor Analysis of Nuclear Explosions in North Korea. Seismological Research Letters 89-6, November/December 2018.

DOI: 10.1785/0220180158

Bobrov D., I. Kitov, and L. Zerbo. 2014. Perspectives of Cross-Correlation in Seismic Monitoring at the International Data Centre. Pure Appl. Geophys. 171 (3–5), 439–468.

Bobrov D. I., I. O. Kitov, M. V. Rozhkov, and P. Friberg. 2016. Towards global seismic monitoring of underground nuclear explosions using waveform cross correlation. Part I: Grand master events. Seismic Instruments 52 (1), 43–59.

Bobrov D., I. Kitov, and M. Rozhkov. 2017a. Absolute and relative body wave magnitudes of five DPRK events as measured by the IDC. 2017 Science and Technology Conference, Hofburg Palace Vienna, Austria, 26- 30 June, 2017. T2.1-P2

Bobrov D., I. Kitov, and M. Rozhkov. 2017b. Absolute and relative location at the IDC: five DPRK events. 2017 Science and Technology Conference, Hofburg Palace Vienna, Austria, 26-30 June, 2017. T2.1-P3

Coyne J., D. Bobrov, P. Bormann, E. Duran, P. Grenard, G. Haralabus, I. Kitov, and Yu. Starovoit. 2012. In New Manual of Seismological Practice Observatory (NMSOP-2) (GFZ German Research Centre for Geosciences, Potsdam, 2012), Chap. 15
.
Kim W.-Y., Richards P.G., Jo E.Y., Ryoo Y.G. 2018. Identification of seismic events on and near the North Korean test site following the underground nuclear test explosion of 2017 September 3. Seismol. Rev Lett. 89:2220–2230

Kitov I.O., O.P. Kuznetsov. 1990. Energy released in aftershock sequence of explosion. Dokl. Akad. Nauk SSSR, 315(4), 839–842.

Kitov I., M. Rozhkov. 2017. Discrimination of the DPRK underground explosions and their aftershocks using the P/S spectral amplitude ratio, arXiv:1712.01819.

Kitov I., Le Bras R., Rozhkov M., Sanina I. 2018. Discrimination of the DPRK underground explosions and their aftershocks using the Pg/Lg spectral amplitude ratio. 2018 Seismology of the Americas Meeting Latin American and Caribbean Seismological Commission Seismological Society of America, 14–17 May 2018, Miami, Florida, Seismological Research Letters Volume 89, Number 2B March/April 2018, p. 869.

Kitov I.O., S.B. Turuntaev, A.V. Konovalov, A. A. Stepnov, and V.V. Pupatenko. 2019. Use of Waveform Cross Correlation to Reconstruct the Aftershock Sequence of the August 14, 2016, Sakhalin Earthquake September 2019. Seismic Instruments 55(5), 544–558. DOI: 10.3103/S0747923919050074

Kitov I., D. Bobrov, and M. Rozhkov. 2021. Seismic events found by waveform cross correlation after the announced underground nuclear tests conducted by the DPRK on 12.02.2013 and 06.01.2016. Aftershocks or hidden nuclear tests?, arXiv:

Schaff D.P., Kim W.-Y., Richards P.G., Jo E., Ryoo Y. 2018. Using waveform cross-correlation for detection, location, and identification of aftershocks of the 2017 nuclear explosion at the North Korea test site. Seismol. Res Lett. 89:2113–2119

Tape W., Tape C. 2015. A uniform parameterization of moment tensors. Geophys. J. Int. 202: 2074–2081. https://doi.org/10.1093/gji/ggv262

Tian D., Yao J., Wen L. 2018. Collapse and earthquake swarm after North Korea's 3 September 2017 nuclear test. Geophys. Res Lett. 5:3976–3983. DOI:/10.1029/2018GL077649

## Supplement 1

Table S1. Principal parameters of the aftershocks found between 01.01.2009 and 01.10.2021. Routine and high-resolution processing are compared. Only the events with Nass≥20 obtained in the high-resolution regime are presented. Individual input of KSRS and USRK is shown separately. The change in Nass from the routine to the high-resolution stage, dNass, demonstrates the valued added.

| N | Date | # | DPRK | Time | Routine | | | Final | | | | |
|---|---|---|---|---|---|---|---|---|---|---|---|---|
| | | | | | Nass | RM | | Nass | KSRS | USRK | RM | dNass |
| 1 | **2016255** | 1 | 5 | 1:50: 48 | 36 | 2.84 | | 37 | 19 | 18 | 2.87 | 1 |
| 2 | 2016257 | 1 | **P** | 03:57:42 | 26 | 2.64 | | 25 | 7 | 18 | 2.61 | -1 |
| 3 | 2017246 | 1 | 6 | 03:38:30 | 36 | 3.84 | | 37 | 17 | 20 | 3.81 | 1 |
| 4 | **2017246** | 2 | 5 | 09:31:28 | 31 | 2.64 | | 41 | 19 | 22 | 2.67 | 10 |
| 5 | **2017266** | 1 | 6 | 4:42: 58 | 42 | 3.05 | | 42 | 21 | 21 | 3.01 | 0 |
| 6 | **2017266** | 2 | 6 | 8:29: 14 | 48 | 3.61 | | 49 | 25 | 24 | 3.61 | 1 |
| 7 | **2017285** | 1 | 6 | 16:41:07 | 43 | 3.26 | | 43 | 18 | 22 | 3.25 | 0 |
| 8 | 2017303 | 1 | **P** | 23:37:48 | 20 | 2.67 | | 27 | 10 | 17 | 2.71 | 7 |
| 9 | **2017304** | 1 | **6/P** | 10:20:12 | 18 | 2.52 | | 20 | 5 | 15 | 2.46 | 2 |
| 10 | **2017335** | 1 | 6 | 22:45:54 | 32 | 2.90 | | 35 | 18 | 17 | 2.90 | 3 |
| 11 | **2017339** | 1 | 6 | 14:40:50 | 42 | 3.14 | | 40 | 19 | 21 | 3.14 | -2 |
| 12 | 2017339 | 2 | 6 | 19:55:53 | 35 | 2.76 | | 40 | 16 | 24 | 2.75 | 5 |
| 13 | 2017339 | 3 | 6 | 23:30:10 | 23 | 2.82 | | 25 | 10 | 15 | 2.83 | 2 |
| 14 | **2017340** | 1 | 6 | 16:20:04 | 23 | 2.61 | | 31 | 14 | 17 | 2.61 | 8 |
| 15 | **2017343** | 1 | 6 | 06:08:39 | 22 | 2.81 | | 30 | 11 | 19 | 2.78 | 8 |
| 16 | **2017343** | 2 | 6 | 6:13: 32 | 41 | 3.41 | | 35 | 17 | 18 | 3.42 | -6 |
| 17 | **2017343** | 3 | 6 | 06:39:59 | 39 | 3.07 | | 37 | 18 | 19 | 3.11 | -2 |
| 18 | **2018036** | 1 | 5 | 10:32:30 | 36 | 2.68 | | 36 | 12 | 24 | 2.66 | 0 |
| 19 | **2018036** | 2 | 5 | 20:07:29 | 28 | 2.73 | | 37 | 14 | 23 | 2.76 | 9 |
| 20 | **2018036** | 3 | 5 | 21:57:35 | 23 | 2.76 | | 27 | 8 | 19 | 2.75 | 4 |
| 21 | **2018037** | 1 | 5 | 04:49:36 | 22 | 2.60 | | 24 | 5 | 19 | 2.64 | 2 |
| 22 | **2018037** | 2 | 5 | 10:12:30 | 17 | 2.59 | | 25 | 5 | 20 | 2.59 | 8 |
| 23 | **2018037** | 3 | 5 | 10:53:52 | 49 | 3.02 | | 47 | 23 | 24 | 3.02 | -2 |
| 24 | **2018038** | 1 | 5 | 21:46:23 | 47 | 3.36 | | 33 | 17 | 16 | 3.38 | -14 |
| 25 | **2018039** | 1 | 5 | 17:39:17 | 41 | 2.73 | | 42 | 19 | 23 | 2.74 | 1 |
| 26 | **2018112** | 1 | 5 | 19:25:09 | 36 | 2.69 | | 47 | 22 | 25 | 2.71 | 11 |
| 27 | **2018112** | 2 | 5 | 19:31:18 | 50 | 3.01 | | 45 | 25 | 20 | 2.99 | -5 |
| 28 | 2018161 | 1 | **P** | 15:53:04 | 17 | 2.57 | | 25 | 11 | 12 | 2.57 | 8 |
| 29 | 2019001 | 1 | 6 | 22:20:27 | 31 | 3.20 | | 35 | 16 | 19 | 3.19 | 4 |
| 30 | 2019011 | 2 | 5 | 21:34:28 | 32 | 2.61 | | 30 | 8 | 22 | 2.60 | -2 |
| 31 | 2019011 | 3 | 5 | 23:15:46 | 41 | 2.80 | | 44 | 19 | 25 | 2.81 | 3 |
| 32 | 2019011 | 4 | **P** | 23:19:21 | 15 | 2.56 | | 20 | 5 | 15 | 2.59 | 5 |
| 33 | 2019078 | 1 | 6 | 08:32:36 | 19 | 2.75 | | 27 | 11 | 16 | 2.78 | 8 |
| 34 | 2019079 | 1 | 6 | 19:41:03 | 31 | 3.13 | | 34 | 17 | 17 | 3.13 | 3 |
| 35 | 2019080 | 1 | 6 | 01:55:35 | 31 | 2.99 | | 35 | 17 | 18 | 2.99 | 4 |
| 36 | 2019080 | 2 | 6 | 10:45:39 | 20 | 2.70 | | 23 | 10 | 13 | 2.69 | 3 |
| 37 | 2019117 | 1 | **P** | 22:37:16 | 14 | 3.63 | | 20 | 10 | 10 | 2.65 | 6 |
| 38 | 2019158 | 1 | 5 | 03:49:30 | 37 | 2.91 | | 38 | 16 | 22 | 2.91 | 1 |
| 39 | 2019158 | 2 | 5 | 05:18:39 | 46 | 3.19 | | 47 | 25 | 22 | 3.20 | 1 |
| 40 | 2019158 | 3 | 5 | 07:45:43 | 31 | 2.89 | | 39 | 17 | 22 | 2.91 | 8 |
| 41 | 2019218 | 4 | 6 | 04:05:31 | 24 | 2.78 | | 26 | 12 | 14 | 2.83 | 2 |
| 42 | 2019233 | 1 | 5 | 03:36:52 | 34 | 2.85 | | 34 | 16 | 17 | 2.86 | 0 |
| 43 | 2019270 | 1 | 6 | 19:01:31 | 45 | 2.65 | | 44 | 22 | 22 | 2.65 | -1 |
| 44 | 2019310 | 1 | 5 | 23:51:34 | 37 | 2.99 | | 40 | 19 | 21 | 3.00 | 3 |
| 45 | 2019323 | 1 | 5 | 21:59:46 | 38 | 2.79 | | 42 | 17 | 25 | 2.81 | 4 |
| 46 | 2019323 | 2 | 5 | 22:02:29 | 41 | 2.93 | | 46 | 23 | 23 | 2.93 | 5 |
| 47 | 2019323 | 3 | **P** | 22:49:36 | 27 | 2.76 | | 28 | 8 | 20 | 2.77 | 1 |
| 48 | 2020029 | 1 | 6 | 00:33:46 | 44 | 3.15 | | 46 | 24 | 22 | 3.15 | 2 |
| 49 | 2020029 | 2 | 6 | 02:48:28 | 27 | 3.00 | | 30 | 11 | 19 | 3.01 | 3 |
| 50 | 2020030 | 1 | 6 | 00:50:00 | 28 | 2.95 | | 35 | 13 | 22 | 2.97 | 7 |
| 51 | 2020046 | 1 | **P** | 17:05:05 | 15 | 2.62 | | 25 | 12 | 13 | 2.63 | 10 |
| 52 | 2020071 | 1 | 5 | 17:09:01 | 22 | 2.51 | | 20 | 12 | 8 | 2.52 | -2 |
| 53 | 2020092 | 1 | 6 | 15:14:59 | 24 | 2.64 | | 26 | 12 | 14 | 2.66 | 2 |
| 54 | 2020220 | 1 | 5 | 12:27:41 | 44 | 3.04 | | 48 | 23 | 25 | 3.06 | 4 |
| 55 | 2020263 | 1 | **P** | 11:22:26 | 15 | 2.64 | | 20 | 12 | 8 | 2.64 | 5 |
| 56 | 2020273 | 1 | 6 | 20:23:29 | 23 | 2.68 | | 30 | 14 | 16 | 2.66 | 7 |
| 57 | 2020282 | 1 | P | 21:49:37 | 24 | 2.72 | | 28 | 7 | 21 | 2.71 | 4 |
| 58 | 2020308 | 1 | 5 | 10:12:17 | 19 | 2.69 | | 27 | 11 | 16 | 2.70 | 8 |
| 59 | 2020308 | 2 | 5 | 10:13:52 | 24 | 2.69 | | 28 | 8 | 20 | 2.66 | 4 |
| 60 | 2020308 | 3 | 5 | 10:30:32 | 38 | 2.72 | | 41 | 20 | 21 | 2.74 | 3 |
| 61 | 2020308 | 4 | 5 | 11:28:06 | 19 | 2.69 | | 27 | 9 | 18 | 2.69 | 8 |
| 62 | 2020309 | 1 | 6 | 10:18:42 | 43 | 2.81 | | 39 | 22 | 17 | 2.85 | -4 |
| 63 | 2020319 | 1 | 6 | 14:45:35 | 30 | 2.99 | | 38 | 18 | 20 | 3.00 | 8 |
| 64 | 2020319 | 2 | 6 | 15:40:26 | 15 | 2.68 | | 23 | 16 | 7 | 2.67 | 8 |
| 65 | 2020321 | 1 | 6 | 9:11: 28 | 14 | 2.93 | | 23 | 13 | 10 | 2.91 | 9 |
| 66 | 2020321 | 2 | 6 | 13:38:34 | 21 | 2.73 | | 21 | 10 | 11 | 2.73 | 0 |
| 67 | 2020321 | 3 | 6 | 18:10:28 | 18 | 3.34 | | 24 | 14 | 10 | 3.35 | 6 |
| 68 | 2021013 | 1 | 6 | 19:29:30 | 20 | 2.66 | | 25 | 9 | 16 | 2.68 | 5 |
| 69 | 2021030 | 1 | 6 | 15:07:45 | 23 | 3.02 | | 25 | 17 | 8 | 3.03 | 2 |
| 70 | 2021031 | 1 | 6 | 5:10: 31 | 15 | 2.77 | | 21 | 11 | 10 | 2.80 | 6 |
| 71 | 2021031 | 2 | 6 | 06:18:12 | 14 | 2.74 | | 20 | 11 | 9 | 2.76 | 6 |
| 72 | 2021078 | 1 | 6 | 12:48:50 | 17 | 2.66 | | 21 | 9 | 12 | 2.67 | 4 |
| 73 | 2021108 | 1 | 5 | 23:20:03 | 27 | 2.75 | | 33 | 12 | 21 | 2.79 | 6 |
| 74 | 2021109 | 1 | 5 | 06:48:48 | 50 | 3.49 | | 46 | 24 | 22 | 3.49 | -4 |
| 75 | 2021110 | 1 | 5 | 17:12:31 | 32 | 2.62 | | 37 | 21 | 16 | 2.64 | 5 |
| 76 | 2021164 | 1 | 6 | 14:57:30 | 41 | 3.40 | | 42 | 20 | 22 | 3.39 | 1 |
| 77 | 2021169 | 1 | 6 | 19:45:57 | 13 | 2.51 | | 22 | 9 | 13 | 2.53 | 9 |
| 78 | 2021175 | 1 | 5 | 17:09:32 | 37 | 2.63 | | 43 | 22 | 13 | 2.65 | 6 |
| 79 | 2021209 | 1 | 6 | 13:20:45 | 27 | 2.66 | | 31 | 13 | 18 | 2.66 | 4 |
| 80 | 2021224 | 1 | 6 | 07:42:59 | 21 | 2.84 | | 20 | 8 | 12 | 2.83 | -1 |
| 81 | 2021225 | 1 | 6 | 04:35:30 | 30 | 2.79 | | 26 | 9 | 17 | 2.81 | -4 |
| 82 | 2021227 | 1 | 6 | 03:46:45 | 35 | 2.62 | | 26 | 10 | 18 | 2.71 | -9 |
| 83 | 2021242 | 1 | 6 | 20:33:25 | 33 | 2.56 | | 33 | 11 | 22 | 2.56 | 0 |
| 84 | 2021242 | 2 | **P** | 20:51:11 | 22 | 2.65 | | 22 | 5 | 17 | 2.65 | 0 |
| 85 | 2021244 | 1 | 5 | 09:39:49 | 50 | 2.95 | | 50 | 23 | 27 | 2.95 | 0 |
| 86 | 2021244 | 2 | 6 | 17:05:32 | 31 | 2.61 | | 30 | 13 | 18 | 2.62 | -1 |
| 87 | 2021253 | 1 | 6 | 18:16:41 | 33 | 2.69 | | 33 | 14 | 19 | 2.69 | 0 |
| 88 | 2021253 | 2 | 6 | 18:47:25 | 33 | 2.80 | | 33 | 16 | 17 | 2.80 | 0 |
| 89 | 2021254 | 1 | 6 | 04:17:02 | 27 | 2.79 | | 27 | 18 | 9 | 2.79 | 0 |

a)

| DPRK5 | 2016255_1 | 2017246_2 | 2018036_1 | 2018036_2 | 2018036_3 | 2018037_1 | 2018037_2 | 2018037_3 | 2018038_1 | 2018039_1 | 2018112_1 | 2018112_2 | 2019011_2 | 2019011_3 | 2019158_1 | 2019158_2 | 2019158_3 | 2019233_1 | 2019310_1 | 2019323_1 | 2019323_2 | 2020071_1 | 2020220_1 | 2020308_1 | 2020308_2 | 2020308_3 | 2020308_4 | 2021108_1 | 2021109_1 | 2021110_1 | 2021175_1 | 2021244_1 | Mean |
|---|---|---|---|---|---|---|---|---|---|---|---|---|---|---|---|---|---|---|---|---|---|---|---|---|---|---|---|---|---|---|---|---|---|
| 20172661 | 3.4 | 3.1 | 4.8 | 4.9 | 3.9 | 3.5 | 4.4 | 7.6 | 4.7 | 6.9 | 6.4 | 9.8 | 4.0 | 5.9 | 4.9 | 9.7 | 4.4 | 4.4 | 3.2 | 6.7 | 8.4 | 3.4 | 6.6 | 2.7 | 3.3 | 4.1 | 3.3 | 6.5 | 5.8 | 3.3 | 4.3 | 16.6 | 5.5 |
| 20172662 | 4.2 | 5.2 | 14.6 | 8.2 | 7.0 | 4.8 | 6.1 | 17.9 | 11.4 | 11.0 | 13.5 | 15.0 | 10.2 | 11.6 | 10.9 | 12.3 | 7.1 | 7.5 | 7.0 | 11.2 | 11.6 | 4.5 | 17.0 | 3.7 | 8.0 | 9.6 | 5.7 | 8.0 | 12.2 | 7.4 | 6.7 | 15.9 | 9.6 |
| 20172851 | 5.2 | 3.4 | 14.2 | 6.6 | 6.8 | 7.1 | 7.0 | 18.0 | 24.7 | 13.2 | 10.8 | 18.9 | 9.6 | 12.3 | 5.7 | 7.7 | 5.3 | 3.9 | 3.8 | 12.6 | 13.0 | 4.9 | 8.4 | 4.1 | 4.3 | 6.2 | 5.4 | 4.3 | 11.4 | 6.5 | 5.0 | 12.2 | 8.8 |
| 20173041 | 2.5 | 3.3 | 3.3 | 3.0 | 2.5 | 2.2 | 2.5 | 5.5 | 3.4 | 3.1 | 3.4 | 4.0 | 4.3 | 3.3 | 2.8 | 4.5 | 2.7 | 3.2 | 2.7 | 2.6 | 3.0 | 2.5 | 3.3 | 2.8 | 2.4 | 3.0 | 3.3 | 3.0 | 5.3 | 3.1 | 3.8 | 5.9 | 3.3 |
| 20173351 | 3.1 | 3.9 | 9.9 | 4.1 | 4.4 | 4.9 | 5.0 | 12.1 | 9.9 | 6.5 | 6.9 | 10.4 | 6.5 | 6.7 | 3.5 | 5.6 | 3.2 | 2.8 | 2.9 | 8.0 | 8.8 | 2.8 | 4.9 | 3.0 | 1.9 | 3.1 | 2.6 | 2.5 | 6.5 | 2.3 | 2.9 | 6.9 | 5.3 |
| 20173391 | 5.7 | 4.7 | 12.8 | 6.1 | 5.2 | 6.1 | 6.4 | 15.9 | 17.3 | 11.3 | 8.6 | 16.7 | 6.9 | 11.2 | 7.0 | 10.7 | 3.9 | 4.6 | 4.2 | 10.5 | 10.6 | 3.3 | 8.3 | 3.1 | 3.5 | 5.2 | 3.5 | 4.3 | 11.2 | 4.7 | 5.4 | 9.1 | 7.7 |
| 20173401 | 3.8 | 3.4 | 6.7 | 2.9 | 2.8 | 3.0 | 2.8 | 5.4 | 8.3 | 5.4 | 3.1 | 6.5 | 3.9 | 5.2 | 3.0 | 4.5 | 2.3 | 3.0 | 2.6 | 5.3 | 4.8 | 3.3 | 3.6 | 2.4 | 2.0 | 2.2 | 1.9 | 2.1 | 5.6 | 2.6 | 2.3 | 3.6 | 3.8 |
| 20173431 | 2.7 | 3.5 | 3.3 | 3.7 | 5.5 | 2.5 | 2.6 | 4.6 | 3.4 | 4.3 | 4.6 | 4.7 | 2.3 | 3.0 | 2.3 | 2.2 | 2.3 | 2.1 | 2.2 | 3.8 | 3.3 | 2.0 | 2.4 | 2.3 | 2.6 | 2.5 | 2.6 | 2.4 | 3.9 | 2.6 | 2.5 | 5.6 | 3.1 |
| 20173432 | 5.3 | 4.4 | 7.0 | 5.5 | 2.9 | 3.6 | 3.2 | 8.9 | 7.2 | 7.0 | 7.5 | 7.5 | 4.2 | 5.6 | 7.3 | 9.0 | 4.3 | 3.8 | 5.0 | 4.9 | 5.5 | 2.3 | 7.1 | 3.0 | 3.8 | 6.5 | 3.8 | 2.9 | 9.0 | 4.7 | 4.8 | 8.0 | 5.5 |
| 20173433 | 4.4 | 4.3 | 6.8 | 6.2 | 3.8 | 3.2 | 3.4 | 7.3 | 6.0 | 7.6 | 8.7 | 7.2 | 5.1 | 5.8 | 4.1 | 5.7 | 4.1 | 3.0 | 4.4 | 5.6 | 4.7 | 2.7 | 5.5 | 3.4 | 3.8 | 4.8 | 3.4 | 2.8 | 6.1 | 3.1 | 4.6 | 8.6 | 5.0 |
| 20162551 |  | 10.4 | 7.6 | 3.7 | 8.0 | 5.3 | 3.8 | 7.5 | 7.4 | 4.0 | 3.8 | 7.6 | 3.8 | 5.5 | 3.9 | 5.5 | 4.2 | 3.5 | 16.3 | 5.2 | 7.3 | 3.8 | 5.6 | 3.3 | 2.8 | 7.0 | 3.8 | 4.9 | 7.0 | 4.5 | 5.3 | 9.7 | 5.9 |
| 20172462 | 9.3 |  | 4.3 | 3.7 | 5.3 | 2.9 | 2.7 | 7.7 | 4.4 | 3.1 | 5.9 | 5.1 | 3.7 | 3.3 | 2.4 | 3.7 | 2.9 | 3.5 | 13.4 | 3.6 | 3.8 | 3.0 | 3.8 | 4.2 | 4.0 | 3.9 | 3.1 | 3.2 | 4.1 | 4.5 | 2.8 | 6.0 | 4.4 |
| 20180361 | 5.4 | 3.8 |  | 8.5 | 14.8 | 7.2 | 11.1 | 23.8 | 18.2 | 15.0 | 14.4 | 25.9 | 10.2 | 16.8 | 11.0 | 17.9 | 11.2 | 10.3 | 4.2 | 14.3 | 17.0 | 5.5 | 20.0 | 12.0 | 8.7 | 16.2 | 6.0 | 11.9 | 15.5 | 7.3 | 8.2 | 12.7 | 12.4 |
| 20180362 | 2.4 | 3.3 | 8.7 |  | 5.8 | 2.5 | 2.9 | 11.4 | 4.0 | 4.7 | 6.7 | 7.2 | 4.8 | 5.9 | 4.5 | 3.5 | 4.2 | 6.6 | 5.0 | 4.8 | 6.8 | 2.1 | 6.3 | 3.9 | 5.3 | 5.8 | 2.5 | 4.8 | 5.0 | 3.6 | 4.3 | 7.2 | 5.0 |
| 20180363 | 3.6 | 2.7 | 16.0 | 4.9 |  | 4.8 | 4.9 | 12.0 | 12.4 | 11.7 | 4.4 | 13.0 | 3.3 | 12.5 | 6.7 | 7.4 | 9.7 | 4.2 | 2.8 | 6.2 | 9.8 | 2.8 | 13.3 | 4.4 | 3.7 | 10.3 | 2.7 | 5.3 | 7.7 | 3.2 | 5.7 | 8.4 | 7.1 |
| 20180371 | 3.8 | 2.4 | 6.9 | 2.4 | 4.7 |  | 6.5 | 8.7 | 12.8 | 6.9 | 3.1 | 11.7 | 3.6 | 8.1 | 6.7 | 9.3 | 6.4 | 3.5 | 3.6 | 5.8 | 6.2 | 2.4 | 9.9 | 2.7 | 2.6 | 10.2 | 3.9 | 4.8 | 10.3 | 3.4 | 3.5 | 9.7 | 6.0 |
| 20180372 | 3.6 | 2.4 | 11.9 | 3.2 | 5.8 | 6.1 |  | 11.1 | 14.6 | 10.2 | 5.1 | 14.3 | 5.7 | 10.7 | 5.5 | 9.6 | 8.0 | 4.3 | 3.7 | 7.0 | 9.9 | 2.9 | 12.3 | 4.8 | 3.4 | 10.2 | 3.5 | 6.4 | 10.3 | 2.3 | 3.9 | 9.5 | 7.2 |
| 20180373 | 7.1 | 6.3 | 24.8 | 13.6 | 13.4 | 8.8 | 13.1 |  | 19.5 | 14.2 | 21.5 | 29.8 | 15.4 | 16.6 | 14.1 | 21.3 | 10.7 | 14.2 | 6.6 | 13.2 | 18.2 | 5.6 | 21.0 | 14.9 | 11.1 | 19.8 | 5.9 | 13.3 | 14.7 | 6.6 | 8.5 | 13.6 | 14.1 |
| 20180381 | 7.2 | 5.1 | 20.4 | 4.9 | 15.4 | 14.6 | 16.0 | 21.2 |  | 20.8 | 8.8 | 29.4 | 10.1 | 21.0 | 12.6 | 23.3 | 13.8 | 9.3 | 5.9 | 14.9 | 16.6 | 7.0 | 27.1 | 7.0 | 4.7 | 21.1 | 10.8 | 10.7 | 21.8 | 10.3 | 9.5 | 26.9 | 14.5 |
| 20180391 | 3.8 | 2.7 | 16.0 | 5.5 | 13.5 | 7.4 | 11.1 | 14.1 | 19.5 |  | 8.3 | 17.6 | 4.5 | 22.5 | 7.9 | 14.7 | 15.4 | 7.5 | 3.7 | 10.6 | 13.2 | 5.4 | 26.2 | 5.3 | 5.8 | 12.1 | 7.2 | 8.2 | 13.9 | 5.7 | 9.8 | 18.9 | 10.9 |
| 20181121 | 3.0 | 5.7 | 13.7 | 8.9 | 4.7 | 3.5 | 4.9 | 20.2 | 6.7 | 7.4 |  | 13.6 | 9.7 | 8.5 | 6.7 | 6.3 | 6.4 | 10.7 | 7.3 | 8.3 | 12.0 | 2.4 | 9.7 | 9.5 | 11.2 | 10.2 | 4.9 | 6.8 | 9.1 | 5.4 | 5.9 | 12.9 | 8.3 |
| 20181122 | 8.6 | 4.8 | 26.8 | 8.3 | 13.3 | 12.3 | 15.1 | 30.5 | 26.2 | 17.7 | 15.2 |  | 12.4 | 19.1 | 13.2 | 20.1 | 13.4 | 13.0 | 6.6 | 18.8 | 21.4 | 6.4 | 25.0 | 13.7 | 9.0 | 25.0 | 9.6 | 13.8 | 17.6 | 9.5 | 10.2 | 19.3 | 15.3 |

b)

| DPRK6 | 2017246_1 | 2017266_1 | 2017266_2 | 2017285_1 | 2017335_1 | 2017339_1 | 2017339_2 | 2017339_3 | 2017340_1 | 2017343_1 | 2017343_2 | 2017343_3 | 2019001_1 | 2019078_1 | 2019079_1 | 2019080_1 | 2019080_2 | 2019218_1 | 2019270_1 | 2020029_1 | 2020029_2 | 2020030_1 | 2020092_1 | 2020273_1 | 2020309_1 | 2020319_1 | 2020319_2 | 2020321_1 | 2020321_2 | 2020321_3 | 2021013_1 | 2021030_1 | 2021031_1 | 2021031_2 | 2021031_1 | 2021164_1 | 2021169_1 | 2021209_1 | 2021224_1 | 2021225_1 | 2021227_1 | 2021242_1 | 2021244_2 | 2021253_1 | 2021253_2 | 2021254_1 | Mean |
|---|---|---|---|---|---|---|---|---|---|---|---|---|---|---|---|---|---|---|---|---|---|---|---|---|---|---|---|---|---|---|---|---|---|---|---|---|---|---|---|---|---|---|---|---|---|---|---|
| 20172661 | 6.1 |  | 39.1 | 41.0 | 22.5 | 13.3 | 15.2 | 14.0 | 11.6 | 9.0 | 13.5 | 15.8 | 8.8 | 10.9 | 15.6 | 14.1 | 4.9 | 4.8 | 5.8 | 9.7 | 3.8 | 5.6 | 5.7 | 6.0 | 5.2 | 6.4 | 4.1 | 5.0 | 3.7 | 10.4 | 5.8 | 9.3 | 7.3 | 6.5 | 3.9 | 9.2 | 3.9 | 6.4 | 6.3 | 7.5 | 13.7 | 9.6 | 9.8 | 6.2 | 7.0 | 4.3 | 10.3 |
| 20172662 | 8.1 | 36.6 |  | 30.3 | 19.8 | 16.8 | 9.3 | 15.2 | 8.8 | 7.3 | 15.7 | 17.3 | 8.7 | 16.9 | 23.8 | 20.9 | 8.7 | 8.6 | 9.8 | 15.4 | 6.8 | 7.6 | 10.1 | 5.2 | 14.5 | 11.7 | 5.2 | 8.0 | 6.4 | 11.4 | 9.9 | 18.5 | 10.6 | 10.1 | 6.1 | 12.7 | 4.0 | 5.3 | 9.4 | 10.7 | 11.7 | 8.8 | 10.7 | 7.1 | 7.9 | 6.4 | 12.2 |
| 20172851 | 8.4 | 32.3 | 30.0 |  | 14.4 | 21.4 | 13.8 | 6.2 | 7.9 | 8.8 | 16.4 | 15.0 | 6.8 | 8.9 | 13.2 | 10.5 | 7.0 | 6.1 | 8.8 | 12.8 | 5.8 | 6.2 | 5.4 | 5.0 | 6.6 | 7.6 | 4.6 | 3.8 | 4.2 | 6.3 | 3.9 | 9.5 | 5.2 | 4.6 | 5.2 | 12.4 | 3.7 | 6.7 | 5.7 | 14.4 | 7.9 | 5.5 | 6.6 | 5.2 | 5.5 | 3.9 | 9.4 |
| 20173041 | 2.8 | 2.5 | 3.2 | 3.7 | 3.2 | 3.4 | 3.1 | 2.4 | 2.3 | 2.7 | 3.1 | 2.7 | 2.6 | 2.7 | 2.2 | 2.2 | 2.4 | 2.4 | 2.2 | 2.6 | 2.3 | 3.1 | 2.4 | 2.8 | 3.6 | 2.0 | 2.9 | 2.6 | 2.8 | 2.6 | 2.8 | 3.2 | 2.6 | 2.1 | 3.1 | 2.9 | 2.6 | 4.0 | 2.9 | 2.5 | 2.3 | 3.5 | 3.1 | 3.7 | 3.1 | 3.5 | 2.8 |
| 20173351 | 7.2 | 12.9 | 10.9 | 3.5 |  | 12.4 | 5.3 | 2.1 | 2.4 | 3.2 | 4.3 | 5.2 | 3.4 | 2.7 | 2.6 | 2.9 | 3.5 | 2.9 | 7.5 | 7.4 | 5.1 | 4.5 | 4.9 | 2.0 | 4.2 | 2.2 | 2.8 | 3.3 | 2.3 | 4.0 | 2.7 | 2.0 | 2.1 | 2.0 | 3.7 | 6.7 | 2.3 | 3.4 | 3.1 | 10.3 | 4.5 | 3.2 | 2.5 | 3.2 | 2.5 | 2.7 | 4.4 |
| 20173391 | 8.5 | 24.3 | 21.9 | 22.8 | 15.3 |  | 21.0 | 7.0 | 13.7 | 9.3 | 19.1 | 16.8 | 9.6 | 6.7 | 11.6 | 9.4 | 8.8 | 4.1 | 12.1 | 10.8 | 7.5 | 8.0 | 5.1 | 4.4 | 5.5 | 7.0 | 2.8 | 3.3 | 2.2 | 5.4 | 5.5 | 7.3 | 4.5 | 4.2 | 5.4 | 12.7 | 5.1 | 5.4 | 7.0 | 16.1 | 8.3 | 3.9 | 10.3 | 3.7 | 4.6 | 4.6 | 9.5 |
| 20173401 | 5.7 | 8.7 | 7.0 | 7.4 | 5.1 | 12.5 | 6.3 | 2.5 |  | 4.4 | 8.4 | 6.0 | 4.2 | 5.0 | 6.5 | 6.0 | 6.7 | 2.2 | 8.0 | 5.5 | 4.7 | 5.7 | 3.5 | 3.9 | 2.4 | 2.9 | 1.8 | 1.9 | 1.7 | 2.8 | 2.4 | 3.3 | 2.2 | 2.2 | 2.6 | 4.3 | 2.5 | 2.4 | 1.8 | 10.7 | 3.7 | 2.0 | 3.9 | 2.2 | 2.6 | 3.1 | 4.6 |
| 20173431 | 4.0 | 5.1 | 3.7 | 2.1 | 1.9 | 7.8 | 6.5 | 1.8 | 1.9 |  | 3.4 | 8.9 | 2.9 | 2.5 | 2.8 | 2.4 | 3.6 | 2.7 | 4.2 | 3.1 | 3.0 | 2.7 | 2.6 | 3.3 | 3.2 | 2.1 | 2.0 | 1.9 | 2.4 | 1.8 | 2.8 | 1.4 | 1.9 | 1.7 | 2.6 | 3.8 | 2.2 | 3.6 | 2.5 | 5.6 | 1.6 | 3.7 | 2.0 | 2.0 | 2.9 | 2.0 | 3.1 |
| 20173432 | 11.1 | 10.6 | 8.9 | 4.4 | 3.1 | 17.0 | 11.5 | 2.2 | 3.0 | 12.1 |  | 18.4 | 5.1 | 2.7 | 4.1 | 2.9 | 4.4 | 4.4 | 7.0 | 5.0 | 3.3 | 4.3 | 5.1 | 3.3 | 6.8 | 3.1 | 2.9 | 3.2 | 2.2 | 4.1 | 5.6 | 2.9 | 2.9 | 3.7 | 5.0 | 9.5 | 3.6 | 3.4 | 6.8 | 7.7 | 4.8 | 2.7 | 4.2 | 5.5 | 6.5 | 3.4 | 5.7 |
| 20173433 | 10.3 | 7.5 | 8.6 | 3.5 | 3.3 | 13.6 | 10.4 | 2.2 | 3.2 | 10.1 | 7.8 |  | 4.9 | 3.0 | 3.6 | 3.4 | 3.6 | 5.0 | 7.0 | 5.5 | 4.2 | 3.8 | 5.2 | 4.1 | 5.5 | 2.7 | 2.4 | 2.9 | 2.1 | 3.2 | 4.3 | 2.1 | 3.1 | 2.3 | 3.4 | 6.0 | 2.5 | 3.4 | 5.3 | 7.0 | 3.3 | 2.5 | 3.2 | 4.8 | 6.7 | 2.9 | 4.8 |
| 20162551 | 11.5 | 4.7 | 5.7 | 5.6 | 3.1 | 6.7 | 5.8 | 2.6 | 4.9 | 4.2 | 8.4 | 6.6 | 4.3 | 3.2 | 5.0 | 5.5 | 4.0 | 3.2 | 6.7 | 3.7 | 4.7 | 5.7 | 2.1 | 3.9 | 4.2 | 6.4 | 3.1 | 3.4 | 2.3 | 4.1 | 2.4 | 4.2 | 3.0 | 2.5 | 3.6 | 5.8 | 2.8 | 3.9 | 2.4 | 3.1 | 3.9 | 3.3 | 4.4 | 5.3 | 5.5 | 5.0 | 4.4 |
| 20172462 | 7.1 | 3.4 | 6.7 | 3.7 | 3.1 | 3.7 | 4.2 | 2.5 | 3.4 | 4.4 | 5.9 | 5.1 | 3.1 | 2.4 | 3.6 | 4.3 | 3.2 | 3.0 | 3.7 | 4.3 | 3.7 | 4.1 | 2.6 | 2.6 | 4.8 | 4.1 | 2.5 | 2.9 | 2.5 | 2.8 | 2.4 | 2.7 | 3.0 | 3.5 | 2.1 | 4.6 | 2.6 | 2.5 | 2.2 | 3.1 | 3.1 | 2.4 | 3.4 | 4.7 | 4.3 | 2.9 | 3.5 |
| 20180361 | 4.4 | 8.6 | 7.4 | 4.0 | 3.6 | 10.8 | 4.6 | 2.8 | 2.3 | 3.2 | 5.7 | 6.1 | 3.5 | 3.2 | 5.2 | 4.2 | 3.0 | 4.9 | 5.4 | 5.9 | 4.3 | 4.6 | 4.5 | 2.3 | 7.5 | 4.1 | 3.3 | 3.0 | 3.7 | 3.7 | 3.4 | 3.8 | 3.2 | 3.4 | 2.6 | 8.1 | 2.8 | 3.2 | 4.2 | 4.9 | 6.2 | 2.7 | 4.8 | 4.9 | 3.6 | 3.2 | 4.5 |
| 20180362 | 4.5 | 3.8 | 3.5 | 2.4 | 1.8 | 4.0 | 3.0 | 2.5 | 3.1 | 2.8 | 2.2 | 4.4 | 2.5 | 2.1 | 2.0 | 2.4 | 2.7 | 2.0 | 3.5 | 2.9 | 2.3 | 3.8 | 2.8 | 1.9 | 4.2 | 2.9 | 2.8 | 2.8 | 1.7 | 1.7 | 2.1 | 2.2 | 2.3 | 1.8 | 2.7 | 2.4 | 2.3 | 2.5 | 2.9 | 2.3 | 1.7 | 3.2 | 2.4 | 2.5 | 3.6 | 2.7 | 2.7 |
| 20180363 | 2.8 | 4.0 | 4.9 | 3.0 | 2.2 | 4.2 | 2.4 | 2.0 | 2.4 | 2.4 | 2.4 | 3.2 | 3.0 | 2.4 | 2.0 | 2.7 | 2.4 | 1.9 | 2.7 | 2.0 | 2.0 | 2.7 | 2.7 | 2.3 | 2.1 | 2.4 | 2.5 | 2.4 | 2.1 | 2.3 | 1.9 | 2.7 | 2.7 | 2.8 | 2.7 | 3.6 | 1.7 | 2.0 | 1.5 | 1.6 | 3.7 | 2.5 | 3.3 | 3.3 | 3.5 | 2.9 | 2.6 |
| 20180371 | 5.0 | 5.0 | 3.1 | 2.4 | 2.3 | 5.4 | 3.2 | 2.7 | 2.6 | 2.5 | 2.1 | 2.4 | 3.1 | 2.4 | 2.8 | 3.0 | 2.4 | 1.9 | 2.5 | 3.0 | 2.5 | 2.8 | 2.1 | 2.1 | 2.4 | 2.9 | 2.0 | 2.5 | 2.6 | 2.1 | 2.1 | 2.5 | 2.8 | 2.2 | 2.2 | 4.0 | 2.3 | 2.0 | 1.7 | 2.5 | 2.6 | 2.4 | 2.8 | 2.4 | 2.9 | 2.5 | 2.7 |
| 20180372 | 4.6 | 4.0 | 3.7 | 3.4 | 2.4 | 5.1 | 3.0 | 2.0 | 2.5 | 2.5 | 2.5 | 3.1 | 2.9 | 2.5 | 2.9 | 2.7 | 2.9 | 2.5 | 2.6 | 3.6 | 2.5 | 3.1 | 2.4 | 2.2 | 3.2 | 2.6 | 2.2 | 3.0 | 2.2 | 2.3 | 2.6 | 2.1 | 2.3 | 2.0 | 2.6 | 5.1 | 2.3 | 2.2 | 2.4 | 2.0 | 2.6 | 3.4 | 3.2 | 3.1 | 3.3 | 2.2 | 2.8 |
| 20180373 | 7.7 | 10.7 | 10.3 | 6.5 | 5.1 | 13.0 | 6.6 | 3.2 | 3.0 | 4.1 | 5.9 | 6.8 | 6.1 | 4.6 | 5.4 | 4.0 | 2.8 | 4.5 | 5.6 | 7.3 | 4.7 | 4.6 | 4.7 | 3.4 | 10.2 | 5.8 | 3.4 | 3.5 | 4.0 | 4.7 | 3.8 | 4.1 | 3.0 | 3.4 | 2.8 | 9.9 | 4.2 | 3.7 | 5.6 | 5.4 | 9.5 | 3.0 | 5.5 | 6.3 | 4.4 | 4.9 | 5.5 |
| 20180381 | 7.8 | 10.7 | 9.1 | 6.1 | 4.4 | 17.0 | 10.2 | 3.0 | 3.2 | 3.6 | 8.6 | 7.4 | 5.5 | 3.5 | 7.0 | 4.5 | 3.8 | 4.3 | 6.4 | 9.6 | 4.5 | 5.4 | 3.7 | 3.5 | 11.3 | 5.9 | 3.1 | 3.3 | 4.2 | 4.5 | 4.1 | 4.2 | 3.4 | 3.4 | 3.6 | 14.1 | 4.8 | 4.5 | 4.7 | 6.1 | 8.2 | 2.8 | 5.9 | 5.9 | 5.4 | 3.5 | 5.9 |
| 20180391 | 5.4 | 6.1 | 5.3 | 3.5 | 2.9 | 8.5 | 3.9 | 2.5 | 2.9 | 3.6 | 4.4 | 4.8 | 3.9 | 2.6 | 3.6 | 3.2 | 2.6 | 3.6 | 4.8 | 4.5 | 3.2 | 3.7 | 3.4 | 2.2 | 4.8 | 3.7 | 2.0 | 2.7 | 2.5 | 3.0 | 2.4 | 3.6 | 2.3 | 2.7 | 2.6 | 5.9 | 2.6 | 2.5 | 4.3 | 4.5 | 5.7 | 2.6 | 4.3 | 5.1 | 4.7 | 2.7 | 3.7 |
| 20181121 | 4.6 | 8.0 | 6.2 | 3.9 | 2.7 | 6.5 | 4.1 | 2.0 | 2.1 | 3.6 | 3.1 | 6.6 | 3.5 | 2.3 | 2.5 | 3.3 | 2.5 | 2.9 | 5.0 | 4.8 | 5.1 | 4.4 | 3.5 | 2.5 | 6.4 | 3.4 | 2.3 | 3.0 | 2.3 | 3.4 | 2.6 | 2.1 | 2.3 | 1.8 | 2.1 | 4.2 | 2.3 | 2.1 | 3.0 | 3.3 | 2.7 | 2.9 | 2.8 | 3.3 | 2.7 | 3.1 | 3.5 |
| 20181122 | 8.3 | 10.1 | 8.8 | 5.3 | 3.8 | 12.3 | 5.9 | 2.5 | 3.1 | 4.2 | 6.8 | 7.2 | 6.9 | 2.5 | 4.2 | 4.2 | 3.3 | 4.1 | 6.2 | 6.2 | 5.0 | 4.9 | 3.8 | 3.8 | 9.5 | 5.7 | 3.1 | 4.1 | 3.0 | 4.4 | 4.5 | 3.9 | 2.8 | 3.6 | 3.8 | 10.5 | 4.6 | 3.4 | 3.7 | 3.5 | 5.4 | 2.8 | 5.4 | 5.8 | 4.9 | 3.6 | 5.1 |

c)

|  | 2016257_1 | 2017303_1 | 2017304_1 | 2018161_1 | 2019011_4 | 2019117_1 | 2019323_3 | 2020046_1 | 2020263_1 | 2020282_1 | 2021242_2 | Mean |
|---|---|---|---|---|---|---|---|---|---|---|---|---|
| 20172661 | 4.4 | 3.0 | 3.0 | 3.6 | 2.5 | 2.6 | 5.7 | 3.8 | 3.8 | 5.9 | 6.7 | 4.1 |
| 20172662 | 5.6 | 7.0 | 4.3 | 4.1 | 2.8 | 5.4 | 9.7 | 3.6 | 4.4 | 10.6 | 6.3 | 5.8 |
| 20172851 | 7.4 | 5.4 | 3.9 | 4.2 | 5.9 | 2.4 | 7.3 | 2.8 | 3.2 | 5.7 | 4.9 | 4.8 |
| 20173041 | 2.8 | 4.4 |  | 2.3 | 2.4 | 2.7 | 2.9 | 2.4 | 2.4 | 3.3 | 3.4 | 2.9 |
| 20173351 | 6.7 | 3.3 | 2.1 | 2.8 | 3.0 | 1.8 | 6.3 | 2.6 | 2.2 | 2.8 | 2.3 | 3.3 |
| 20173391 | 7.1 | 4.4 | 3.2 | 5.6 | 5.4 | 2.6 | 6.2 | 2.0 | 3.4 | 5.5 | 3.2 | 4.4 |
| 20173401 | 6.0 | 3.1 | 2.3 | 3.0 | 3.8 | 1.5 | 4.4 | 2.8 | 2.1 | 2.2 | 2.5 | 3.1 |
| 20173431 | 3.2 | 3.1 | 3.1 | 2.2 | 3.2 | 2.2 | 4.0 | 3.3 | 2.4 | 2.8 | 2.1 | 2.9 |
| 20173432 | 4.3 | 3.8 | 3.8 | 2.4 | 3.1 | 2.5 | 7.0 | 2.3 | 2.4 | 5.9 | 2.8 | 3.7 |
| 20173433 | 4.4 | 4.3 | 3.9 | 2.7 | 3.3 | 2.3 | 7.2 | 2.5 | 2.2 | 4.3 | 2.5 | 3.6 |
| 20162551 | 4.0 | 2.9 | 2.7 | 3.1 | 2.7 | 4.0 | 3.2 | 3.6 | 3.0 | 2.7 | 2.8 | 3.2 |
| 20172462 | 3.5 | 2.2 | 3.0 | 3.2 | 2.4 | 2.6 | 3.5 | 3.0 | 2.2 | 3.7 | 2.6 | 2.9 |
| 20180361 | 4.9 | 3.1 | 3.0 | 2.5 | 5.3 | 4.4 | 6.4 | 2.3 | 3.3 | 7.0 | 2.4 | 4.1 |
| 20180362 | 4.3 | 2.3 | 2.3 | 2.2 | 2.4 | 3.1 | 5.5 | 3.8 | 2.0 | 4.4 | 1.9 | 3.1 |
| 20180363 | 2.2 | 2.8 | 3.4 | 2.6 | 2.9 | 3.1 | 3.2 | 3.2 | 2.7 | 2.5 | 2.2 | 2.8 |
| 20180371 | 3.5 | 2.7 | 2.4 | 2.5 | 3.5 | 2.7 | 2.3 | 2.8 | 2.9 | 3.2 | 2.3 | 2.8 |
| 20180372 | 2.4 | 2.1 | 2.9 | 2.3 | 4.3 | 3.4 | 2.5 | 2.1 | 2.4 | 2.7 | 2.4 | 2.7 |
| 20180373 | 5.2 | 4.0 | 5.3 | 2.3 | 4.7 | 6.0 | 8.9 | 3.2 | 4.0 | 9.0 | 2.7 | 5.0 |
| 20180381 | 4.4 | 5.0 | 3.5 | 3.0 | 11.0 | 4.7 | 8.0 | 2.4 | 3.9 | 6.9 | 2.6 | 5.0 |
| 20180391 | 3.2 | 3.3 | 2.8 | 2.6 | 8.3 | 2.5 | 5.0 | 2.1 | 2.4 | 6.0 | 3.7 | 3.8 |
| 20181121 | 4.4 | 3.3 | 3.3 | 3.9 | 4.8 | 5.0 | 9.0 | 2.4 | 2.6 | 7.8 | 2.9 | 4.5 |
| 20181122 | 5.4 | 2.9 | 3.6 | 2.5 | 7.3 | 7.0 | 8.2 | 2.9 | 3.3 | 8.1 | 2.7 | 4.9 |

Figure S1. SNRcc values at station USRK for aftershocks associated with a) DPRK5, b) DPRK6, c) no specific association is possible because of low SNRcc values. Column 1 presents master events, other columns present aftershocks found by these master events. SNRcc values in each column are formatted with colour scale to highlight the difference of mutual correlation between aftershocks associated with the DPRK5 and DPRK6. Panel a) demonstrates that the events in the bottom part better correlate (green cells) with the immediate aftershock of the DPRK5: 2016255_1. Panel b) illustrates the better correlation between events observed after the DPRK6 test. Panel c) shows relatively reliable aftershocks without significant difference in correlation between events associated with DPRK5 and DPRK6. These aftershocks are characterized by lower SNRcc values, which makes discrimination difficult. The column with the mean value shows the average performance of a given template for all found aftershocks in a given cluster.

a)

| DPRK5 | 2016255_1 | 2017246_2 | 2018036_1 | 2018036_2 | 2018036_3 | 2018037_1 | 2018037_2 | 2018037_3 | 2018038_1 | 2018039_1 | 2018112_1 | 2018112_2 | 2019011_2 | 2019011_3 | 2019158_1 | 2019158_2 | 2019158_3 | 2019233_1 | 2019310_1 | 2019323_1 | 2019323_2 | 2020071_1 | 2020220_1 | 2020308_1 | 2020308_2 | 2020308_3 | 2020308_4 | 2021108_1 | 2021109_1 | 2021110_1 | 2021175_1 | 2021244_1 | Mean |
|---|---|---|---|---|---|---|---|---|---|---|---|---|---|---|---|---|---|---|---|---|---|---|---|---|---|---|---|---|---|---|---|---|---|
| 20172661 | 7.3 | 5.1 | 4.5 | 3.3 | 4.6 | 3.0 | 2.5 | 10.8 | 16.4 | 7.7 | 4.9 | 11.5 | 4.8 | 7.0 | 5.5 | 14.9 | 4.0 | 6.1 | 6.5 | 4.1 | 7.3 | 3.7 | 12.3 | 2.6 | 3.2 | 4.8 | 2.6 | 2.9 | 13.6 | 7.6 | 11.6 | 10.2 | 6.8 |
| 20172662 | 8.7 | 7.5 | 7.4 | 4.3 | 4.9 | 4.3 | 3.4 | 13.9 | 22.5 | 10.8 | 8.3 | 17.8 | 5.9 | 10.5 | 8.2 | 23.1 | 5.7 | 8.2 | 8.3 | 6.4 | 10.1 | 4.3 | 19.2 | 3.7 | 2.9 | 6.4 | 4.5 | 3.6 | 17.6 | 10.6 | 14.9 | 15.5 | 9.5 |
| 20172851 | 7.3 | 6.4 | 5.1 | 2.8 | 3.0 | 3.0 | 2.5 | 10.3 | 13.7 | 6.3 | 5.9 | 12.0 | 4.1 | 6.4 | 6.0 | 13.9 | 4.4 | 6.5 | 7.0 | 3.7 | 7.7 | 3.1 | 11.1 | 3.3 | 3.0 | 6.3 | 3.6 | 3.5 | 12.1 | 6.8 | 10.8 | 11.0 | 6.6 |
| 20173041 | 2.6 | 2.6 | 2.5 | 2.7 | 2.1 | 2.5 | 1.9 | 2.8 | 3.2 | 3.1 | 2.9 | 3.0 | 2.7 | 3.3 | 2.4 | 2.6 | 2.9 | 2.3 | 2.3 | 2.2 | 2.8 | 2.7 | 3.3 | 2.8 | 2.2 | 2.8 | 2.6 | 2.5 | 3.3 | 2.7 | 2.9 | 3.0 | 2.7 |
| 20173351 | 3.9 | 4.1 | 2.7 | 2.5 | 3.4 | 2.5 | 2.9 | 5.9 | 7.7 | 4.3 | 3.5 | 7.8 | 2.8 | 4.6 | 4.7 | 7.4 | 2.3 | 2.8 | 4.2 | 3.5 | 4.9 | 2.5 | 6.2 | 2.2 | 2.7 | 2.6 | 2.4 | 2.5 | 8.6 | 3.7 | 6.9 | 6.4 | 4.2 |
| 20173391 | 5.2 | 5.4 | 3.7 | 3.8 | 2.8 | 2.6 | 2.9 | 7.6 | 10.8 | 5.2 | 4.9 | 10.7 | 3.9 | 4.9 | 4.6 | 10.3 | 4.0 | 6.5 | 4.9 | 3.7 | 6.1 | 2.2 | 8.0 | 2.5 | 2.5 | 3.9 | 3.1 | 3.0 | 10.3 | 5.3 | 8.8 | 9.1 | 5.4 |
| 20173401 | 2.7 | 2.4 | 2.5 | 2.7 | 2.0 | 1.8 | 2.0 | 3.8 | 4.7 | 2.0 | 2.7 | 4.2 | 2.4 | 2.3 | 3.2 | 4.6 | 2.5 | 2.8 | 2.3 | 2.3 | 2.2 | 2.4 | 3.3 | 2.6 | 2.4 | 2.5 | 2.2 | 1.8 | 4.7 | 2.1 | 3.1 | 2.9 | 2.8 |
| 20173431 | 2.8 | 2.9 | 2.4 | 3.8 | 2.0 | 2.9 | 2.3 | 2.3 | 2.2 | 2.1 | 3.0 | 2.3 | 3.0 | 2.5 | 3.9 | 2.3 | 2.9 | 2.7 | 2.0 | 2.3 | 2.5 | 2.6 | 2.4 | 2.3 | 2.1 | 2.3 | 2.4 | 2.1 | 2.9 | 2.6 | 2.4 | 3.4 | 2.6 |
| 20173432 | 3.8 | 3.5 | 3.3 | 3.2 | 2.4 | 2.3 | 2.2 | 5.8 | 9.1 | 3.6 | 3.7 | 6.9 | 3.0 | 4.0 | 3.6 | 4.6 | 2.0 | 3.4 | 3.7 | 2.8 | 4.2 | 2.9 | 4.9 | 2.7 | 2.7 | 3.2 | 2.3 | 2.7 | 4.5 | 3.0 | 2.9 | 6.8 | 3.7 |
| 20173433 | 2.5 | 2.3 | 2.9 | 2.6 | 2.5 | 2.2 | 2.0 | 3.9 | 3.3 | 2.7 | 2.8 | 3.7 | 2.6 | 2.6 | 2.9 | 3.3 | 2.6 | 3.0 | 2.9 | 2.8 | 3.0 | 2.4 | 3.4 | 2.2 | 2.3 | 2.7 | 2.6 | 2.2 | 3.4 | 2.2 | 2.7 | 4.6 | 2.8 |
| 20162551 |  | 14.0 | 5.1 | 5.7 | 3.3 | 2.4 | 2.8 | 14.3 | 12.7 | 6.3 | 8.6 | 14.3 | 4.3 | 5.7 | 5.5 | 14.2 | 4.6 | 6.6 | 16.2 | 4.3 | 8.0 | 3.6 | 11.6 | 4.2 | 3.8 | 5.5 | 3.2 | 3.9 | 12.4 | 4.8 | 7.8 | 7.8 | 7.3 |
| 20172462 | 14.5 |  | 4.5 | 6.3 | 2.2 | 2.5 | 3.0 | 13.7 | 10.9 | 5.5 | 6.9 | 12.8 | 4.9 | 5.8 | 5.5 | 12.8 | 3.1 | 7.1 | 12.8 | 4.7 | 8.3 | 2.5 | 10.1 | 3.7 | 3.8 | 5.8 | 2.7 | 3.7 | 12.4 | 3.4 | 7.9 | 7.8 | 6.8 |
| 20180361 | 5.4 | 4.6 |  | 5.1 | 4.5 | 3.2 | 4.6 | 14.9 | 13.7 | 5.9 | 6.7 | 15.9 | 3.0 | 7.1 | 7.4 | 13.9 | 5.0 | 8.4 | 4.3 | 4.5 | 7.7 | 3.0 | 14.3 | 6.5 | 4.9 | 6.9 | 4.0 | 3.7 | 14.4 | 4.8 | 6.2 | 11.3 | 7.3 |
| 20180362 | 5.2 | 5.4 | 5.4 |  | 2.4 | 2.7 | 2.1 | 10.5 | 7.9 | 3.2 | 7.4 | 11.6 | 3.3 | 5.1 | 5.9 | 9.5 | 3.1 | 6.7 | 5.1 | 4.1 | 7.2 | 2.8 | 10.4 | 5.5 | 4.4 | 6.5 | 3.1 | 3.1 | 11.2 | 3.5 | 4.6 | 8.8 | 5.7 |
| 20180363 | 3.6 | 2.9 | 4.3 | 2.0 |  | 2.5 | 2.3 | 7.6 | 7.1 | 4.5 | 3.4 | 10.5 | 3.0 | 6.5 | 4.1 | 8.2 | 3.4 | 5.3 | 2.7 | 3.8 | 5.9 | 3.4 | 13.1 | 3.5 | 2.9 | 3.9 | 2.9 | 3.1 | 10.3 | 4.5 | 6.4 | 7.1 | 5.0 |
| 20180371 | 2.5 | 2.1 | 3.2 | 2.7 | 2.2 |  | 2.3 | 4.3 | 5.2 | 3.0 | 2.1 | 5.1 | 3.3 | 2.8 | 2.6 | 4.7 | 2.6 | 2.1 | 2.5 | 2.4 | 3.4 | 2.5 | 5.0 | 2.6 | 2.8 | 4.1 | 2.8 | 2.9 | 5.3 | 2.3 | 3.4 | 4.4 | 3.2 |
| 20180372 | 2.8 | 2.9 | 3.3 | 2.2 | 1.9 | 2.2 |  | 4.9 | 7.0 | 3.1 | 2.5 | 6.8 | 3.1 | 4.0 | 3.1 | 6.8 | 2.6 | 3.5 | 2.8 | 2.3 | 3.6 | 2.5 | 6.9 | 3.3 | 2.2 | 4.2 | 2.2 | 2.6 | 7.6 | 3.1 | 3.3 | 6.1 | 3.7 |
| 20180373 | 13.7 | 14.4 | 13.7 | 12.3 | 9.3 | 4.4 | 5.7 |  | 30.2 | 15.7 | 19.8 | 36.9 | 9.4 | 18.1 | 16.7 | 34.4 | 12.1 | 17.1 | 14.6 | 13.9 | 22.6 | 7.4 | 30.3 | 13.6 | 7.4 | 16.1 | 7.0 | 8.2 | 26.2 | 10.6 | 15.1 | 26.9 | 16.3 |
| 20180381 | 12.9 | 11.4 | 16.2 | 9.5 | 8.9 | 5.6 | 8.8 | 32.1 |  | 20.1 | 14.4 | 38.3 | 9.7 | 19.4 | 15.9 | 39.6 | 14.0 | 17.9 | 11.7 | 13.0 | 20.9 | 9.5 | 34.3 | 9.0 | 6.0 | 15.7 | 7.4 | 8.0 | 31.3 | 17.1 | 21.7 | 35.3 | 17.3 |
| 20180391 | 6.3 | 5.3 | 6.1 | 3.6 | 3.2 | 3.2 | 2.9 | 15.4 | 18.5 |  | 7.4 | 20.4 | 4.6 | 14.8 | 6.5 | 16.7 | 7.9 | 7.1 | 5.1 | 5.4 | 10.4 | 4.0 | 26.0 | 3.8 | 2.5 | 5.2 | 3.7 | 4.8 | 20.0 | 6.9 | 12.9 | 13.7 | 8.8 |
| 20181121 | 9.1 | 7.0 | 6.3 | 7.6 | 2.9 | 2.2 | 2.5 | 19.8 | 14.4 | 6.7 |  | 17.3 | 5.6 | 7.7 | 7.3 | 15.8 | 4.4 | 8.4 | 7.6 | 5.5 | 10.2 | 2.5 | 13.6 | 7.0 | 5.5 | 7.3 | 3.7 | 3.9 | 13.4 | 4.6 | 6.2 | 10.0 | 7.9 |
| 20181122 | 13.4 | 14.1 | 16.8 | 12.6 | 10.6 | 5.5 | 8.0 | 38.6 | 35.5 | 20.8 | 18.9 |  | 10.6 | 17.8 | 17.9 | 39.1 | 12.8 | 18.3 | 14.7 | 17.3 | 24.0 | 7.5 | 31.2 | 13.5 | 8.6 | 18.6 | 8.3 | 8.4 | 27.7 | 12.4 | 14.7 | 33.2 | 17.8 |

b)

| DPRK6 | 2017246_1 | 2017266_1 | 2017266_2 | 2017285_1 | 2017335_1 | 2017339_1 | 2017339_2 | 2017339_3 | 2017340_1 | 2017343_1 | 2017343_2 | 2017343_3 | 2019001_1 | 2019078_1 | 2019079_1 | 2019080_1 | 2019080_2 | 2019218_1 | 2019270_1 | 2020029_1 | 2020029_2 | 2020030_1 | 2020092_1 | 2020273_1 |
|---|---|---|---|---|---|---|---|---|---|---|---|---|---|---|---|---|---|---|---|---|---|---|---|---|
| 20172661 | 6.8 |  | 53.7 | 46.6 | 26.2 | 33.1 | 14.4 | 11.0 | 12.6 | 8.3 | 23.6 | 15.6 | 8.3 | 10.0 | 15.5 | 12.4 | 5.8 | 7.1 | 12.6 | 17.9 | 5.4 | 7.2 | 8.6 | 5.7 |
| 20172662 | 11.0 | 45.0 |  | 48.5 | 27.7 | 35.1 | 14.0 | 12.3 | 11.5 | 6.8 | 24.3 | 16.8 | 11.8 | 12.0 | 25.1 | 16.7 | 7.2 | 9.8 | 15.9 | 21.2 | 7.4 | 10.6 | 11.7 | 11.3 |
| 20172851 | 7.9 | 46.0 | 55.4 |  | 26.3 | 43.3 | 18.1 | 11.4 | 15.4 | 10.8 | 32.4 | 24.1 | 8.8 | 10.3 | 15.1 | 12.7 | 5.4 | 9.5 | 13.4 | 22.7 | 5.4 | 7.9 | 11.4 | 8.2 |
| 20173041 | 2.6 | 2.0 | 2.1 | 2.5 | 2.5 | 2.0 | 1.9 | 2.6 | 2.3 | 4.0 | 2.1 | 2.9 | 2.2 | 2.0 | 2.3 | 2.1 | 3.0 | 2.4 | 3.0 | 2.4 | 3.1 | 3.0 | 2.8 | 2.2 |
| 20173351 | 3.6 | 25.2 | 31.7 | 26.8 |  | 24.4 | 9.4 | 8.8 | 11.3 | 4.8 | 20.1 | 11.4 | 6.1 | 9.2 | 13.1 | 9.2 | 5.8 | 4.2 | 8.1 | 9.3 | 3.7 | 3.7 | 5.6 | 5.8 |
| 20173391 | 7.3 | 34.8 | 38.1 | 40.9 | 23.5 |  | 14.4 | 8.6 | 17.2 | 10.8 | 35.0 | 21.2 | 8.4 | 11.1 | 15.1 | 12.5 | 6.4 | 5.9 | 10.1 | 15.1 | 3.8 | 6.2 | 8.9 | 7.6 |
| 20173401 | 3.7 | 11.9 | 12.6 | 14.8 | 9.9 | 16.0 | 8.1 | 3.7 |  | 4.0 | 16.4 | 8.5 | 3.3 | 3.7 | 7.0 | 5.6 | 3.1 | 2.4 | 3.6 | 4.5 | 2.1 | 2.5 | 3.1 | 3.8 |
| 20173431 | 2.8 | 5.8 | 6.5 | 3.9 | 4.4 | 8.8 | 5.5 | 2.6 | 3.8 |  | 11.6 | 9.8 | 2.6 | 3.1 | 4.4 | 4.9 | 3.3 | 2.4 | 3.8 | 2.8 | 1.6 | 2.3 | 2.9 | 2.2 |
| 20173432 | 8.6 | 16.1 | 20.3 | 4.8 | 10.0 | 33.6 | 17.4 | 7.6 | 14.3 | 14.5 |  | 39.1 | 9.1 | 11.2 | 19.3 | 17.8 | 5.9 | 4.3 | 6.0 | 4.2 | 3.3 | 3.4 | 3.5 | 5.2 |
| 20173433 | 5.5 | 6.5 | 7.7 | 3.4 | 3.5 | 14.2 | 7.5 | 5.0 | 5.4 | 10.0 | 19.8 |  | 4.1 | 7.5 | 10.3 | 9.9 | 3.0 | 2.8 | 2.7 | 2.8 | 3.3 | 2.8 | 1.9 | 2.4 |
| 20162551 | 7.9 | 6.3 | 8.5 | 7.2 | 4.0 | 5.4 | 3.2 | 2.5 | 2.7 | 3.3 | 7.4 | 4.7 | 3.8 | 2.6 | 4.0 | 3.4 | 2.4 | 4.3 | 6.0 | 8.1 | 3.6 | 2.9 | 2.4 | 2.8 |
| 20172462 | 4.4 | 5.0 | 7.4 | 6.0 | 4.3 | 4.7 | 3.2 | 2.8 | 2.4 | 3.0 | 6.1 | 3.5 | 3.3 | 2.3 | 2.9 | 2.5 | 2.9 | 4.5 | 5.8 | 7.1 | 2.7 | 3.0 | 2.3 | 1.5 |
| 20180361 | 2.4 | 3.5 | 7.0 | 4.7 | 2.9 | 2.7 | 2.5 | 2.5 | 2.2 | 2.4 | 3.9 | 3.1 | 2.7 | 2.7 | 3.4 | 3.1 | 2.5 | 2.5 | 4.2 | 5.9 | 2.5 | 2.9 | 2.4 | 2.6 |
| 20180362 | 2.7 | 3.1 | 3.6 | 2.0 | 2.2 | 3.2 | 2.8 | 2.9 | 2.7 | 2.3 | 2.3 | 2.7 | 2.2 | 1.9 | 2.8 | 2.5 | 2.7 | 2.6 | 3.1 | 4.9 | 2.6 | 2.4 | 1.5 | 1.8 |
| 20180363 | 2.7 | 3.1 | 4.5 | 2.6 | 2.6 | 2.2 | 2.7 | 2.0 | 2.2 | 2.4 | 3.4 | 2.6 | 2.1 | 2.4 | 1.9 | 2.7 | 1.7 | 3.1 | 4.7 | 4.0 | 2.5 | 2.2 | 2.3 | 2.1 |
| 20180371 | 2.5 | 2.6 | 4.2 | 3.1 | 2.5 | 2.9 | 2.3 | 1.6 | 1.8 | 2.3 | 2.0 | 2.2 | 1.7 | 2.2 | 1.9 | 2.2 | 2.6 | 1.9 | 2.5 | 2.5 | 2.3 | 2.5 | 2.0 | 2.6 |
| 20180372 | 2.5 | 2.2 | 3.5 | 2.3 | 2.0 | 2.5 | 1.7 | 2.2 | 2.1 | 2.6 | 2.3 | 2.7 | 2.2 | 2.0 | 3.2 | 2.1 | 3.2 | 2.8 | 3.4 | 2.7 | 2.2 | 2.7 | 2.8 | 2.3 |
| 20180373 | 5.9 | 9.5 | 15.9 | 10.6 | 5.6 | 8.0 | 4.0 | 3.5 | 4.1 | 2.5 | 9.5 | 5.3 | 4.0 | 2.8 | 7.0 | 5.3 | 2.9 | 5.7 | 11.6 | 15.7 | 4.5 | 4.9 | 4.1 | 3.4 |
| 20180381 | 9.2 | 14.3 | 24.4 | 14.2 | 7.4 | 10.6 | 4.7 | 5.1 | 4.6 | 2.3 | 9.3 | 5.9 | 5.0 | 4.2 | 9.8 | 6.4 | 2.8 | 6.4 | 16.3 | 19.9 | 6.8 | 8.4 | 4.4 | 3.9 |
| 20180391 | 3.8 | 5.8 | 10.9 | 5.7 | 4.0 | 3.3 | 2.3 | 2.5 | 2.4 | 2.5 | 3.8 | 2.9 | 3.0 | 1.9 | 3.2 | 3.9 | 2.1 | 2.5 | 8.0 | 9.3 | 2.7 | 3.1 | 4.3 | 2.6 |
| 20181121 | 4.1 | 4.7 | 8.0 | 5.3 | 3.3 | 4.0 | 3.1 | 2.3 | 2.9 | 2.6 | 4.8 | 3.0 | 3.0 | 2.1 | 3.0 | 2.9 | 2.1 | 4.6 | 5.0 | 7.8 | 2.3 | 2.1 | 2.9 | 2.0 |
| 20181122 | 5.9 | 10.7 | 16.8 | 12.2 | 7.6 | 10.7 | 5.3 | 4.0 | 4.6 | 2.3 | 11.3 | 6.4 | 4.9 | 3.9 | 7.7 | 5.6 | 3.1 | 6.8 | 12.9 | 17.8 | 5.3 | 6.2 | 4.8 | 4.5 |

| DPRK6 | 2020309_1 | 2020319_1 | 2020319_2 | 2020321_1 | 2020321_2 | 2020321_3 | 2021013_1 | 2021030_1 | 2021031_1 | 2021031_2 | 2021031_1 | 2021164_1 | 2021169_1 | 2021209_1 | 2021224_1 | 2021225_1 | 2021227_1 | 2021242_1 | 2021244_2 | 2021253_1 | 2021253_2 | 2021254_1 | Mean |
|---|---|---|---|---|---|---|---|---|---|---|---|---|---|---|---|---|---|---|---|---|---|---|---|
| 20172661 | 14.3 | 9.6 | 7.1 | 4.9 | 3.4 | 7.7 | 6.3 | 9.3 | 4.0 | 3.9 | 4.5 | 19.3 | 5.5 | 11.2 | 6.7 | 5.0 | 9.5 | 4.4 | 4.3 | 7.9 | 8.0 | 7.2 | 12.1 |
| 20172662 | 16.3 | 20.4 | 17.8 | 4.7 | 6.5 | 8.1 | 7.6 | 11.3 | 4.8 | 3.3 | 6.4 | 26.4 | 5.8 | 12.6 | 7.1 | 2.4 | 10.2 | 6.0 | 6.9 | 11.5 | 12.6 | 9.3 | 14.2 |
| 20172851 | 17.2 | 12.4 | 12.2 | 7.0 | 3.9 | 9.7 | 8.1 | 12.4 | 4.7 | 4.1 | 5.5 | 25.0 | 7.1 | 12.7 | 6.8 | 3.3 | 10.2 | 4.0 | 5.2 | 10.5 | 11.1 | 9.4 | 14.1 |
| 20173041 | 2.9 | 2.6 | 2.3 | 2.2 | 2.0 | 2.0 | 2.4 | 2.4 | 2.4 | 2.2 | 2.4 | 2.1 | 2.6 | 2.5 | 2.7 | 10.9 | 3.1 | 2.6 | 2.6 | 3.1 | 2.6 | 3.0 | 2.7 |
| 20173351 | 7.2 | 11.0 | 11.6 | 3.5 | 3.9 | 6.7 | 5.0 | 8.6 | 4.0 | 3.6 | 2.8 | 12.6 | 5.3 | 9.7 | 5.4 | 4.0 | 11.1 | 3.5 | 4.4 | 8.3 | 8.1 | 6.1 | 9.2 |
| 20173391 | 10.8 | 11.5 | 10.2 | 6.9 | 4.2 | 8.1 | 7.4 | 10.4 | 4.4 | 3.6 | 4.3 | 24.1 | 5.9 | 12.4 | 7.1 | 5.5 | 13.2 | 3.3 | 5.0 | 10.4 | 11.2 | 11.7 | 12.4 |
| 20173401 | 4.7 | 6.1 | 5.4 | 2.9 | 4.2 | 4.0 | 2.8 | 4.9 | 2.8 | 2.4 | 2.3 | 8.9 | 2.5 | 5.5 | 3.1 | 2.9 | 6.6 | 2.9 | 3.6 | 7.1 | 5.7 | 4.2 | 5.5 |
| 20173431 | 3.4 | 3.7 | 2.2 | 2.2 | 1.9 | 3.5 | 2.9 | 2.9 | 1.9 | 2.3 | 2.6 | 5.7 | 3.0 | 2.4 | 3.2 | 2.2 | 3.2 | 3.1 | 3.1 | 4.6 | 3.7 | 3.3 | 3.7 |
| 20173432 | 7.3 | 10.6 | 2.3 | 9.0 | 2.8 | 11.7 | 3.5 | 6.0 | 4.2 | 4.2 | 3.5 | 18.4 | 8.3 | 5.6 | 8.2 | 7.4 | 15.8 | 3.2 | 10.2 | 15.1 | 16.3 | 10.9 | 10.0 |
| 20173433 | 3.0 | 5.2 | 2.4 | 6.2 | 3.3 | 4.6 | 2.3 | 3.1 | 2.6 | 2.9 | 2.3 | 8.7 | 3.9 | 3.1 | 4.2 | 5.0 | 8.0 | 3.3 | 6.2 | 12.7 | 14.6 | 6.9 | 5.3 |
| 20162551 | 6.0 | 4.1 | 3.3 | 2.6 | 2.3 | 3.2 | 2.2 | 4.5 | 2.5 | 2.4 | 3.8 | 6.8 | 2.9 | 2.9 | 2.8 | 2.7 | 2.5 | 4.3 | 2.8 | 3.4 | 3.3 | 3.3 | 4.0 |
| 20172462 | 5.2 | 3.0 | 2.2 | 2.1 | 2.3 | 3.2 | 3.0 | 3.7 | 2.6 | 1.9 | 2.8 | 4.8 | 2.5 | 3.1 | 2.6 | 2.9 | 2.6 | 4.5 | 2.8 | 3.0 | 3.5 | 2.9 | 3.5 |
| 20180361 | 4.6 | 3.5 | 2.3 | 2.1 | 2.7 | 2.6 | 2.0 | 2.6 | 2.6 | 2.4 | 2.6 | 3.7 | 1.7 | 2.7 | 2.3 | 2.5 | 2.3 | 3.2 | 3.2 | 2.7 | 2.7 | 3.9 | 3.0 |
| 20180362 | 4.8 | 2.3 | 2.4 | 3.4 | 2.6 | 2.7 | 2.1 | 2.0 | 1.9 | 2.7 | 2.1 | 4.2 | 2.1 | 2.0 | 2.3 | 2.2 | 2.5 | 3.0 | 2.8 | 3.2 | 2.5 | 4.3 | 2.6 |
| 20180363 | 3.8 | 2.6 | 3.8 | 2.6 | 2.2 | 3.5 | 2.7 | 2.8 | 2.6 | 1.8 | 2.5 | 2.4 | 2.3 | 2.8 | 2.5 | 2.6 | 2.5 | 3.4 | 3.1 | 3.1 | 2.5 | 5.4 | 2.7 |
| 20180371 | 3.0 | 3.0 | 2.4 | 2.2 | 2.6 | 2.3 | 2.7 | 2.2 | 2.8 | 1.9 | 2.1 | 2.3 | 2.8 | 2.3 | 2.6 | 2.6 | 2.3 | 2.8 | 2.4 | 4.0 | 3.1 | 3.1 | 2.4 |
| 20180372 | 2.6 | 2.5 | 2.1 | 2.2 | 2.6 | 2.5 | 2.4 | 1.8 | 2.1 | 2.3 | 2.3 | 2.3 | 2.1 | 2.8 | 2.8 | 2.2 | 2.4 | 3.0 | 3.8 | 2.8 | 3.0 | 4.3 | 2.5 |
| 20180373 | 11.4 | 6.4 | 3.4 | 3.7 | 2.6 | 6.4 | 3.7 | 5.3 | 3.8 | 3.0 | 4.7 | 11.1 | 3.6 | 4.5 | 3.6 | 2.6 | 5.7 | 9.8 | 3.8 | 5.0 | 5.6 | 4.6 | 5.9 |
| 20180381 | 11.0 | 8.1 | 4.5 | 5.1 | 3.2 | 6.8 | 2.8 | 7.2 | 3.6 | 3.7 | 4.2 | 12.0 | 2.9 | 5.1 | 2.8 | 3.4 | 5.8 | 9.9 | 3.7 | 5.4 | 6.0 | 4.5 | 7.3 |
| 20180391 | 4.5 | 4.2 | 2.5 | 2.8 | 2.2 | 4.6 | 2.5 | 2.9 | 1.9 | 2.6 | 2.4 | 5.5 | 3.0 | 2.4 | 2.9 | 3.2 | 3.5 | 4.1 | 3.2 | 3.4 | 4.1 | 3.6 | 3.7 |
| 20181121 | 4.8 | 2.8 | 2.0 | 2.3 | 2.1 | 3.4 | 2.4 | 3.0 | 1.8 | 2.1 | 2.5 | 4.3 | 2.8 | 2.5 | 2.1 | 2.6 | 3.3 | 4.5 | 2.9 | 3.3 | 2.2 | 3.3 | 3.3 |
| 20181122 | 13.5 | 6.9 | 4.0 | 5.0 | 3.2 | 7.8 | 3.4 | 6.4 | 3.2 | 4.1 | 4.8 | 11.4 | 3.5 | 4.5 | 3.0 | 2.5 | 6.2 | 10.0 | 4.3 | 5.8 | 6.7 | 4.6 | 6.7 |

c)

|  | 2016257_1 | 2017303_1 | 2017304_1 | 2018161_1 | 2019011_4 | 2019117_1 | 2019323_3 | 2020046_1 | 2020263_1 | 2020282_1 | 2021242_2 | Mean |
|---|---|---|---|---|---|---|---|---|---|---|---|---|
| 20172661 | 4.6 | 2.5 | 2.6 | 3.6 | 2.4 | 2.5 | 3.0 | 4.3 | 4.5 | 2.4 | 2.7 | 3.2 |
| 20172662 | 4.5 | 3.2 | 2.3 | 4.6 | 3.7 | 2.6 | 3.3 | 6.1 | 6.0 | 2.8 | 3.3 | 3.9 |
| 20172851 | 4.7 | 2.7 | 2.5 | 4.2 | 2.2 | 2.1 | 2.7 | 4.3 | 4.6 | 3.1 | 3.0 | 3.3 |
| 20173041 | 2.3 | 2.3 |  | 2.1 | 3.5 | 2.7 | 2.8 | 2.7 | 2.8 | 2.4 | 3.1 | 2.7 |
| 20173351 | 2.7 | 2.2 | 2.7 | 2.7 | 2.8 | 2.5 | 2.2 | 2.9 | 4.5 | 2.4 | 3.1 | 2.8 |
| 20173391 | 4.4 | 2.3 | 2.4 | 3.1 | 2.0 | 2.3 | 3.3 | 2.9 | 5.1 | 2.4 | 3.3 | 3.1 |
| 20173401 | 4.3 | 2.6 | 2.2 | 2.4 | 2.3 | 2.0 | 3.2 | 2.1 | 2.2 | 2.1 | 3.0 | 2.6 |
| 20173431 | 3.9 | 4.4 | 3.8 | 2.3 | 3.1 | 2.1 | 2.7 | 2.6 | 2.6 | 2.7 | 4.4 | 3.2 |
| 20173432 | 2.5 | 2.9 | 2.3 | 2.7 | 2.8 | 2.4 | 3.2 | 2.6 | 2.9 | 2.2 | 2.7 | 2.6 |
| 20173433 | 3.2 | 2.4 | 1.9 | 2.5 | 2.4 | 2.7 | 2.7 | 2.5 | 2.0 | 2.7 | 3.7 | 2.6 |
| 20162551 | 2.3 | 4.7 | 2.7 | 2.3 | 2.9 | 4.7 | 4.3 | 3.9 | 2.5 | 4.2 | 2.7 | 3.4 |
| 20172462 | 2.8 | 2.7 | 3.1 | 2.0 | 2.4 | 3.9 | 4.6 | 3.1 | 2.1 | 4.5 | 3.2 | 3.1 |
| 20180361 | 2.8 | 3.5 | 4.0 | 2.5 | 2.0 | 3.7 | 3.0 | 4.6 | 2.6 | 3.3 | 3.7 | 3.3 |
| 20180362 | 2.0 | 2.4 | 2.4 | 3.0 | 2.3 | 3.1 | 3.8 | 4.0 | 2.8 | 2.9 | 3.1 | 2.9 |
| 20180363 | 2.9 | 2.5 | 2.2 | 3.1 | 2.2 | 2.2 | 2.1 | 3.0 | 2.6 | 2.5 | 3.1 | 2.6 |
| 20180371 | 2.5 | 2.1 | 2.5 | 2.3 | 2.3 | 2.1 | 3.6 | 2.4 | 2.0 | 3.0 | 3.8 | 2.6 |
| 20180372 | 2.7 | 2.5 | 2.0 | 2.0 | 2.0 | 2.2 | 2.4 | 2.9 | 1.7 | 2.3 | 3.9 | 2.4 |
| 20180373 | 3.8 | 4.7 | 2.7 | 2.9 | 4.8 | 7.7 | 8.5 | 9.6 | 2.6 | 7.0 | 5.9 | 5.5 |
| 20180381 | 4.1 | 5.5 | 3.0 | 5.3 | 5.7 | 6.4 | 6.1 | 14.1 | 3.5 | 4.4 | 7.2 | 5.9 |
| 20180391 | 2.4 | 2.3 | 3.0 | 2.8 | 3.4 | 2.3 | 3.4 | 4.3 | 3.0 | 2.7 | 3.7 | 3.0 |
| 20181121 | 2.6 | 3.1 | 3.6 | 2.1 | 2.5 | 3.2 | 5.0 | 3.2 | 2.8 | 3.8 | 4.3 | 3.3 |
| 20181122 | 4.4 | 5.5 | 3.0 | 3.5 | 5.2 | 7.7 | 8.6 | 10.3 | 2.6 | 5.9 | 6.0 | 5.7 |

Figure S2. Same as in Figure S1 for KSRS

a)

| USRK | 2016255_1 | 2017246_2 | 2018036_1 | 2018036_2 | 2018036_3 | 2018037_1 | 2018037_2 | 2018037_3 | 2018038_1 | 2018039_1 | 2018112_1 | 2018112_2 | 2019011_2 | 2019011_3 | 2019158_1 | 2019158_2 | 2019158_3 | 2019233_1 | 2019310_1 | 2019323_1 | 2019323_2 | 2020071_1 | 2020220_1 | 2020308_1 | 2020308_2 | 2020308_3 | 2020308_4 | 2021108_1 | 2021109_1 | 2021110_1 | 2021175_1 | 2021244_1 |
|---|---|---|---|---|---|---|---|---|---|---|---|---|---|---|---|---|---|---|---|---|---|---|---|---|---|---|---|---|---|---|---|---|
| 20172661 | 3.4 | 3.1 | 4.8 | 4.9 | 3.9 | 3.5 | 4.4 | 7.6 | 4.7 | 6.9 | 6.4 | 9.8 | 4.0 | 5.9 | 4.9 | 9.7 | 4.4 | 4.4 | 3.2 | 6.7 | 8.4 | 3.4 | 6.6 | 2.7 | 3.3 | 4.1 | 3.3 | 6.5 | 5.8 | 3.3 | 4.3 | 16.6 |
| 20172662 | 4.2 | 5.2 | 14.6 | 8.2 | 7.0 | 4.8 | 6.1 | 17.9 | 11.4 | 11.0 | 13.5 | 15.0 | 10.2 | 11.6 | 10.9 | 12.3 | 7.1 | 7.5 | 7.0 | 11.2 | 11.6 | 4.5 | 17.0 | 3.7 | 8.0 | 9.6 | 5.7 | 8.0 | 12.2 | 7.4 | 6.7 | 15.9 |
| 20172851 | 5.2 | 3.4 | 14.2 | 6.6 | 6.8 | 7.1 | 7.0 | 18.0 | 24.7 | 13.2 | 10.8 | 18.9 | 9.6 | 12.3 | 5.7 | 7.7 | 5.3 | 3.9 | 3.8 | 12.6 | 13.0 | 4.9 | 8.4 | 4.1 | 4.3 | 6.2 | 5.4 | 4.3 | 11.4 | 6.5 | 5.0 | 12.2 |
| 20173041 | 2.5 | 3.3 | 3.3 | 3.0 | 2.5 | 2.2 | 2.5 | 5.5 | 3.4 | 3.1 | 3.4 | 4.0 | 4.3 | 3.3 | 2.8 | 4.5 | 2.7 | 3.2 | 2.7 | 2.6 | 3.0 | 2.5 | 3.3 | 2.8 | 2.4 | 3.0 | 3.3 | 3.0 | 5.3 | 3.1 | 3.8 | 5.9 |
| 20173351 | 3.1 | 3.9 | 9.9 | 4.1 | 4.4 | 4.9 | 5.0 | 12.1 | 9.9 | 6.5 | 6.9 | 10.4 | 6.5 | 6.7 | 3.5 | 5.6 | 3.2 | 2.8 | 2.9 | 8.0 | 8.8 | 2.8 | 4.9 | 3.0 | 1.9 | 3.1 | 2.6 | 2.5 | 6.5 | 2.3 | 2.9 | 6.9 |
| 20173391 | 5.7 | 4.7 | 12.8 | 6.1 | 5.2 | 6.1 | 6.4 | 15.9 | 17.3 | 11.3 | 8.6 | 16.7 | 6.9 | 11.2 | 7.0 | 10.7 | 3.9 | 4.6 | 4.2 | 10.5 | 10.6 | 3.3 | 8.3 | 3.1 | 3.5 | 5.2 | 3.5 | 4.3 | 11.2 | 4.7 | 5.4 | 9.1 |
| 20173401 | 3.8 | 3.4 | 6.7 | 2.9 | 2.8 | 3.0 | 2.8 | 5.4 | 8.3 | 5.4 | 3.1 | 6.5 | 3.9 | 5.2 | 3.0 | 4.5 | 2.3 | 3.0 | 2.6 | 5.3 | 4.8 | 3.3 | 3.6 | 2.4 | 2.0 | 2.2 | 1.9 | 2.1 | 5.6 | 2.6 | 2.3 | 3.6 |
| 20173431 | 2.7 | 3.5 | 3.3 | 3.7 | 5.5 | 2.5 | 2.6 | 4.6 | 3.4 | 4.3 | 4.6 | 4.7 | 2.3 | 3.0 | 2.3 | 2.2 | 2.3 | 2.1 | 2.2 | 3.8 | 3.3 | 2.0 | 2.4 | 2.3 | 2.6 | 2.5 | 2.6 | 2.4 | 3.9 | 2.6 | 2.5 | 5.6 |
| 20173432 | 5.3 | 4.4 | 7.0 | 5.5 | 2.9 | 3.6 | 3.2 | 8.9 | 7.2 | 7.0 | 7.5 | 7.5 | 4.2 | 5.6 | 7.3 | 9.0 | 4.3 | 3.8 | 5.0 | 4.9 | 5.5 | 2.3 | 7.1 | 3.0 | 3.8 | 6.5 | 3.8 | 2.9 | 9.0 | 4.7 | 4.8 | 8.0 |
| 20173433 | 4.4 | 4.3 | 6.8 | 6.2 | 3.8 | 3.2 | 3.4 | 7.3 | 6.0 | 7.6 | 8.7 | 7.2 | 5.1 | 5.8 | 4.1 | 5.7 | 4.1 | 3.0 | 4.4 | 5.6 | 4.7 | 2.7 | 5.5 | 3.4 | 3.8 | 4.8 | 3.4 | 2.8 | 6.1 | 3.1 | 4.6 | 8.6 |
| 20162551 |  | 10.4 | 7.6 | 3.7 | 8.0 | 5.3 | 3.8 | 7.5 | 7.4 | 4.0 | 3.8 | 7.6 | 3.8 | 5.5 | 3.9 | 5.5 | 4.2 | 3.5 | 16.3 | 5.2 | 7.3 | 3.8 | 5.6 | 3.3 | 2.8 | 7.0 | 3.8 | 4.9 | 7.0 | 4.5 | 5.3 | 9.7 |
| 20172462 | 9.3 |  | 4.3 | 3.7 | 5.3 | 2.9 | 2.7 | 7.7 | 4.4 | 3.1 | 5.9 | 5.1 | 3.7 | 3.3 | 2.4 | 3.7 | 2.9 | 3.5 | 13.4 | 3.6 | 3.8 | 3.0 | 3.8 | 4.2 | 4.0 | 3.9 | 3.1 | 3.2 | 4.1 | 4.5 | 2.8 | 6.0 |
| 20180361 | 5.4 | 3.8 |  | 8.5 | 14.8 | 7.2 | 11.1 | 23.8 | 18.2 | 15.0 | 14.4 | 25.9 | 10.2 | 16.8 | 11.0 | 17.9 | 11.2 | 10.3 | 4.2 | 14.3 | 17.0 | 5.5 | 20.0 | 12.0 | 8.7 | 16.2 | 6.0 | 11.9 | 15.5 | 7.3 | 8.2 | 12.7 |
| 20180362 | 2.4 | 3.3 | 8.7 |  | 5.8 | 2.5 | 2.9 | 11.4 | 4.0 | 4.7 | 6.7 | 7.2 | 4.8 | 5.9 | 4.5 | 3.5 | 4.2 | 6.6 | 5.0 | 4.8 | 6.8 | 2.1 | 6.3 | 3.9 | 5.3 | 5.8 | 2.5 | 4.8 | 5.0 | 3.6 | 4.3 | 7.2 |
| 20180363 | 3.6 | 2.7 | 16.0 | 4.9 |  | 4.8 | 4.9 | 12.0 | 12.4 | 11.7 | 4.4 | 13.0 | 3.3 | 12.5 | 6.7 | 7.4 | 9.7 | 4.2 | 2.8 | 6.2 | 9.8 | 2.8 | 13.3 | 4.4 | 3.7 | 10.3 | 2.7 | 5.3 | 7.7 | 3.2 | 5.7 | 8.4 |
| 20180371 | 3.8 | 2.4 | 6.9 | 2.4 | 4.7 |  | 6.5 | 8.7 | 12.8 | 6.9 | 3.1 | 11.7 | 3.6 | 8.1 | 6.7 | 9.3 | 6.4 | 3.5 | 3.6 | 5.8 | 6.2 | 2.4 | 9.9 | 2.7 | 2.6 | 10.2 | 3.9 | 4.8 | 10.3 | 3.4 | 3.5 | 9.7 |
| 20180372 | 3.6 | 2.4 | 11.9 | 3.2 | 5.8 | 6.1 |  | 11.1 | 14.6 | 10.2 | 5.1 | 14.3 | 5.7 | 10.7 | 5.5 | 9.6 | 8.0 | 4.3 | 3.7 | 7.0 | 9.9 | 2.9 | 12.3 | 4.8 | 3.4 | 10.2 | 3.5 | 6.4 | 10.3 | 2.3 | 3.9 | 9.5 |
| 20180373 | 7.1 | 6.3 | 24.8 | 13.6 | 13.4 | 8.8 | 13.1 |  | 19.5 | 14.2 | 21.5 | 29.8 | 15.4 | 16.6 | 14.1 | 21.3 | 10.7 | 14.2 | 6.6 | 13.2 | 18.2 | 5.6 | 21.0 | 14.9 | 11.1 | 19.8 | 5.9 | 13.3 | 14.7 | 6.6 | 8.5 | 13.6 |
| 20180381 | 7.2 | 5.1 | 20.4 | 4.9 | 15.4 | 14.6 | 16.0 | 21.2 |  | 20.8 | 8.8 | 29.4 | 10.1 | 21.0 | 12.6 | 23.3 | 13.8 | 9.3 | 5.9 | 14.9 | 16.6 | 7.0 | 27.1 | 7.0 | 4.7 | 21.1 | 10.8 | 10.7 | 21.8 | 10.3 | 9.5 | 26.9 |
| 20180391 | 3.8 | 2.7 | 16.0 | 5.5 | 13.5 | 7.4 | 11.1 | 14.1 | 19.5 |  | 8.3 | 17.6 | 4.5 | 22.5 | 7.9 | 14.7 | 15.4 | 7.5 | 3.7 | 10.6 | 13.2 | 5.4 | 26.2 | 5.3 | 5.8 | 12.1 | 7.2 | 8.2 | 13.9 | 5.7 | 9.8 | 18.9 |
| 20181121 | 3.0 | 5.7 | 13.7 | 8.9 | 4.7 | 3.5 | 4.9 | 20.2 | 6.7 | 7.4 |  | 13.6 | 9.7 | 8.5 | 6.7 | 6.3 | 6.4 | 10.7 | 7.3 | 8.3 | 12.0 | 2.4 | 9.7 | 9.5 | 11.2 | 10.2 | 4.9 | 6.8 | 9.1 | 5.4 | 5.9 | 12.9 |
| 20181122 | 8.6 | 4.8 | 26.8 | 8.3 | 13.3 | 12.3 | 15.1 | 30.5 | 26.2 | 17.7 | 15.2 |  | 12.4 | 19.1 | 13.2 | 20.1 | 13.4 | 13.0 | 6.6 | 18.8 | 21.4 | 6.4 | 25.0 | 13.7 | 9.0 | 25.0 | 9.6 | 13.8 | 17.6 | 9.5 | 10.2 | 19.3 |
| KSRS |  |  |  |  |  |  |  |  |  |  |  |  |  |  |  |  |  |  |  |  |  |  |  |  |  |  |  |  |  |  |  |  |
| 20172661 | 7.3 | 5.1 | 4.5 | 3.3 | 4.6 | 3.0 | 2.5 | 10.8 | 16.4 | 7.7 | 4.9 | 11.5 | 4.8 | 7.0 | 5.5 | 14.9 | 4.0 | 6.1 | 6.5 | 4.1 | 7.3 | 3.7 | 12.3 | 2.6 | 3.2 | 4.8 | 2.6 | 2.9 | 13.6 | 7.6 | 11.6 | 10.2 |
| 20172662 | 8.7 | 7.5 | 7.4 | 4.3 | 4.9 | 4.3 | 3.4 | 13.9 | 22.5 | 10.8 | 8.3 | 17.8 | 5.9 | 10.5 | 8.2 | 23.1 | 5.7 | 8.2 | 8.3 | 6.4 | 10.1 | 4.3 | 19.2 | 3.7 | 2.9 | 6.4 | 4.5 | 3.6 | 17.6 | 10.6 | 14.9 | 15.5 |
| 20172851 | 7.3 | 6.4 | 5.1 | 2.8 | 3.0 | 3.0 | 2.5 | 10.3 | 13.7 | 6.3 | 5.9 | 12.0 | 4.1 | 6.4 | 6.0 | 13.9 | 4.4 | 6.5 | 7.0 | 3.7 | 7.7 | 3.1 | 11.1 | 3.3 | 3.0 | 6.3 | 3.6 | 3.5 | 12.1 | 6.8 | 10.8 | 11.0 |
| 20173041 | 2.6 | 2.6 | 2.5 | 2.7 | 2.1 | 2.5 | 1.9 | 2.8 | 3.2 | 3.1 | 2.9 | 3.0 | 2.7 | 3.3 | 2.4 | 2.6 | 2.9 | 2.3 | 2.3 | 2.2 | 2.8 | 2.7 | 3.3 | 2.8 | 2.2 | 2.8 | 2.6 | 2.5 | 3.3 | 2.7 | 2.9 | 3.0 |
| 20173351 | 3.9 | 4.1 | 2.7 | 2.5 | 3.4 | 2.5 | 2.9 | 5.9 | 7.7 | 4.3 | 3.5 | 7.8 | 2.8 | 4.6 | 4.7 | 7.4 | 2.3 | 2.8 | 4.2 | 3.5 | 4.9 | 2.5 | 6.2 | 2.2 | 2.7 | 2.6 | 2.4 | 2.5 | 8.6 | 3.7 | 6.9 | 6.4 |
| 20173391 | 5.2 | 5.4 | 3.7 | 3.8 | 2.8 | 2.6 | 2.9 | 7.6 | 10.8 | 5.2 | 4.9 | 10.7 | 3.9 | 4.9 | 4.6 | 10.3 | 4.0 | 6.5 | 4.9 | 3.7 | 6.1 | 2.2 | 8.0 | 2.5 | 2.5 | 3.9 | 3.1 | 3.0 | 10.3 | 5.3 | 8.8 | 9.1 |
| 20173401 | 2.7 | 2.4 | 2.5 | 2.7 | 2.0 | 1.8 | 2.0 | 3.8 | 4.7 | 2.0 | 2.7 | 4.2 | 2.4 | 2.3 | 3.2 | 4.6 | 2.5 | 2.8 | 2.3 | 2.3 | 2.2 | 2.4 | 3.3 | 2.6 | 2.4 | 2.5 | 2.2 | 1.8 | 4.7 | 2.1 | 3.1 | 2.9 |
| 20173431 | 2.8 | 2.9 | 2.4 | 3.8 | 2.0 | 2.9 | 2.3 | 2.3 | 2.2 | 2.1 | 3.0 | 2.3 | 3.0 | 2.5 | 3.9 | 2.3 | 2.9 | 2.7 | 2.0 | 2.3 | 2.5 | 2.6 | 2.4 | 2.3 | 2.1 | 2.3 | 2.4 | 2.1 | 2.9 | 2.6 | 2.4 | 3.4 |
| 20173432 | 3.8 | 3.5 | 3.3 | 3.2 | 2.4 | 2.3 | 2.2 | 5.8 | 9.1 | 3.6 | 3.7 | 6.9 | 3.0 | 4.0 | 3.6 | 4.6 | 2.0 | 3.4 | 3.7 | 2.8 | 4.2 | 2.9 | 4.9 | 2.7 | 2.7 | 3.2 | 2.3 | 2.7 | 4.5 | 3.0 | 2.9 | 6.8 |
| 20173433 | 2.5 | 2.3 | 2.9 | 2.6 | 2.5 | 2.2 | 2.0 | 3.9 | 3.3 | 2.7 | 2.8 | 3.7 | 2.6 | 2.6 | 2.9 | 3.3 | 2.6 | 3.0 | 2.9 | 2.8 | 3.0 | 2.4 | 3.4 | 2.2 | 2.3 | 2.7 | 2.6 | 2.2 | 3.4 | 2.2 | 2.7 | 4.6 |
| 20162551 |  | 14.0 | 5.1 | 5.7 | 3.3 | 2.4 | 2.8 | 14.3 | 12.7 | 6.3 | 8.6 | 14.3 | 4.3 | 5.7 | 5.5 | 14.2 | 4.6 | 6.6 | 16.2 | 4.3 | 8.0 | 3.6 | 11.6 | 4.2 | 3.8 | 5.5 | 3.2 | 3.9 | 12.4 | 4.8 | 7.8 | 7.8 |
| 20172462 | 14.5 |  | 4.5 | 6.3 | 2.2 | 2.5 | 3.0 | 13.7 | 10.9 | 5.5 | 6.9 | 12.8 | 4.9 | 5.8 | 5.5 | 12.8 | 3.1 | 7.1 | 12.8 | 4.7 | 8.3 | 2.5 | 10.1 | 3.7 | 3.8 | 5.8 | 2.7 | 3.7 | 12.4 | 3.4 | 7.9 | 7.8 |
| 20180361 | 5.4 | 4.6 |  | 5.1 | 4.5 | 3.2 | 4.6 | 14.9 | 13.7 | 5.9 | 6.7 | 15.9 | 3.0 | 7.1 | 7.4 | 13.9 | 5.0 | 8.4 | 4.3 | 4.5 | 7.7 | 3.0 | 14.3 | 6.5 | 4.9 | 6.9 | 4.0 | 3.7 | 14.4 | 4.8 | 6.2 | 11.3 |
| 20180362 | 5.2 | 5.4 | 5.4 |  | 2.4 | 2.7 | 2.1 | 10.5 | 7.9 | 3.2 | 7.4 | 11.6 | 3.3 | 5.1 | 5.9 | 9.5 | 3.1 | 6.7 | 5.1 | 4.1 | 7.2 | 2.8 | 10.4 | 5.5 | 4.4 | 6.5 | 3.1 | 3.1 | 11.2 | 3.5 | 4.6 | 8.8 |
| 20180363 | 3.6 | 2.9 | 4.3 | 2.0 |  | 2.5 | 2.3 | 7.6 | 7.1 | 4.5 | 3.4 | 10.5 | 3.0 | 6.5 | 4.1 | 8.2 | 3.4 | 5.3 | 2.7 | 3.8 | 5.9 | 3.4 | 13.1 | 3.5 | 2.9 | 3.9 | 2.9 | 3.1 | 10.3 | 4.5 | 6.4 | 7.1 |
| 20180371 | 2.5 | 2.1 | 3.2 | 2.7 | 2.2 |  | 2.3 | 4.3 | 5.2 | 3.0 | 2.1 | 5.1 | 3.3 | 2.8 | 2.6 | 4.7 | 2.6 | 2.1 | 2.5 | 2.4 | 3.4 | 2.5 | 5.0 | 2.6 | 2.8 | 4.1 | 2.8 | 2.9 | 5.3 | 2.3 | 3.4 | 4.4 |
| 20180372 | 2.8 | 2.9 | 3.3 | 2.2 | 1.9 | 2.2 |  | 4.9 | 7.0 | 3.1 | 2.5 | 6.8 | 3.1 | 4.0 | 3.1 | 6.8 | 2.6 | 3.5 | 2.8 | 2.3 | 3.6 | 2.5 | 6.9 | 3.3 | 2.2 | 4.2 | 2.2 | 2.6 | 7.6 | 3.1 | 3.3 | 6.1 |
| 20180373 | 13.7 | 14.4 | 13.7 | 12.3 | 9.3 | 4.4 | 5.7 |  | 30.2 | 15.7 | 19.8 | 36.9 | 9.4 | 18.1 | 16.7 | 34.4 | 12.1 | 17.1 | 14.6 | 13.9 | 22.6 | 7.4 | 30.3 | 13.6 | 7.4 | 16.1 | 7.0 | 8.2 | 26.2 | 10.6 | 15.1 | 26.9 |
| 20180381 | 12.9 | 11.4 | 16.2 | 9.5 | 8.9 | 5.6 | 8.8 | 32.1 |  | 20.1 | 14.4 | 38.3 | 9.7 | 19.4 | 15.9 | 39.6 | 14.0 | 17.9 | 11.7 | 13.0 | 20.9 | 9.5 | 34.3 | 9.0 | 6.0 | 15.7 | 7.4 | 8.0 | 31.3 | 17.1 | 21.7 | 35.3 |
| 20180391 | 6.3 | 5.3 | 6.1 | 3.6 | 3.2 | 3.2 | 2.9 | 15.4 | 18.5 |  | 7.4 | 20.4 | 4.6 | 14.8 | 6.5 | 16.7 | 7.9 | 7.1 | 5.1 | 5.4 | 10.4 | 4.0 | 26.0 | 3.8 | 2.5 | 5.2 | 3.7 | 4.8 | 20.0 | 6.9 | 12.9 | 13.7 |
| 20181121 | 9.1 | 7.0 | 6.3 | 7.6 | 2.9 | 2.2 | 2.5 | 19.8 | 14.4 | 6.7 |  | 17.3 | 5.6 | 7.7 | 7.3 | 15.8 | 4.4 | 8.4 | 7.6 | 5.5 | 10.2 | 2.5 | 13.6 | 7.0 | 5.5 | 7.3 | 3.7 | 3.9 | 13.4 | 4.6 | 6.2 | 10.0 |
| 20181122 | 13.4 | 14.1 | 16.8 | 12.6 | 10.6 | 5.5 | 8.0 | 38.6 | 35.5 | 20.8 | 18.9 |  | 10.6 | 17.8 | 17.9 | 39.1 | 12.8 | 18.3 | 14.7 | 17.3 | 24.0 | 7.5 | 31.2 | 13.5 | 8.6 | 18.6 | 8.3 | 8.4 | 27.7 | 12.4 | 14.7 | 33.2 |

b)

| USRK | 2017246_1 | 2017266_1 | 2017266_2 | 2017285_1 | 2017335_1 | 2017339_1 | 2017339_2 | 2017339_3 | 2017340_1 | 2017343_1 | 2017343_2 | 2017343_3 | 2019001_1 | 2019078_1 | 2019079_1 | 2019080_1 | 2019080_2 | 2019218_1 | 2019270_1 | 2020029_1 | 2020029_2 | 2020030_1 | 2020092_1 | 2020273_1 | 2020309_1 | 2020319_1 | 2020319_2 | 2020321_1 | 2020321_2 | 2020321_3 | 2021013_1 | 2021030_1 | 2021031_1 | 2021031_2 | 2021031_1 | 2021164_1 | 2021169_1 | 2021209_1 | 2021224_1 | 2021225_1 | 2021227_1 | 2021242_1 | 2021244_2 | 2021253_1 | 2021253_2 | 2021254_1 |
|---|---|---|---|---|---|---|---|---|---|---|---|---|---|---|---|---|---|---|---|---|---|---|---|---|---|---|---|---|---|---|---|---|---|---|---|---|---|---|---|---|---|---|---|---|---|---|
| 20172661 | 6.1 |  | 39.1 | 6.1 | 22.5 | 13.3 | 15.2 | 14.0 | 11.6 | 9.0 | 13.5 | 15.8 | 8.8 | 10.9 | 15.6 | 14.1 | 4.9 | 4.8 | 5.8 | 9.7 | 3.8 | 5.6 | 5.7 | 6.0 | 5.2 | 6.4 | 4.1 | 5.0 | 3.7 | 10.4 | 5.8 | 9.3 | 7.3 | 6.5 | 3.9 | 9.2 | 3.9 | 6.4 | 6.3 | 7.5 | 13.7 | 9.6 | 9.8 | 6.2 | 7.0 | 4.3 |
| 20172662 | 8.1 | 36.6 |  | 8.1 | 19.8 | 16.8 | 9.3 | 15.2 | 8.8 | 7.3 | 15.7 | 17.3 | 8.7 | 16.9 | 23.8 | 20.9 | 8.7 | 8.6 | 9.8 | 15.4 | 6.8 | 7.6 | 10.1 | 5.2 | 14.5 | 11.7 | 5.2 | 8.0 | 6.4 | 11.4 | 9.9 | 18.5 | 10.6 | 10.1 | 6.1 | 12.7 | 4.0 | 5.3 | 9.4 | 10.7 | 11.7 | 8.8 | 10.7 | 7.1 | 7.9 | 6.4 |
| 20172851 | 8.4 | 32.3 | 30.0 | 8.4 | 14.4 | 21.4 | 13.8 | 6.2 | 7.9 | 8.8 | 16.4 | 15.0 | 6.8 | 8.9 | 13.2 | 10.5 | 7.0 | 6.1 | 8.8 | 12.8 | 5.8 | 6.2 | 5.4 | 5.0 | 6.6 | 7.6 | 4.6 | 3.8 | 4.2 | 6.3 | 3.9 | 9.5 | 5.2 | 4.6 | 5.2 | 12.4 | 3.7 | 6.7 | 5.7 | 14.4 | 7.9 | 5.5 | 6.6 | 5.2 | 5.5 | 3.9 |
| 20173041 | 2.8 | 2.5 | 3.2 | 2.8 | 3.2 | 3.4 | 3.1 | 2.4 | 2.3 | 2.7 | 3.1 | 2.7 | 2.6 | 2.7 | 2.2 | 2.2 | 2.4 | 2.4 | 2.2 | 2.6 | 2.3 | 3.1 | 2.4 | 2.8 | 3.6 | 2.0 | 2.9 | 2.6 | 2.8 | 2.6 | 2.8 | 3.2 | 2.6 | 2.1 | 3.1 | 2.9 | 2.6 | 4.0 | 2.9 | 2.5 | 2.3 | 3.5 | 3.1 | 3.7 | 3.1 | 3.5 |
| 20173351 | 7.2 | 12.9 | 10.9 | 7.2 |  | 12.4 | 5.3 | 2.1 | 2.4 | 3.2 | 4.3 | 5.2 | 3.4 | 2.7 | 2.6 | 2.9 | 3.5 | 2.9 | 7.5 | 7.4 | 5.1 | 4.5 | 4.9 | 2.0 | 4.2 | 2.2 | 2.8 | 3.3 | 2.3 | 4.0 | 2.7 | 2.0 | 2.1 | 2.0 | 3.7 | 6.7 | 2.3 | 3.4 | 3.1 | 10.3 | 4.5 | 3.2 | 2.5 | 3.2 | 2.5 | 2.7 |
| 20173391 | 8.5 | 24.3 | 21.9 | 8.5 | 15.3 |  | 21.0 | 7.0 | 13.7 | 9.3 | 19.1 | 16.8 | 9.6 | 6.7 | 11.6 | 9.4 | 8.8 | 4.1 | 12.1 | 10.8 | 7.5 | 8.0 | 5.1 | 4.4 | 5.5 | 7.0 | 2.8 | 3.3 | 2.2 | 5.4 | 5.5 | 7.3 | 4.5 | 4.2 | 5.4 | 12.7 | 5.1 | 5.4 | 7.0 | 16.1 | 8.3 | 3.9 | 10.3 | 3.7 | 4.6 | 4.6 |
| 20173401 | 5.7 | 8.7 | 7.0 | 5.7 | 5.1 | 12.5 | 6.3 | 2.5 |  | 4.4 | 8.4 | 6.0 | 4.2 | 5.0 | 6.5 | 6.0 | 6.7 | 2.2 | 8.0 | 5.5 | 4.7 | 5.7 | 3.5 | 3.9 | 2.4 | 2.9 | 1.8 | 1.9 | 1.7 | 2.8 | 2.4 | 3.3 | 2.2 | 2.2 | 2.6 | 4.3 | 2.5 | 2.4 | 1.8 | 10.7 | 3.7 | 2.0 | 3.9 | 2.2 | 2.6 | 3.1 |
| 20173431 | 4.0 | 5.1 | 3.7 | 4.0 | 1.9 | 7.8 | 6.5 | 1.8 | 1.9 |  | 3.4 | 8.9 | 2.9 | 2.5 | 2.8 | 2.4 | 3.6 | 2.7 | 4.2 | 3.1 | 3.0 | 2.7 | 2.6 | 3.3 | 3.2 | 2.1 | 2.0 | 1.9 | 2.4 | 1.8 | 2.8 | 1.4 | 1.9 | 1.7 | 2.6 | 3.8 | 2.2 | 3.6 | 2.5 | 5.6 | 1.6 | 3.7 | 2.0 | 2.0 | 2.9 | 2.0 |
| 20173432 | 11.1 | 10.6 | 8.9 | 11.1 | 3.1 | 17.0 | 11.5 | 2.2 | 3.0 | 12.1 |  | 18.4 | 5.1 | 2.7 | 4.1 | 2.9 | 4.4 | 4.4 | 7.0 | 5.0 | 3.3 | 4.3 | 5.1 | 3.3 | 6.8 | 3.1 | 2.9 | 3.2 | 2.2 | 4.1 | 5.6 | 2.9 | 2.9 | 3.7 | 5.0 | 9.5 | 3.6 | 3.4 | 6.8 | 7.7 | 4.8 | 2.7 | 4.2 | 5.5 | 6.5 | 3.4 |
| 20173433 | 10.3 | 7.5 | 8.6 | 10.3 | 3.3 | 13.6 | 10.4 | 2.2 | 3.2 | 10.1 | 7.8 |  | 4.9 | 3.0 | 3.6 | 3.4 | 3.6 | 5.0 | 7.0 | 5.5 | 4.2 | 3.8 | 5.2 | 4.1 | 5.5 | 2.7 | 2.4 | 2.9 | 2.1 | 3.2 | 4.3 | 2.1 | 3.1 | 2.3 | 3.4 | 6.0 | 2.5 | 3.4 | 5.3 | 7.0 | 3.3 | 2.5 | 3.2 | 4.8 | 6.7 | 2.9 |
| 20162551 | 11.5 | 4.7 | 5.7 | 11.5 | 3.1 | 6.7 | 5.8 | 2.6 | 4.9 | 4.2 | 8.4 | 6.6 | 4.3 | 3.2 | 5.0 | 5.5 | 4.0 | 3.2 | 6.7 | 3.7 | 4.7 | 5.7 | 2.1 | 3.9 | 4.2 | 6.4 | 3.1 | 3.4 | 2.3 | 4.1 | 2.4 | 4.2 | 3.0 | 2.5 | 3.6 | 5.8 | 2.8 | 3.9 | 2.4 | 3.1 | 3.9 | 3.3 | 4.4 | 5.3 | 5.5 | 5.0 |
| 20172462 | 7.1 | 3.4 | 6.7 | 7.1 | 3.1 | 3.7 | 4.2 | 2.5 | 3.4 | 4.4 | 5.9 | 5.1 | 3.1 | 2.4 | 3.6 | 4.3 | 3.2 | 3.0 | 3.7 | 4.3 | 3.7 | 4.1 | 2.6 | 2.6 | 4.8 | 4.1 | 2.5 | 2.9 | 2.5 | 2.8 | 2.4 | 2.7 | 3.0 | 3.5 | 2.1 | 4.6 | 2.6 | 2.5 | 2.2 | 3.1 | 3.1 | 2.4 | 3.4 | 4.7 | 4.3 | 2.9 |
| 20180361 | 4.4 | 8.6 | 7.4 | 4.4 | 3.6 | 10.8 | 4.6 | 2.8 | 2.3 | 3.2 | 5.7 | 6.1 | 3.5 | 3.2 | 5.2 | 4.2 | 3.0 | 4.9 | 5.4 | 5.9 | 4.3 | 4.6 | 4.5 | 2.3 | 7.5 | 4.1 | 3.3 | 3.0 | 3.7 | 3.7 | 3.4 | 3.8 | 3.2 | 3.4 | 2.6 | 8.1 | 2.8 | 3.2 | 4.2 | 4.9 | 6.2 | 2.7 | 4.8 | 4.9 | 3.6 | 3.2 |
| 20180362 | 4.5 | 3.8 | 3.5 | 4.5 | 1.8 | 4.0 | 3.0 | 2.5 | 3.1 | 2.8 | 2.2 | 4.4 | 2.5 | 2.1 | 2.0 | 2.4 | 2.7 | 2.0 | 3.5 | 2.9 | 2.3 | 3.8 | 2.8 | 1.9 | 4.2 | 2.9 | 2.8 | 2.8 | 1.7 | 1.7 | 2.1 | 2.2 | 2.3 | 1.8 | 2.7 | 2.4 | 2.3 | 2.5 | 2.9 | 2.3 | 1.7 | 3.2 | 2.4 | 2.5 | 3.6 | 2.7 |
| 20180363 | 2.8 | 4.0 | 4.9 | 2.8 | 2.2 | 4.2 | 2.4 | 2.0 | 2.4 | 2.4 | 2.4 | 3.2 | 3.0 | 2.4 | 2.0 | 2.7 | 2.4 | 1.9 | 2.7 | 2.0 | 2.0 | 2.7 | 2.7 | 2.3 | 2.1 | 2.4 | 2.5 | 2.4 | 2.1 | 2.3 | 1.9 | 2.7 | 2.7 | 2.8 | 2.7 | 3.6 | 1.7 | 2.0 | 1.5 | 1.6 | 3.7 | 2.5 | 3.3 | 3.3 | 3.5 | 2.9 |
| 20180371 | 5.0 | 5.0 | 3.1 | 5.0 | 2.3 | 5.4 | 3.2 | 2.7 | 2.6 | 2.5 | 2.1 | 2.4 | 3.1 | 2.4 | 2.8 | 3.0 | 2.4 | 1.9 | 2.5 | 3.0 | 2.5 | 2.8 | 2.1 | 2.1 | 2.4 | 2.9 | 2.0 | 2.5 | 2.6 | 2.1 | 2.1 | 2.5 | 2.8 | 2.2 | 2.2 | 4.0 | 2.3 | 2.0 | 1.7 | 2.5 | 2.6 | 2.4 | 2.8 | 2.4 | 2.9 | 2.5 |
| 20180372 | 4.6 | 4.0 | 3.7 | 4.6 | 2.4 | 5.1 | 3.0 | 2.0 | 2.5 | 2.5 | 2.5 | 3.1 | 2.9 | 2.5 | 2.9 | 2.7 | 2.9 | 2.5 | 2.6 | 3.6 | 2.5 | 3.1 | 2.4 | 2.2 | 3.2 | 2.6 | 2.2 | 3.0 | 2.2 | 2.3 | 2.6 | 2.1 | 2.3 | 2.0 | 2.6 | 5.1 | 2.3 | 2.2 | 2.4 | 2.0 | 2.6 | 3.4 | 3.2 | 3.1 | 3.3 | 2.2 |
| 20180373 | 7.7 | 10.7 | 10.3 | 7.7 | 5.1 | 13.0 | 6.6 | 3.2 | 3.0 | 4.1 | 5.9 | 6.8 | 6.1 | 4.6 | 5.4 | 4.0 | 2.8 | 4.5 | 5.6 | 7.3 | 4.7 | 4.6 | 4.7 | 3.4 | 10.2 | 5.8 | 3.4 | 3.5 | 4.0 | 4.7 | 3.8 | 4.1 | 3.0 | 3.4 | 2.8 | 9.9 | 4.2 | 3.7 | 5.6 | 5.4 | 9.5 | 3.0 | 5.5 | 6.3 | 4.4 | 4.9 |
| 20180381 | 7.8 | 10.7 | 9.1 | 7.8 | 4.4 | 17.0 | 10.2 | 3.0 | 3.2 | 3.6 | 8.6 | 7.4 | 5.5 | 3.5 | 7.0 | 4.5 | 3.8 | 4.3 | 6.4 | 9.6 | 4.5 | 5.4 | 3.7 | 3.5 | 11.3 | 5.9 | 3.1 | 3.3 | 4.2 | 4.5 | 4.1 | 4.2 | 3.4 | 3.4 | 3.6 | 14.1 | 4.8 | 4.5 | 4.7 | 6.1 | 8.2 | 2.8 | 5.9 | 5.9 | 5.4 | 3.5 |
| 20180391 | 5.4 | 6.1 | 5.3 | 5.4 | 2.9 | 8.5 | 3.9 | 2.5 | 2.9 | 3.6 | 4.4 | 4.8 | 3.9 | 2.6 | 3.6 | 3.2 | 2.6 | 3.6 | 4.8 | 4.5 | 3.2 | 3.7 | 3.4 | 2.2 | 4.8 | 3.7 | 2.0 | 2.7 | 2.5 | 3.0 | 2.4 | 3.6 | 2.3 | 2.7 | 2.6 | 5.9 | 2.6 | 2.5 | 4.3 | 4.5 | 5.7 | 2.6 | 4.3 | 5.1 | 4.7 | 2.7 |
| 20181121 | 4.6 | 8.0 | 6.2 | 4.6 | 2.7 | 6.5 | 4.1 | 2.0 | 2.1 | 3.6 | 3.1 | 6.6 | 3.5 | 2.3 | 2.5 | 3.3 | 2.5 | 2.9 | 5.0 | 4.8 | 5.1 | 4.4 | 3.5 | 2.5 | 6.4 | 3.4 | 2.3 | 3.0 | 2.3 | 3.4 | 2.6 | 2.1 | 2.3 | 1.8 | 2.1 | 4.2 | 2.3 | 2.1 | 3.0 | 3.3 | 2.7 | 2.9 | 2.8 | 3.3 | 2.7 | 3.1 |
| 20181122 | 8.3 | 10.1 | 8.8 | 8.3 | 3.8 | 12.3 | 5.9 | 2.5 | 3.1 | 4.2 | 6.8 | 7.2 | 6.9 | 2.5 | 4.2 | 4.2 | 3.3 | 4.1 | 6.2 | 6.2 | 5.0 | 4.9 | 3.8 | 3.8 | 9.5 | 5.7 | 3.1 | 4.1 | 3.0 | 4.4 | 4.5 | 3.9 | 2.8 | 3.6 | 3.8 | 10.5 | 4.6 | 3.4 | 3.7 | 3.5 | 5.4 | 2.8 | 5.4 | 5.8 | 4.9 | 3.6 |
| KSRS |  |  |  |  |  |  |  |  |  |  |  |  |  |  |  |  |  |  |  |  |  |  |  |  |  |  |  |  |  |  |  |  |  |  |  |  |  |  |  |  |  |  |  |  |  |  |
| 20172661 | 6.8 |  | 53.7 | 6.8 | 26.2 | 33.1 | 14.4 | 11.0 | 12.6 | 8.3 | 23.6 | 15.6 | 8.3 | 10.0 | 15.5 | 12.4 | 5.8 | 7.1 | 12.6 | 17.9 | 5.4 | 7.2 | 8.6 | 5.7 | 14.3 | 9.6 | 7.1 | 4.9 | 3.4 | 7.7 | 6.3 | 9.3 | 4.0 | 3.9 | 4.5 | 19.3 | 5.5 | 11.2 | 6.7 | 5.0 | 9.5 | 4.4 | 4.3 | 7.9 | 8.0 | 7.2 |
| 20172662 | 11.0 | 45.0 |  | 11.0 | 27.7 | 35.1 | 14.0 | 12.3 | 11.5 | 6.8 | 24.3 | 16.8 | 11.8 | 12.0 | 25.1 | 16.7 | 7.2 | 9.8 | 15.9 | 21.2 | 7.4 | 10.6 | 11.7 | 11.3 | 16.3 | 20.4 | 17.8 | 4.7 | 6.5 | 8.1 | 7.6 | 11.3 | 4.8 | 3.3 | 6.4 | 26.4 | 5.8 | 12.6 | 7.1 | 2.4 | 10.2 | 6.0 | 6.9 | 11.5 | 12.6 | 9.3 |
| 20172851 | 7.9 | 46.0 | 55.4 | 7.9 | 26.3 | 43.3 | 18.1 | 11.4 | 15.4 | 10.8 | 32.4 | 24.1 | 8.8 | 10.3 | 15.1 | 12.7 | 5.4 | 9.5 | 13.4 | 22.7 | 5.4 | 7.9 | 11.4 | 8.2 | 17.2 | 12.4 | 12.2 | 7.0 | 3.9 | 9.7 | 8.1 | 12.4 | 4.7 | 4.1 | 5.5 | 25.0 | 7.1 | 12.7 | 6.8 | 3.3 | 10.2 | 4.0 | 5.2 | 10.5 | 11.1 | 9.4 |
| 20173041 | 2.6 | 2.0 | 2.1 | 2.6 | 2.5 | 2.0 | 1.9 | 2.6 | 2.3 | 4.0 | 2.1 | 2.9 | 2.2 | 2.0 | 2.3 | 2.1 | 3.0 | 2.4 | 3.0 | 2.4 | 3.1 | 3.0 | 2.8 | 2.2 | 2.9 | 2.6 | 2.3 | 2.2 | 2.0 | 2.0 | 2.4 | 2.4 | 2.4 | 2.2 | 2.4 | 2.1 | 2.6 | 2.5 | 2.7 | 10.9 | 3.1 | 2.6 | 2.6 | 3.1 | 2.6 | 3.0 |
| 20173351 | 3.6 | 25.2 | 31.7 | 3.6 |  | 24.4 | 9.4 | 8.8 | 11.3 | 4.8 | 20.1 | 11.4 | 6.1 | 9.2 | 13.1 | 9.2 | 5.8 | 4.2 | 8.1 | 9.3 | 3.7 | 3.7 | 5.6 | 5.8 | 7.2 | 11.0 | 11.6 | 3.5 | 3.9 | 6.7 | 5.0 | 8.6 | 4.0 | 3.6 | 2.8 | 12.6 | 5.3 | 9.7 | 5.4 | 4.0 | 11.1 | 3.5 | 4.4 | 8.3 | 8.1 | 6.1 |
| 20173391 | 7.3 | 34.8 | 38.1 | 7.3 | 23.5 |  | 14.4 | 8.6 | 17.2 | 10.8 | 35.0 | 21.2 | 8.4 | 11.1 | 15.1 | 12.5 | 6.4 | 5.9 | 10.1 | 15.1 | 3.8 | 6.2 | 8.9 | 7.6 | 10.8 | 11.5 | 10.2 | 6.9 | 4.2 | 8.1 | 7.4 | 10.4 | 4.4 | 3.6 | 4.3 | 24.1 | 5.9 | 12.4 | 7.1 | 5.5 | 13.2 | 3.3 | 5.0 | 10.4 | 11.2 | 11.7 |
| 20173401 | 3.7 | 11.9 | 12.6 | 3.7 | 9.9 | 16.0 | 8.1 | 3.7 |  | 4.0 | 16.4 | 8.5 | 3.3 | 3.7 | 7.0 | 5.6 | 3.1 | 2.4 | 3.6 | 4.5 | 2.1 | 2.5 | 3.1 | 3.8 | 4.7 | 6.1 | 5.4 | 2.9 | 4.2 | 4.0 | 2.8 | 4.9 | 2.8 | 2.4 | 2.3 | 8.9 | 2.5 | 5.5 | 3.1 | 2.9 | 6.6 | 2.9 | 3.6 | 7.1 | 5.7 | 4.2 |
| 20173431 | 2.8 | 5.8 | 6.5 | 2.8 | 4.4 | 8.8 | 5.5 | 2.6 | 3.8 |  | 11.6 | 9.8 | 2.6 | 3.1 | 4.4 | 4.9 | 3.3 | 2.4 | 3.8 | 2.8 | 1.6 | 2.3 | 2.9 | 2.2 | 3.4 | 3.7 | 2.2 | 2.2 | 1.9 | 3.5 | 2.9 | 2.9 | 1.9 | 2.3 | 2.6 | 5.7 | 3.0 | 2.4 | 3.2 | 2.2 | 3.2 | 3.1 | 3.1 | 4.6 | 3.7 | 3.3 |
| 20173432 | 8.6 | 16.1 | 20.3 | 8.6 | 10.0 | 33.6 | 17.4 | 7.6 | 14.3 | 14.5 |  | 39.1 | 9.1 | 11.2 | 19.3 | 17.8 | 5.9 | 4.3 | 6.0 | 4.2 | 3.3 | 3.4 | 3.5 | 5.2 | 7.3 | 10.6 | 2.3 | 9.0 | 2.8 | 11.7 | 3.5 | 6.0 | 4.2 | 4.2 | 3.5 | 18.4 | 8.3 | 5.6 | 8.2 | 7.4 | 15.8 | 3.2 | 10.2 | 15.1 | 16.3 | 10.9 |
| 20173433 | 5.5 | 6.5 | 7.7 | 5.5 | 3.5 | 14.2 | 7.5 | 5.0 | 5.4 | 10.0 | 19.8 |  | 4.1 | 7.5 | 10.3 | 9.9 | 3.0 | 2.8 | 2.7 | 2.8 | 3.3 | 2.8 | 1.9 | 2.4 | 3.0 | 5.2 | 2.4 | 6.2 | 3.3 | 4.6 | 2.3 | 3.1 | 2.6 | 2.9 | 2.3 | 8.7 | 3.9 | 3.1 | 4.2 | 5.0 | 8.0 | 3.3 | 6.2 | 12.7 | 14.6 | 6.9 |
| 20162551 | 7.9 | 6.3 | 8.5 | 7.9 | 4.0 | 5.4 | 3.2 | 2.5 | 2.7 | 3.3 | 7.4 | 4.7 | 3.8 | 2.6 | 4.0 | 3.4 | 2.4 | 4.3 | 6.0 | 8.1 | 3.6 | 2.9 | 2.4 | 2.8 | 6.0 | 4.1 | 3.3 | 2.6 | 2.3 | 3.2 | 2.2 | 4.5 | 2.5 | 2.4 | 3.8 | 6.8 | 2.9 | 2.9 | 2.8 | 2.7 | 2.5 | 4.3 | 2.8 | 3.4 | 3.3 | 3.3 |
| 20172462 | 4.4 | 5.0 | 7.4 | 4.4 | 4.3 | 4.7 | 3.2 | 2.8 | 2.4 | 3.0 | 6.1 | 3.5 | 3.3 | 2.3 | 2.9 | 2.5 | 2.9 | 4.5 | 5.8 | 7.1 | 2.7 | 3.0 | 2.3 | 1.5 | 5.2 | 3.0 | 2.2 | 2.1 | 2.3 | 3.2 | 3.0 | 3.7 | 2.6 | 1.9 | 2.8 | 4.8 | 2.5 | 3.1 | 2.6 | 2.9 | 2.6 | 4.5 | 2.8 | 3.0 | 3.5 | 2.9 |
| 20180361 | 2.4 | 3.5 | 7.0 | 2.4 | 2.9 | 2.7 | 2.5 | 2.5 | 2.2 | 2.4 | 3.9 | 3.1 | 2.7 | 2.7 | 3.4 | 3.1 | 2.5 | 2.5 | 4.2 | 5.9 | 2.5 | 2.9 | 2.4 | 2.6 | 4.6 | 3.5 | 2.3 | 2.1 | 2.7 | 2.6 | 2.0 | 2.6 | 2.6 | 2.4 | 2.6 | 3.7 | 1.7 | 2.7 | 2.3 | 2.5 | 2.3 | 3.2 | 3.2 | 2.7 | 2.7 | 3.9 |
| 20180362 | 2.7 | 3.1 | 3.6 | 2.7 | 2.2 | 3.2 | 2.8 | 2.9 | 2.7 | 2.3 | 2.3 | 2.7 | 2.2 | 1.9 | 2.8 | 2.5 | 2.7 | 2.6 | 3.1 | 4.9 | 2.6 | 2.4 | 1.5 | 1.8 | 4.8 | 2.3 | 2.4 | 3.4 | 2.6 | 2.7 | 2.1 | 2.0 | 1.9 | 2.7 | 2.1 | 4.2 | 2.1 | 2.0 | 2.3 | 2.2 | 2.5 | 3.0 | 2.8 | 3.2 | 2.5 | 4.3 |
| 20180363 | 2.7 | 3.1 | 4.5 | 2.7 | 2.6 | 2.2 | 2.7 | 2.0 | 2.2 | 2.4 | 3.4 | 2.6 | 2.1 | 2.4 | 1.9 | 2.7 | 1.7 | 3.1 | 4.7 | 4.0 | 2.5 | 2.2 | 2.3 | 2.1 | 3.8 | 2.6 | 3.8 | 2.6 | 2.2 | 3.5 | 2.7 | 2.8 | 2.6 | 1.8 | 2.5 | 2.4 | 2.3 | 2.8 | 2.5 | 2.6 | 2.5 | 3.4 | 3.1 | 3.1 | 2.5 | 5.4 |
| 20180371 | 2.5 | 2.6 | 4.2 | 2.5 | 2.5 | 2.9 | 2.3 | 1.6 | 1.8 | 2.3 | 2.0 | 2.2 | 1.7 | 2.2 | 1.9 | 2.2 | 2.6 | 1.9 | 2.5 | 2.5 | 2.3 | 2.5 | 2.0 | 2.6 | 3.0 | 3.0 | 2.4 | 2.2 | 2.6 | 2.3 | 2.7 | 2.2 | 2.8 | 1.9 | 2.1 | 2.3 | 2.8 | 2.3 | 2.6 | 2.6 | 2.3 | 2.8 | 2.4 | 4.0 | 3.1 | 3.1 |
| 20180372 | 2.5 | 2.2 | 3.5 | 2.5 | 2.0 | 2.5 | 1.7 | 2.2 | 2.1 | 2.6 | 2.3 | 2.7 | 2.2 | 2.0 | 3.2 | 2.1 | 3.2 | 2.8 | 3.4 | 2.7 | 2.2 | 2.7 | 2.8 | 2.3 | 2.6 | 2.5 | 2.1 | 2.2 | 2.6 | 2.5 | 2.4 | 1.8 | 2.1 | 2.3 | 2.3 | 2.3 | 2.1 | 2.8 | 2.8 | 2.2 | 2.4 | 3.0 | 3.8 | 2.8 | 3.0 | 4.3 |
| 20180373 | 5.9 | 9.5 | 15.9 | 5.9 | 5.6 | 8.0 | 4.0 | 3.5 | 4.1 | 2.5 | 9.5 | 5.3 | 4.0 | 2.8 | 7.0 | 5.3 | 2.9 | 5.7 | 11.6 | 15.7 | 4.5 | 4.9 | 4.1 | 3.4 | 11.4 | 6.4 | 3.4 | 3.7 | 2.6 | 6.4 | 3.7 | 5.3 | 3.8 | 3.0 | 4.7 | 11.1 | 3.6 | 4.5 | 3.6 | 2.6 | 5.7 | 9.8 | 3.8 | 5.0 | 5.6 | 4.6 |
| 20180381 | 9.2 | 14.3 | 24.4 | 9.2 | 7.4 | 10.6 | 4.7 | 5.1 | 4.6 | 2.3 | 9.3 | 5.9 | 5.0 | 4.2 | 9.8 | 6.4 | 2.8 | 6.4 | 16.3 | 19.9 | 6.8 | 8.4 | 4.4 | 3.9 | 11.0 | 8.1 | 4.5 | 5.1 | 3.2 | 6.8 | 2.8 | 7.2 | 3.6 | 3.7 | 4.2 | 12.0 | 2.9 | 5.1 | 2.8 | 3.4 | 5.8 | 9.9 | 3.7 | 5.4 | 6.0 | 4.5 |
| 20180391 | 3.8 | 5.8 | 10.9 | 3.8 | 4.0 | 3.3 | 2.3 | 2.5 | 2.4 | 2.5 | 3.8 | 2.9 | 3.0 | 1.9 | 3.2 | 3.9 | 2.1 | 2.5 | 8.0 | 9.3 | 2.7 | 3.1 | 4.3 | 2.6 | 4.5 | 4.2 | 2.5 | 2.8 | 2.2 | 4.6 | 2.5 | 2.9 | 1.9 | 2.6 | 2.4 | 5.5 | 3.0 | 2.4 | 2.9 | 3.2 | 3.5 | 4.1 | 3.2 | 3.4 | 4.1 | 3.6 |
| 20181121 | 4.1 | 4.7 | 8.0 | 4.1 | 3.3 | 4.0 | 3.1 | 2.3 | 2.9 | 2.6 | 4.8 | 3.0 | 3.0 | 2.1 | 3.0 | 2.9 | 2.1 | 4.6 | 5.0 | 7.8 | 2.3 | 2.1 | 2.9 | 2.0 | 4.8 | 2.8 | 2.0 | 2.3 | 2.1 | 3.4 | 2.4 | 3.0 | 1.8 | 2.1 | 2.5 | 4.3 | 2.8 | 2.5 | 2.1 | 2.6 | 3.3 | 4.5 | 2.9 | 3.3 | 2.2 | 3.3 |
| 20181122 | 5.9 | 10.7 | 16.8 | 5.9 | 7.6 | 10.7 | 5.3 | 4.0 | 4.6 | 2.3 | 11.3 | 6.4 | 4.9 | 3.9 | 7.7 | 5.6 | 3.1 | 6.8 | 12.9 | 17.8 | 5.3 | 6.2 | 4.8 | 4.5 | 13.5 | 6.9 | 4.0 | 5.0 | 3.2 | 7.8 | 3.4 | 6.4 | 3.2 | 4.1 | 4.8 | 11.4 | 3.5 | 4.5 | 3.0 | 2.5 | 6.2 | 10.0 | 4.3 | 5.8 | 6.7 | 4.6 |

Figure S3. SNRcc in a) DPRK5 (32 aftershocks), and b) DPRK6 (46 aftershocks) clusters. Colour formatting for a given aftershock (vertical column) at two stations together. The level of similarity almost the same at both stations.